\documentclass{aa}
\usepackage{graphicx}
\usepackage{natbib}
\usepackage{textcomp}
\usepackage{amsmath} 
\newcommand{\angstrom}{\textup{\AA}}
\bibpunct{(}{)}{;}{a}{}{,} 
\usepackage{amsmath}	
\usepackage{amssymb}	

\begin{document}

\title{What is the origin of the stacked radio emission in radio-undetected
quasars? }
\subtitle{Insights from a radio-infrared image stacking analysis}

\author{E. Retana-Montenegro\inst{1,2} }

\offprints{E. Retana-Montenegro}

\institute{Leiden Observatory, Leiden University, P.O. Box 9513, 2300 RA, Leiden,
The Netherlands \and Astrophysics and Cosmology Research Unit, School of Mathematics, Statistics and Computer Science, University of KwaZulu-Natal, Durban 4041, South Africa\\
\email{edwinretana@gmail.com}\\
}

\date{Received June xx, xxxx; accepted March xx, xxxx}


\abstract{Radio emission in the brightest radio quasars can be attributed to
processes inherent to active galactic nuclei (AGN) powered by super
massive black holes (SMBHs), while the physical origins of the radio
fluxes in quasars without radio detections have not been established
with full certainly. Deep radio surveys carried out with the Low Frequency
ARray (LOFAR) are at least one order of magnitude more sensitive for
objects with typical synchrotron spectra than previous wide-area high-frequency
surveys $(>1.0\,\textrm{GHz})$. With the enhanced sensitivity that
LOFAR offers, we investigate the radio-infrared continuum of LOFAR
radio-detected quasars (RDQs) and LOFAR radio-undetected quasars (RUQs)
in the $9.3\:\textrm{deg}^{2}$ NOAO Deep Wide-field survey (NDWFS)
of the Bo\"otes field; RUQs are quasars that are individually undetected
at a level of $\geq5\sigma$ in the LOFAR observations. To probe the
nature of the radio and infrared emission, where direct detection
is not possible due to the flux density limits, we used a median image
stacking procedure. This was done in the radio frequencies of 150
MHz, 325 MHz, 1.4 GHz and 3.0 GHz, and in nine infrared bands between
$8$ and $500\;\mu\textrm{m}$. The stacking analysis allows us to
probe the radio-luminosity for quasars that are up to one order of
magnitude fainter than the ones detected directly. The radio and infrared
photometry allow us to derive the median spectral energy distributions
of RDQs and RUQs in four contiguous redshift bins between $0<z<6.15$.
The infrared photometry is used to derive the infrared star-formation
rate (SFR) through SED fitting, and is compared with two independent
radio-based star-formation (SF) tracers using the far-infrared radio
correlation (FIRC) of star-forming galaxies. We find a good agreement between
our radio and infrared SFR measurements and the predictions of the
FIRC. Moreover, we use the FIRC predictions to establish the level
of the contribution due to SMBH accretion to the total radio-luminosity.
We show that SMBH accretion can account for $\sim5-41\%$ of the total
radio-luminosity in median RUQs, while for median RDQs the contribution
is $\sim50-84\%$. This implies that vigorous SF activity is coeval
with SMBH growth in our median stacked quasars. We find that median
RDQs have higher SFRs that agree well with those of massive star-forming main
sequence galaxies, while median RUQs present lower SFRs than RDQs.
Furthermore, the behavior of the radio-loudness parameter
${\displaystyle \left(R=\log_{\textrm{10}}\left(L_{\textrm{rad}}/L_{\textrm{AGN}}\right)\right)}$
is investigated. For quasars with $R\geq-4.5$, the radio-emission
is consistent with being dominated by SMBH accretion, while for low
radio luminosity quasars with $R<-4.5$ the relative contribution
of SF to the radio fluxes increases as the SMBH component becomes
weaker. We also find signatures of SF suppression due to negative
AGN feedback in the brightest median RDQs at 150 MHz. Finally, taking
advantage of our broad spectral coverage, we studied the radio spectra
of median RDQs and RUQs. The spectral indices of RUQs and RDQs do
not evolve significantly with redshift, but they become flatter towards
lower frequencies. }
\maketitle

\section{Introduction \label{sec:intro}}

Even though more than five decades have passed since the discovery
of the first radio quasar by \citet{1963Natur.197.1040S}, the physical
processes responsible for the radio emission in quasars are not fully
understood. The radio emission in the brightest radio quasars, known
as radio-loud quasars (RLQs), is associated with the accretion onto
supermassive black holes (SMBHs). The accretion energy output often
manifests as large-scale structures (radio jets and lobes). The origins
of the weaker radio emission in the fainter radio quasars, called
radio-quiet quasars (RQQs), are still uncertain. It has been suggested
that the radio activity in RQQs is produced by a starburst. It is
thought that the starburst radio-emission has a synchrotronic component
produced by electrons accelerated by supernovae remnants, and a thermal
free-free component arising from the ionization of hydrogen clouds
(HII regions) by hot massive stars \citep{1991MNRAS.251..112S,1993MNRAS.262..491T}.
Conversely, it had been proposed that the radio emission in RQQs is
caused by small-scale radio jets with a kinetic power that is significantly
lower $(\sim1000)$ than those of RLQs \citep{1993MNRAS.263..425M}.
This scenario is supported by high-resolution observations of RQQs,
where the brightness temperatures and jet-like structure of the radio
emission suggest that the radio emission in RQQs has a non-thermal
origin \citep{Blundell_1996,2005ApJ...621..123U,2006A&A...455..161L,2016A&A...589L...2H}.
Finally, in recent years the possibility that the radio emission in
RQQs is driven by magnetically heated accretion disk coronae \citep{2008MNRAS.390..847L,2019MNRAS.482.5513L}
and winds from active galactic nuclei (AGN) \citep{2014MNRAS.442..784Z,2016MNRAS.455.4191Z,2018MNRAS.477..830H}
has been explored.

The stacking technique combines the signal of many individual sources
that have been previously identified in other wavelength observations,
some of these sources may be below the noise level (and therefore
have gone undetected) in a particular survey. The co-addition brings
down the statistical noise, enabling statistical flux measurements
of sources below the original detection threshold. Radio image stacking
has previously been carried out to study quasars that remain silent
in flux density limited surveys due to their low radio power. For
example, \citet{2007ApJ...654...99W} analyzed the stacked FIRST $1.4\,\textrm{GHz}$
radio images of $\sim41000$ RQQs positions from the SDSS DR3 survey
\citep{2005AJ....130..367S} with $0<z<5$, which resulted in a median
flux density of $110\:\mu\textrm{Jy}$. \citet{2005MNRAS.360..453W}
co-added FIRST images of 8000 RQQs of the 2QZ survey \citep{2004MNRAS.349.1397C},
and found median fluxes of $20-40\:\mu\textrm{Jy}$. \citet{2018MNRAS.477..830H}
found similar flux values by stacking VLA maps of red quasars at $1.4\,\textrm{GHz}$
and $6.0\,\textrm{GHz}$. \citet{2015MNRAS.448.2665W} using a stacking
analysis of photometrically and spectroscopically selected quasars
with $0<z<3.1$ in the VISTA Deep Extragalactic Observations (VIDEO,
\citealt{2013MNRAS.428.1281J}) survey found that the radio emission
of the quasars in their sample is likely caused by accretion activity.
Finally, \citet{2019MNRAS.490.2542P} found a $1.4\,\textrm{GHz}$
flux density of $52\:\mu\textrm{Jy}$ for 2229 RQQs with $4.0<z<7.5$
using FIRST radio maps.

An important point to consider in the study of the radio emission
of RQQs is the observed frequency. Most of the radio stacking analyses
of RQQs have been carried out using high-frequency observations $(>1.0\,\textrm{GHz})$.
With new low-frequency radio interferometer arrays such as the Low
Frequency Array (LOFAR; \citealt{2013A&A...556A...2V}), we are able
to investigate the radio emission of RQQs at very low-frequencies
exploring a new parameter space. A number of quasar studies using
LOFAR observations have been done in the last few years. For instance,
\citet{2019A&A...622A..11G} found that the radio emission at $150\,\textrm{MHz}$
in low-luminosity quasars is consistent with being dominated by star
formation (SF). \citet{2019A&A...622A..15M} studied the low-frequency
radio properties of broad absorption line quasars. \citet{2018FrASS...5....5R}
investigated the selection of high-z quasars using LOFAR observations.
\citet{2020AA} investigated the optical luminosity function of LOFAR
radio-selected quasars (RSQs) at $1.4<z<5.0$ in the NDWFS-Bo\"otes
field using deep LOFAR observations of the NDWFS-Bo\"otes field \citep{2018A&A...620A..74R}.
They found that RSQs show an evolution that is very similar to the
exhibited by faint quasars $(M_{\textrm{1450}}\leq-22)$, and that
RSQs may compose up to $\sim20\%$ of the whole faint quasar population. 

Another point of intense study in quasars is to understand the physical
origin of the difference between RLQs and RQQs. For this purpose,
the radio-loudness parameter, defined as the ratio of radio to optical
quantities (fluxes or luminosities), was introduced to classify quasars
as radio-loud or radio-quiet (e.g., \citealt{1989AJ.....98.1195K,2002AJ....124.2364I}).
Across the literature, there is no clear consensus on the radio-loudness
limit for classifying quasars. Moreover, the calculation of the ratio
depends on the radio and optical bands available, and these bands
tend to vary between the studies. This has led to discrepancies between
the radio-loudness studies: some authors found that radio-loudness
distribution for optical-selected quasars is bimodal \citep{1990MNRAS.244..207M,2007ApJ...656..680J,2007ApJ...654...99W},
while others have confirmed a very broad range for the radio-loudness
parameter, and question its bimodal nature \citep{2003MNRAS.341..993C,2011ApJ...743..104S,2012ApJ...759...30B}.
In particular, recent LOFAR studies have found that the radio properties
of quasars show a continuous distribution rather than a bimodal distribution
\citep{2019A&A...622A..11G,2020MNRAS.494.4802F}. 

The study of the radio emission in quasars is particularly important
for understanding the role of AGN feedback in suppressing or, alternatively,
enhancing of SF. The hosts of luminous quasars $(L_{\textrm{X}}>10^{44}\,\textrm{erg}\textrm{s}^{-1})$
at $z>1$ are known to be sites of intense SF \citep{2016MNRAS.457.4179H,2016MNRAS.462.4067P,2016ApJ...824...70D,2017A&A...604A..67D},
however, these studies contain an important number of non-detections
and are restricted only to objects with the highest star formation
rates (SFRs). On the other hand, the results provided by studies of
moderate luminosity AGN $(L_{\textrm{bol}}\sim10^{43}-10^{44}\,\textrm{erg}\textrm{s}^{-1})$
provide no clear evidence for the AGN influence on the SF of the host
galaxy (e.g., \citealt{2012A&A...546A..58R,2015MNRAS.453..591S,2017ApJ...835...27A}).
Recent deep observations with ALMA and SCUBA have found that AGN hosts
have SFRs lower than those predicted by the SF main sequence \citep{2007ApJ...660L..43N,2011A&A...533A.119E,2012ApJ...754L..29W,2015A&A...575A..74S}.
This might be an indication for the suppression of SF due to AGN feedback,
namely, quenching. The investigation of the properties of radio-detected
quasars could provide useful insights of the relationships between
SMBHs and the SF on their host galaxies. 

In this paper, we use radio and infrared observations of the NDWFS-Bo\"otes
field to investigate the origins of the radio emission in LOFAR radio-detected
quasars (RDQs) and LOFAR radio-undetected quasars (RUQs). RUQs are
defined as quasars that are individually undetected at $\geq5\sigma$
on the LOFAR map. The NDWFS-Bo\"otes field has a wealth of multi-wavelength
datasets with infrared coverage provided by Spitzer, WISE, and Herschel
observations along with GMRT, WSRT, and VLA radio imaging at 325 MHz
\citep{2015MNRAS.450.1477C}, 1.4 GHz \citep{2002AJ....123.1784D},
and 3.0 GHz \citep{2020PASP..132c5001L}, respectively. This is complemented
with deep LOFAR observations at 150 MHz \citep{2018A&A...620A..74R}.
The key part of our work is that we are using radio maps at low- and
high- frequencies that are deep enough to obtain a statistical measurement
of the flux densities of RUQs using a stacking analysis. This work
will help to address the following questions regarding the radio emission
in quasars: 1) is the radio-emission of quasars powered by SF or SMBH
accretion?; 2) how similar are the radio spectra indices of RDQs and
RUQs?; 3) what is the behaviour of the radio-loudness parameter at
low radio luminosities?; 4) can we find signatures of AGN feedback
with our data?.

This paper is organized as follows. In Section \ref{sec:bootes_data},
we present the data used in this work. In Section \ref{sec:quasar_sample},
we introduce our spectroscopic quasar sample, and we discuss our stacking
analysis in Section \ref{sec:stacking_method}. Section \ref{sec:results}
presents the results of our stacking analysis. In Section \ref{subsec:discussion},
we discuss our findings. Finally, we summarize our conclusions in
Section \ref{subsec:conclusions}. Through this paper, we use a $\Lambda$
cosmology with the matter density $\Omega_{m}=0.30$, and the cosmological
constant $\Omega_{\Lambda}=0.70$, the Hubble constant $H_{0}=70\,\textrm{km}\,\textrm{s}^{-1}\,\textrm{Mpc}^{-1}$.
We assume a definition of the form $S_{\nu}\propto\nu^{-\alpha}$,
where $S_{\nu}$ is the source flux, $\nu$ the observing frequency,
and $\alpha$ the spectral index. To estimate the radio luminosities,
we adopted a radio spectral index of $\alpha=0.7$. The optical luminosities
were calculated using a power-law continuum index of $\epsilon_{\textrm{opt}}=0.5$.
All the magnitudes are expressed in the AB magnitude system \citep{1983ApJ...266..713O}
and are corrected for Galactic extinction using the prescription by
\citet{2011ApJ...737..103S}.

\section{Data \label{sec:bootes_data}}

In this section, we introduce the NDWFS-Bo\"otes datasets that will
be utilized in our analysis. A summary of the radio and infrared data
used in this work is provided in Table \ref{fig:summary_data_table}.
\begin{table*}[htp]
\begin{centering}
\begin{tabular}{ccc}
\hline 
Radio-Telescope & Frequency {[}MHz{]} & Resolution {[}arcsec{]}\tabularnewline
\hline 
LOFAR & 150 & $3.98\times6.45$\tabularnewline
VLA-PB & 325  & $5.6\times5.1$\tabularnewline
WSRT & 1400  & $13\times27$\tabularnewline
VLASS & 3000  & $2.5\times2.5$\tabularnewline
\hline 
Space Telescope & Wavelength {[}$\mu$m{]} & Resolution {[}arcsec{]}\tabularnewline
\hline 
Spitzer/IRAC-Ch1 & 3.6 & 2\tabularnewline
Spitzer/IRAC-Ch2 & 4.5 & 2\tabularnewline
Spitzer/IRAC-Ch3 & 5.8 & 2\tabularnewline
Spitzer/IRAC-Ch4 & 8.0 & 2\tabularnewline
WISE/W3 & 12 & 6.5\tabularnewline
Spitzer/MIPS-Band1 & 24 & 6\tabularnewline
Spitzer/MIPS-Band2 & 70 & 20\tabularnewline
Herschel/PACS-Band1 & 100 & 6.8\tabularnewline
Herschel/PACS-Band2 & 160 & 11\tabularnewline
Herschel/SPIRE-Band1 & 250 & 18\tabularnewline
Herschel/SPIRE-Band2 & 350 & 25\tabularnewline
Herschel/SPIRE-Band3 & 500 & 37\tabularnewline
\hline 
\end{tabular}
\par\end{centering}
\centering{}\centering\caption{Summary of the radio and infrared data used in this work. \label{fig:summary_data_table}}
\end{table*}

\subsection{Optical data}

\noindent The NOAO Deep Wide-field survey (NDWFS) is a deep imaging
survey in the northern sky \citep{1999ASPC..191..111J}. A region
of $9.3\;\textrm{deg}^{2}$ towards the Bo\"otes constellation, the
Bo\"otes field, was originally observed in the $B_{W}$, $R$, and
$I$ optical bands, and subsequently imaging in the $U_{\textrm{spec}}$
\citep{2013ApJ...774...28B} and $Z_{\textrm{Subaru}}$ bands was
obtained. We use the I-band matched photometry catalog presented by
\citet{2007ApJ...654..858B}. The $5\sigma$ limiting AB magnitudes
for the optical filters are $U_{\textrm{spec}}=25.2$, $B_{w}=25.4$,
$R=25.0$, $I=24.9$, and $Z_{\textrm{Subaru}}=24.1$, respectively.
Additionally, the Bo\"otes field has also targeted as part of the
observations of the $3\pi$ survey performed with the 1.8m Pan-STARRS1
telescope \citep{2004SPIE.5489..667H}. The $3\pi$ survey \citep{2016arXiv161205560C}
observed for almost four years the sky north of $-30^{\circ}$ declination,
reaching $5\sigma$ limiting magnitudes in the $g_{\textrm{PS}},r_{\textrm{PS}},i_{\textrm{PS}},z_{\textrm{PS}},y_{\textrm{PS}}$
bands of 23.3, 23.2, 23.1, 22.3, 21.3, respectively.

\subsection{Radio }

The NDWFS-Bo\"otes field has been surveyed with various radio-telescopes
(e.g., VLA, WSRT, LOFAR) providing an excellent coverage at low- and
high- frequencies. The $150\,\textrm{MHz}$ LOFAR observations have
been described in \citet{2018A&A...620A..74R}. These observations
reach a central rms noise of $55\:\mu\textrm{Jy/beam}$ with an angular
resolution of $3.98^{''}\times6.45^{''}$. The NDWFS-Bo\"otes region
was also observed with the VLA in the P-band (VLA-PB) with a central
frequency of $\sim325\,\textrm{MHz}$ \citep{2015MNRAS.450.1477C}.
The VLA-PB mosaic has an angular resolution $5.6^{''}\times5.1^{''}$
with a rms noise of $0.2\:\textrm{mJy}$/beam. The NDWFS-Bo\"otes
region was also observed with the VLA at $3000\,\textrm{MHz}$, as
part of the VLA Sky Survey (VLASS, \citealt{2020PASP..132c5001L}).
These observations correspond only to one of the three planned VLASS
epochs to be observed. The VLASS mosaic has a resolution of $2.5^{''}$
with a sensitivity of $140\:\mu\textrm{Jy/beam}$. We also use the
$1400\,\textrm{MHz}$ WSRT observations of the NDWFS-Bo\"otes presented
by \citet{2002AJ....123.1784D}. The WSRT maps cover approximately
$7.0\;\textrm{deg}^{2}$ of the NDWFS-Bo\"otes field, and have a
sensitivity of $28\:\mu\textrm{Jy/beam}$ with an angular resolution
of $13^{''}\times27^{''}$. Due to the smaller size of the WSRT map,
we limit our stacking analysis only to quasars that are located within
the footprint of the WSRT-Bo\"otes observations. The footprints of
the different radio surveys are shown in Figure \ref{fig:footprint_surveys}.

\subsection{Mid-infrared and far-infrared data \label{sec:fir_ir_data}}

The mid-infrared Spitzer Infrared Array Camera (IRAC, \citealt{2004ApJS..154...10F})
imaging used in this paper comes from the publicly available images\footnote{https://irsa.ipac.caltech.edu/data/SPITZER/SDWFS/}
from the Spitzer Deep, Wide-field Survey (SDWFS, \citealt{2009ApJ...701..428A})
of the NDWFS-Bo\"otes field. The IRAC channels have a field of view
of ${\sim}5^{\textdegree}$ and a spatial resolution of
$2^{\prime\prime}$. Additionally, we employ Spitzer Multiband Imaging
Photometer (MIPS, \citealt{2004ApJS..154...25R}) imaging at 24 and
70 $\mu\textrm{m}$. At MIPS-$24\,\mu$m, the resolution is about
$6^{\prime\prime}$, and at MIPS-$70\,\mu$m is about $20^{\prime\prime}$.
We retrieved the MIPS data from the Spitzer Heritage Archive\footnote{https://sha.ipac.caltech.edu/applications/Spitzer/SHA/}
(Program ID: 50148), and the individual pointings were reduced and
mosaiced using the MOPEX package \citep{2006SPIE.6274E..0CM} following
a standard calibration (e.g., \citealt{2007ApJS..172...86S}). The
IRAC-$8\,\mu$m band reaches a $5\sigma$ sensitivity of $30\:\mu\textrm{Jy}$,
while the MIPS data reach depths ($1\sigma$) of $51\:\mu\textrm{Jy}$
and 5 mJy, at 24 and 70 $\mu\textrm{m}$, respectively. To cover the
gap between the IRAC-$8\,\mu$m and MIPS-$24\,\mu$m filters, we added
to our analysis the WISE-W3 band \citep{2010AJ....140.1868W} with
a central wavelength of $\lambda_{c}=12\:\mu\textrm{m}$. We downloaded
the WISE-W3 images that cover the WSRT-Bo\"otes region from the IRSA
archive\footnote{https://irsa.ipac.caltech.edu/applications/wise/}
and created a mosaic using the Montage\footnote{http://montage.ipac.caltech.edu/}
package.

\begin{figure}[tp]
\centering{}\includegraphics[clip,scale=0.3]{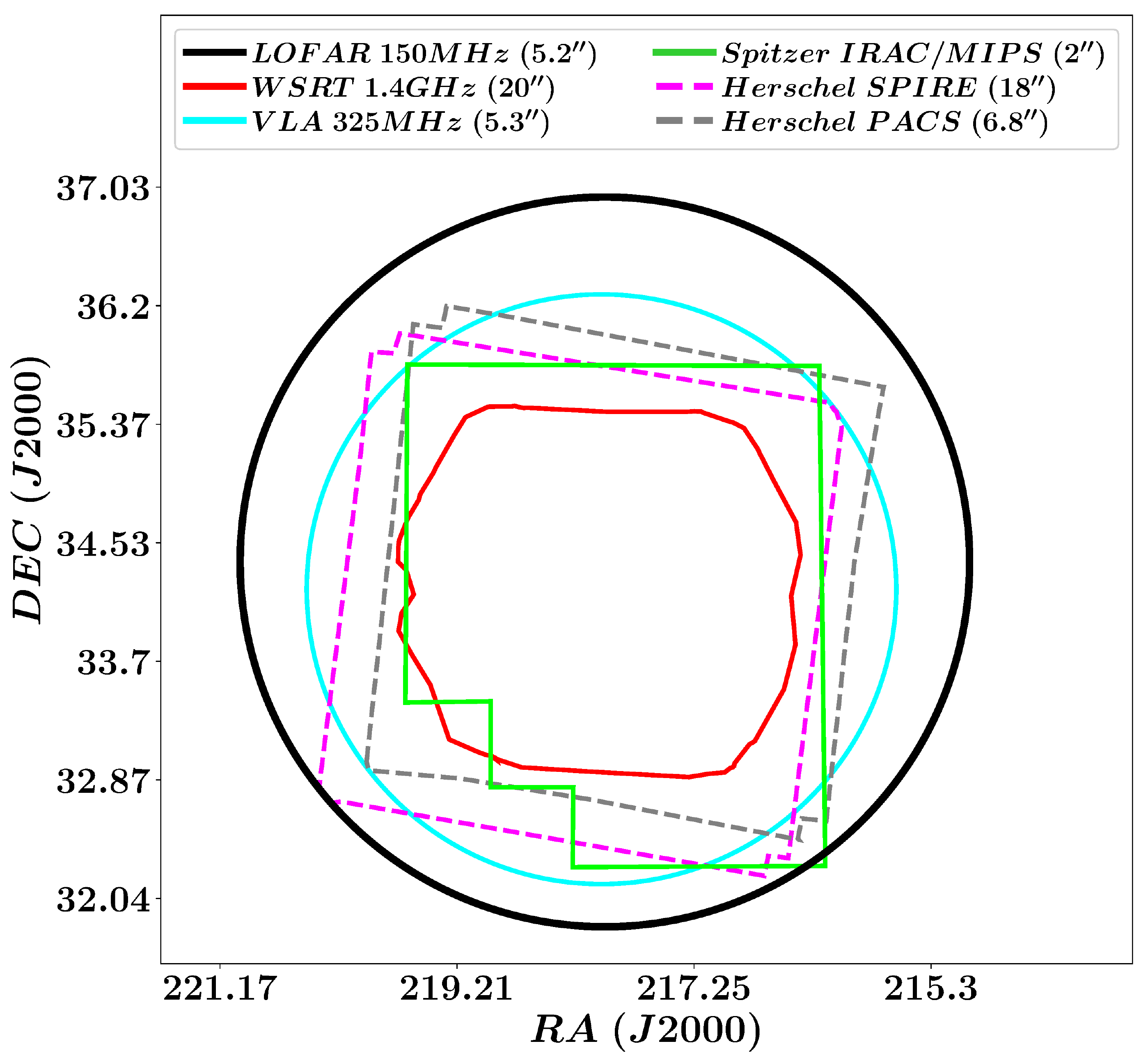}\centering\caption{\label{fig:footprint_surveys} Footprints of the various surveys of
the NDWFS-Bo\"otes field used in this work. The beam size for each
instrument is also indicated. The 3.0 GHz VLASS observations overlap
with the LOFAR region. The analysis in this work is limited to the
region with WSRT coverage. }
\end{figure}

Our far-infrared photometry comes from the Level 6 Herschel Multi-tiered
Extragalactic Survey\footnote{https://hedam.lam.fr/HerMES/} (HerMES
\citealt{2012MNRAS.424.1614O}). in the NDWFS-Bo\"otes field. HerMES
has a $5\sigma$ sensitivity of about 25.8, 21.2 and 30.8 mJy at 250,
350 and 500 $\mu\textrm{m}$, respectively with the Spectral and Photometric
Imaging Receiver (SPIRE, \citealt{2010A&A...518L...3G}). The Photodetector
Array Camera and Spectrometer (PACS, \citealt{2010A&A...518L...2P})
has a $5\sigma$ sensitivity of 49.9 and 95.1 mJy at 100 and 160 $\mu\textrm{m}$..
SPIRE provides images having angular resolution $\sim18^{\prime\prime}$,
$25^{''}$ and $37^{''}$ at 250, 350 and 500 $\mu\textrm{m}$, respectively.
The angular resolution of PACS is $6.8^{''}$ and $11^{''}$ at 100
and 160 $\mu\textrm{m}$, respectively. The footprints of the HerMES
SPIRE/PACS observations are shown in Figure \ref{fig:footprint_surveys}.
We used the HerMES xID24 catalogs, which provide SPIRE and PACS photometry
for objects whose positions are taken from catalogs extracted from
MIPS-$24\,\mu$m maps as described by \citet{2010MNRAS.409...48R}. 

The far-infrared maps available for NDWFS-Bo\"otes field are shallower
in comparison with other extragalactic fields such as COSMOS and GOODs
\citep{2012MNRAS.424.1614O}. In the  $250\,\mu\textrm{m}$band, $59\%$
($25\%$) of the RDQs (RUQs) are detected at $\geq3\sigma$, while
$92\%$ ($78\%$) are detected with a $2\sigma$ significance. To
avoid overly complicating the analysis, we set the detection threshold
in the SPIRE-$250\,\mu\textrm{m}$ band to $2\sigma$, and classify
these quasars as SPIRE-$250\,\mu\textrm{m}$ detected in Sections
\ref{sec:results}  and \ref{subsec:discussion}, respectively. This
threshold allow us to maximize the number of quasars with FIR detections,
while simultaneously keeping the number of likely non-detections low.

\noindent 

\section{Quasar sample \label{sec:quasar_sample}}

\begin{figure}[tp]
\centering{}\includegraphics[bb=0bp 0bp 788bp 771bp,clip,scale=0.33]{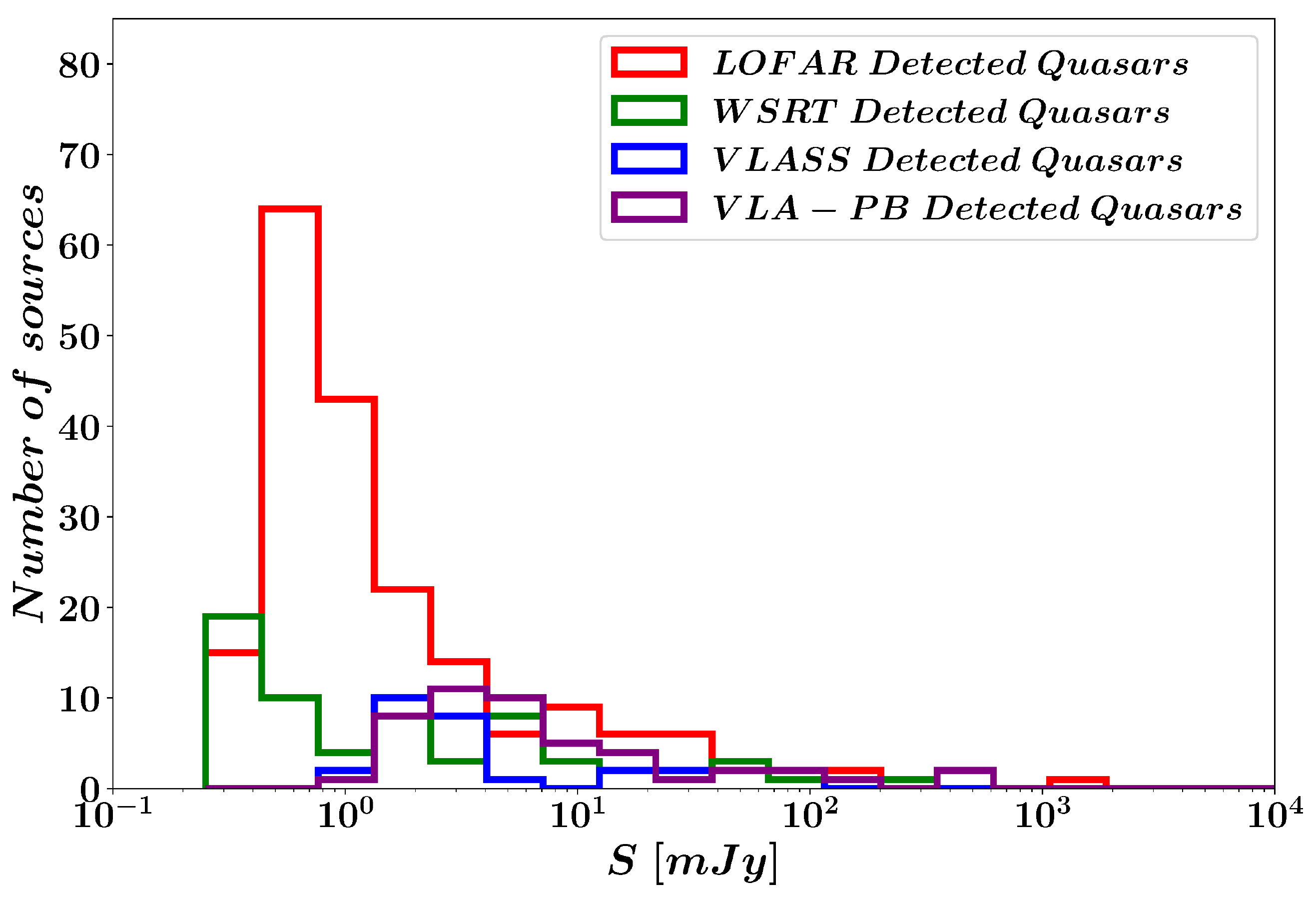}\centering\caption{\label{fig:radio_flux_distribution} Total flux $S_{\textrm{150MHz}}$
distributions of LOFAR radio-detected quasars.}
\end{figure}

\begin{figure}[tp]
\begin{centering}
\includegraphics[clip,scale=0.33]{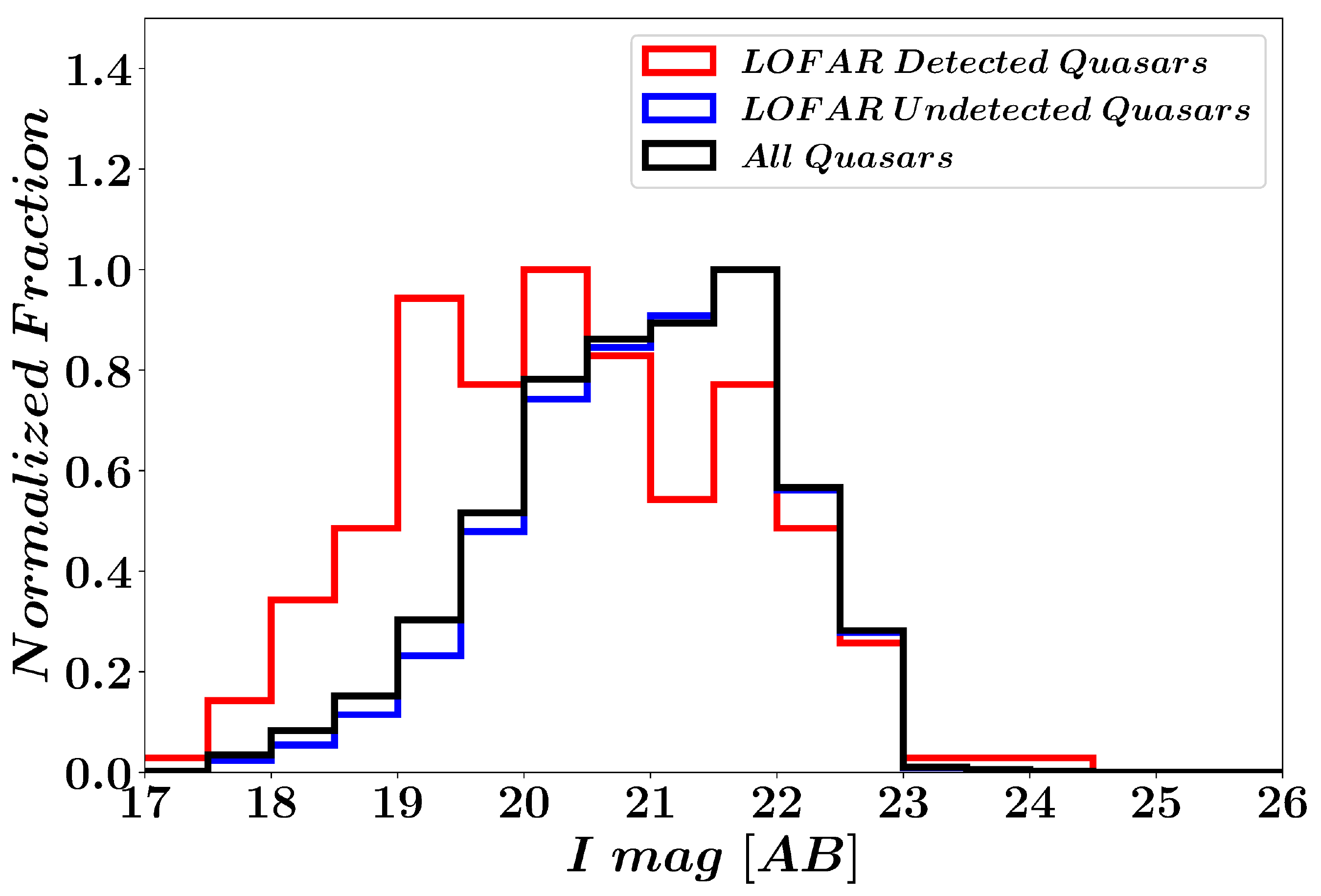}
\par\end{centering}
\centering{}\centering\caption{\label{fig:mag_flux_distribution} Maximum-value normalized $I$-band
distributions of LOFAR radio-detected quasars (RDQs, red) and LOFAR
radio-undetected quasars (RUQs, blue). Also, the total combined redshift
distribution of RDQs and RUQs (black) is plotted.}
\end{figure}

\begin{figure}[tp]
\centering{}\includegraphics[bb=0bp 0bp 772bp 764bp,clip,scale=0.33]{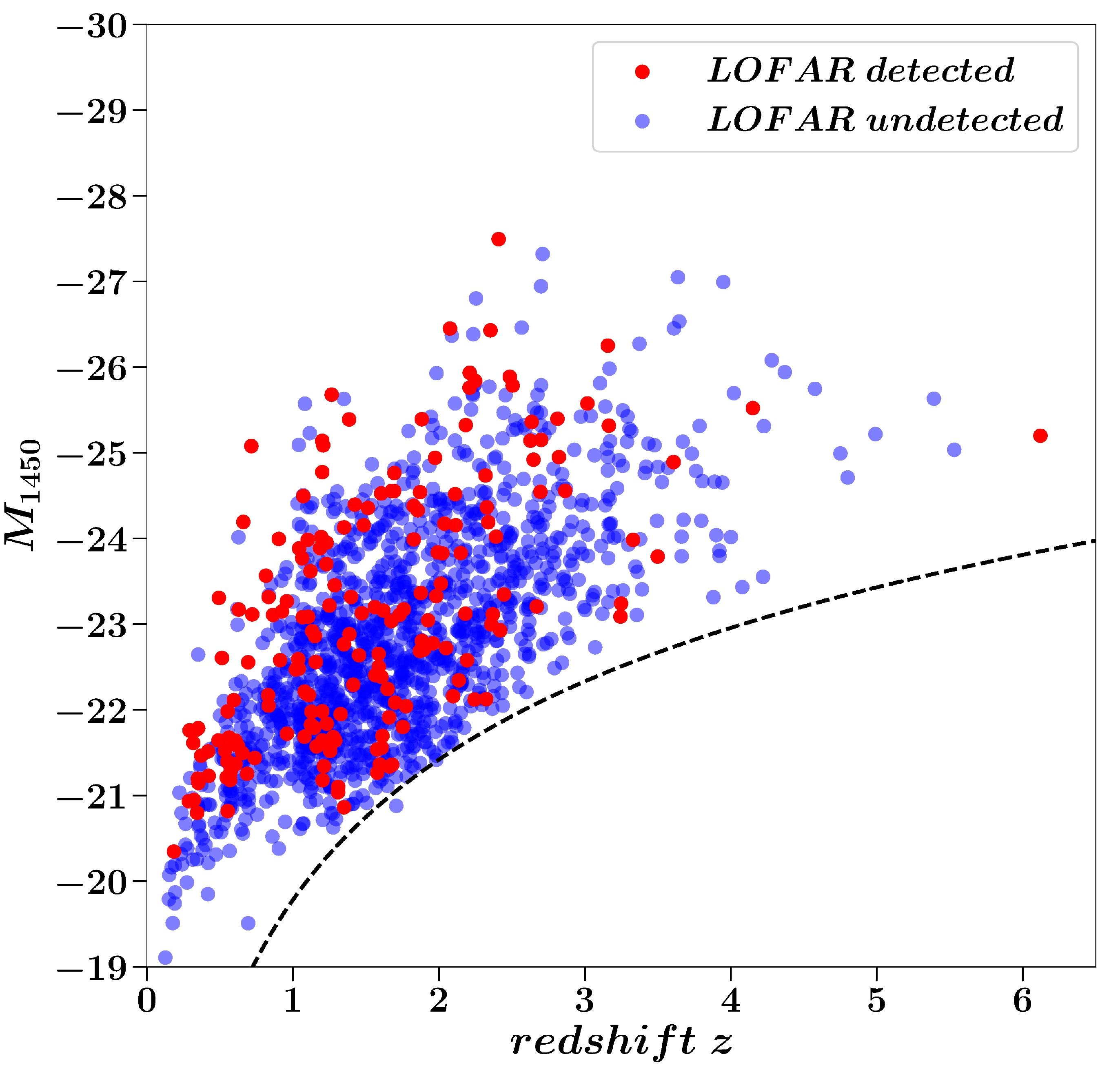}
\centering\caption{\label{fig:M1450_redshift} Absolute magnitude as rest-frame $1450 \> {\angstrom}$,
$M_{\textrm{1450}}$, versus redshift for the LOFAR radio-detected
quasars (red) and LOFAR radio-undetected quasars (blue) samples. The
dashed line denotes the magnitude limit $i_{\textrm{PS}}=23.0$. This
limit is calculated assuming a quasar continuum described by a power-law
with slope $\alpha=-0.5$, and with no emission line contribution
or $\textrm{Ly}_{\alpha}$ forest blanketing.}
\end{figure}

\begin{figure}[tp]
\centering{}\includegraphics[bb=0bp 0bp 788bp 771bp,clip,scale=0.33]{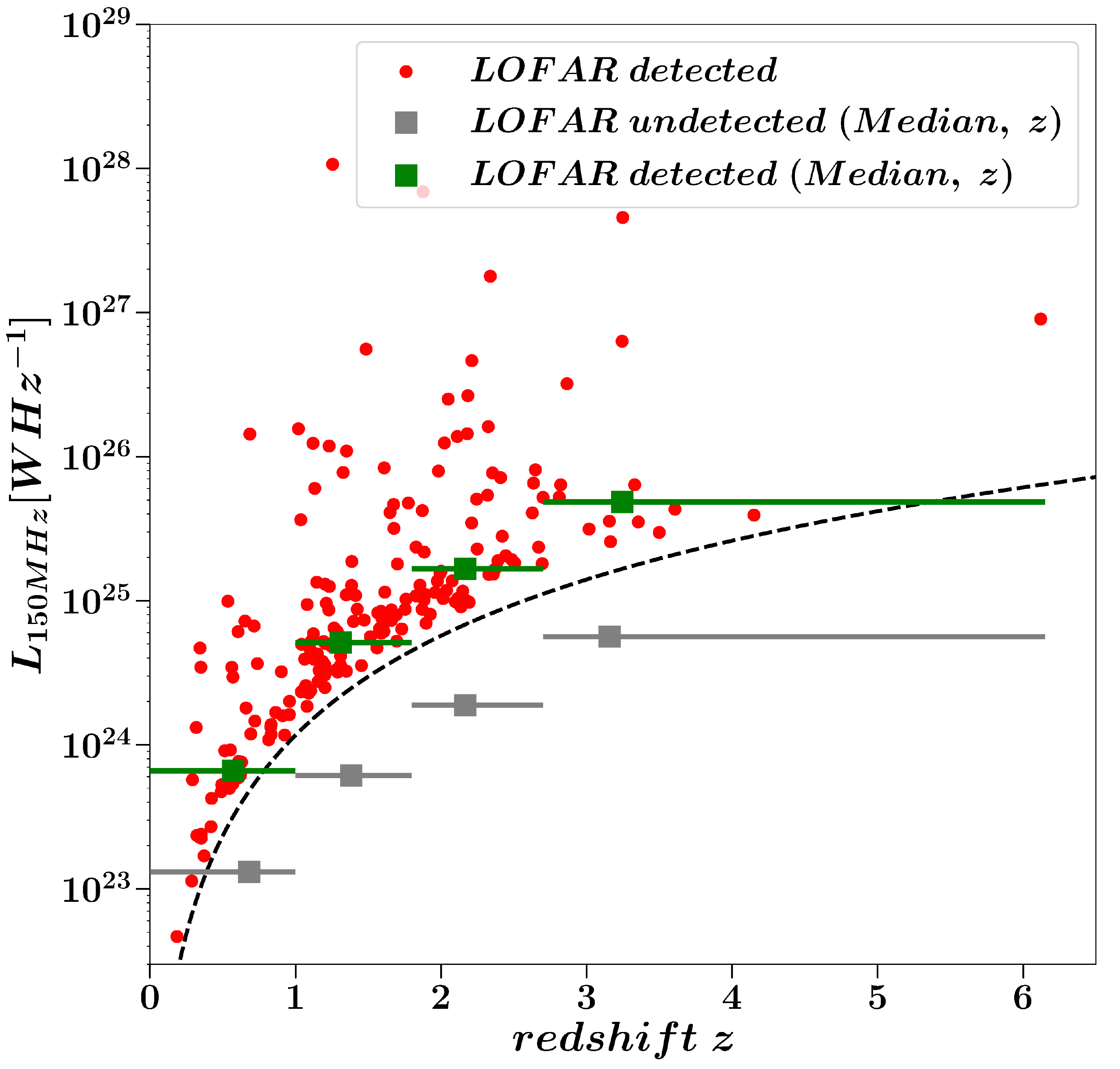}\centering\caption{\label{fig:L150_redshift} Rest-frame radio luminosity density at
150 MHz versus redshift for LOFAR radio-detected quasars (red) in
our sample. The green (LOFAR radio-detected quasars) and gray squares (LOFAR radio-undetected quasars)
are the radio luminosities obtained from the stacking analysis in
Section \ref{sec:stacking_method}. The dashed line denotes the $5\sigma$
flux limit ($275\:\mu\textrm{Jy}$) of the NDWFS-Bo\"{o}tes observations
presented by \citet{2018A&A...620A..74R}. The stacking analysis allows
us to probe the radio-luminosity for quasars that are up to one order
of magnitude fainter than the ones detected individually.}
\end{figure}

The starting point to create our sample is the Milliquas catalog from
\citet{2015PASA...32...10F}. This catalog contains all the known
NDWFS-Bo\"otes spectroscopic quasars from the literature (e.g., \citealt{2012ApJS..200....8K,2018A&A...613A..51P,2019ApJS..240....6Y}).
We remark that this catalog is not uniform as it is composed of various
surveys with different selection criteria. The majority of quasars
in the NDWFS-Bo\"otes sample are type-1, thus, for this reason we
use only type-1 spectroscopic quasars in our analysis. We restrict
the sample to quasars with positions falling into the WSRT-Bo\"otes
footprint. To establish whether a quasar is detected in the radio
or not, we use the $150\,\textrm{MHz}$ LOFAR catalog by \citet{2018A&A...620A..74R}.
We cross-matched both catalogs using a matching radius of $2^{\prime\prime}$,
and found that 235 of 1574 quasars are detected at $5\sigma$ significance
by LOFAR, while 1339 quasars are not detected in the LOFAR maps. From
this point, we call the LOFAR detected quasars as radio-detected quasars
(RDQs), while LOFAR undetected quasars are denoted as radio-undetected
quasars (RUQs). These quasars are not detected individually at $\geq5\sigma$
in the LOFAR map. We also crossmatch the Milliquas catalog with the
WSRT, and VLA 325Mhz/3.0GHz catalogs using a matching radius of $15^{\prime\prime}$,
$2^{\prime\prime}$, and $1^{\prime\prime}$, respectively. Figure
\ref{fig:radio_flux_distribution} shows the histograms of the quasars
detected by each radio telescope. It is clear that the number of quasars
detected by LOFAR is significantly larger in comparison with the other
radio-telescopes. A total of 21 quasars are detected by WSRT and the
VLA, but not by LOFAR. These quasars might not be detected by LOFAR
due to having an inverted radio spectra. We omitted the radio flux
histogram for these quasars due to their low numbers, but we consider
their WSRT/VLA radio fluxes for all our calculations in the rest of
the paper. These quasars are undetected by LOFAR, but detected by
the other radio-telescopes, therefore according to our requirement,
those with a LOFAR $5\sigma$ detection are classified as RUQs. The
maximum-value normalized $I$-band magnitude distribution of the quasar
samples is presented in Figure \ref{fig:mag_flux_distribution}. The
absolute magnitude and radio luminosity are displayed in Figures \ref{fig:M1450_redshift}
and \ref{fig:L150_redshift}, respectively. The absolute magnitudes
at rest-frame $1450  \> {\angstrom}$, $M_{\textrm{1450}}$, and K-corrections
are calculated following the procedure described by \citet{2020AA}.
This is done using the $R$, $I$, or $Z_{\textrm{Subaru}}$ band
magnitudes depending on the spectroscopic redshift. Figure \ref{fig:mag_flux_distribution}
shows the normalized absolute magnitude $M_{\textrm{1450}}$ distribution
of the quasar samples. The RDQs are slightly brighter than RUQs with
a median absolute magnitude of $M_{\textrm{1450}}=-23.0$, while RUQs
present a median absolute magnitude of $M_{\textrm{1450}}=-22.69$, 

\noindent 

\noindent 

\noindent 

\noindent 

\noindent 
\begin{figure}[tp]
\begin{centering}
\includegraphics[clip,scale=0.33]{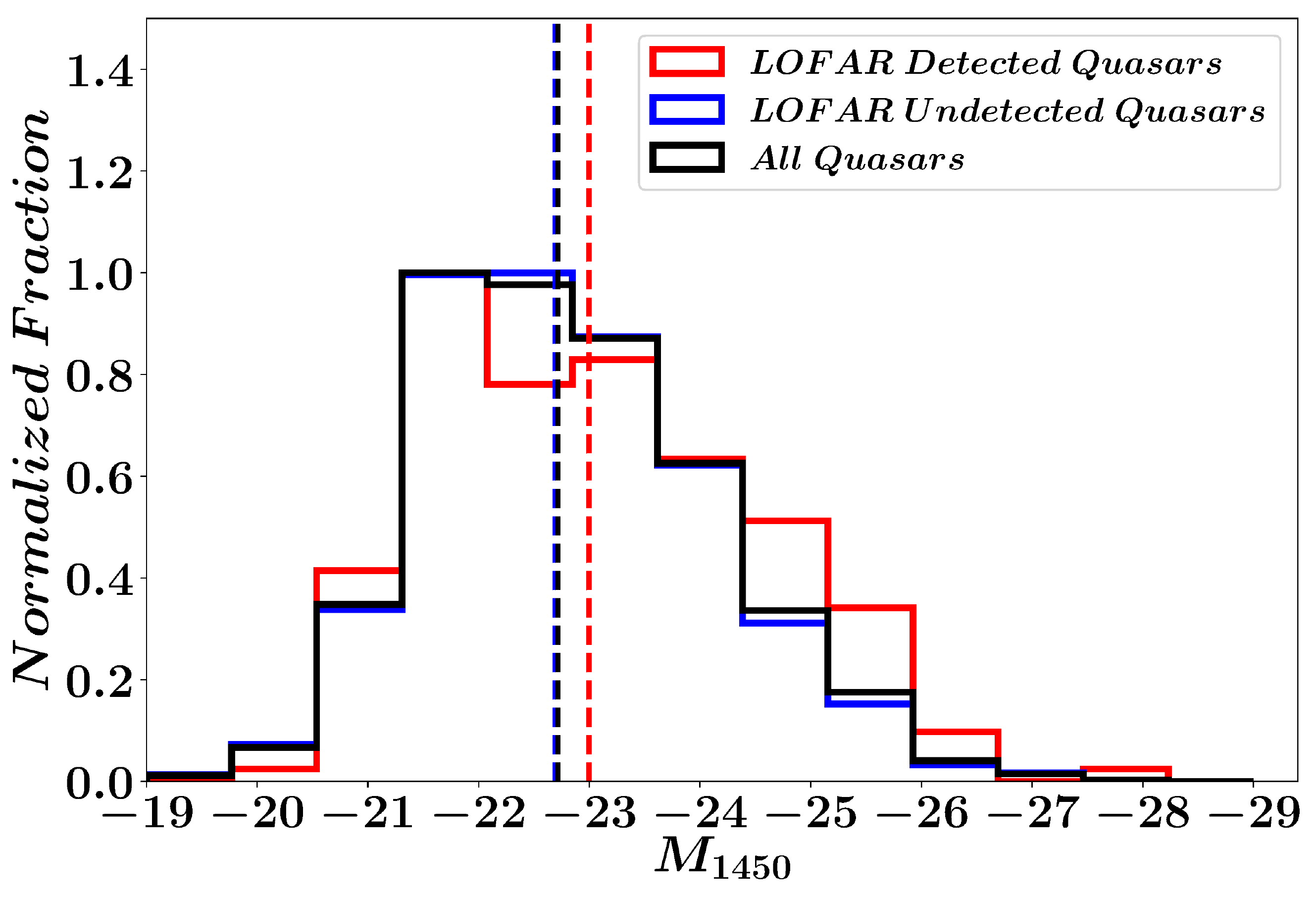}
\par\end{centering}
\centering{}\centering\caption{\label{fig:M1450_distribution} Normalized absolute magnitude $M_{\textrm{1450}}$
distributions of LOFAR radio-detected quasars (RDQs, red) and LOFAR
radio-undetected quasars (RUQs, blue). Also, the combined redshift
distribution of RDQs and RUQs (black) is plotted. The median $M_{\textrm{1450}}$
values of the samples are indicated by the dashed lines}
\end{figure}

\section{Methods}

\subsection{Stacking analysis \label{sec:stacking_method}}

The stacking technique permits the extraction of the signal of sources
below the detection threshold of a particular survey. We divided our
quasar samples into four redshift bins. This allows to have a reasonable
number of sources in each bin to achieve a high signal-to-noise ratio
(S/N). From the original mosaics, we extracted $50\times50$ pixel
cutouts around each quasar position, we also extracted the noise maps.
All the cutouts in the redshift bin are grouped together to form a
pixel cube. The size of the stacked maps is $50\times50$ pixels.
This is ten times larger than the beam size of the radio map with
the lowest angular resolution (WSRT). The next step is to collapse
each pixel cube into a single image by combining the cutouts together.
The two most common ways to do this is either to compute the mean
or the median flux of all the cutouts in a given pixel cube. The main
advantage of the median is that it is more robust against outliers
and it is not sensitive to bright off-center sources \citep{2007ApJ...654...99W,2009MNRAS.394..105G}. 

\noindent We compute the median using a noise weighted average:
\noindent \begin{center}
\begin{equation}
S_{\lambda}^{\textrm{med}}=\dfrac{{\textstyle \sum_{i=1}^{N_{\textrm{QSO}}}}w_{\lambda}^{i}\times S_{\lambda}^{i}}{{\textstyle \sum_{i=1}^{N_{\textrm{QSO}}}}w_{\lambda}^{i}},\label{eq:noise_weg_median}
\end{equation}
\par\end{center}

\noindent where $N_{\textrm{QSO }}$ is the number of quasars within
a redshift bin and grouped into a pixel cube, $S_{\lambda}^{i}$ is
the pixel flux density in the cutout $i$ at a wavelength $\lambda$,
and $w_{\lambda}^{i}$ is the noise flux density in the same pixel.
The fluxes for the different stacked maps are calculated as follows.
The flux densities in the IRAC-$3.6\,\mu\textrm{m}$, WISE-W3, MIPS-$24\,\mu$m,
MIPS-$70\,\mu$m, PACS-$100\,\mu$m, and PACS-$160\,\mu$m maps are
estimated using aperture photometry using radii of $2.5^{\prime\prime}$,$8.5^{\prime\prime}$,
$22^{\prime\prime}$, $30^{\prime\prime}$, $5.6^{\prime\prime}$,
and $11.2^{\prime\prime}$; respectively. The sky background is determined
using inner and outer radii between $30^{\prime\prime}$ and $60^{\prime\prime}$.
For the SPIRE maps, we fit a 2D Gaussian in addition to a constant
background to the stack map and we considered the peak flux as the
flux density estimate.

\noindent The density fluxes in the stacked radio maps are estimated
using the relations by \citet{Condon_1998}. We applied a scaling
factor of $12$ percent to the $150\,\textrm{MHz}$ LOFAR fluxes to
make them consistent with the \citet{2012MNRAS.423L..30S} (hereafter,
SH12) flux scale \citep{2018A&A...620A..74R}. The $1.4\,\textrm{GHz}$
WSRT flux scale is consistent with the SH12 scale \citep{2016MNRAS.460.2385W},
while the $325\,\textrm{MHz}$ VLA scale is multiplied by a factor
of $0.91$ to make it consistent with the SH12 scale \citep{2017MNRAS.469.3468C}.
To investigate the consistency of the VLASS fluxes with the SH12 scale,
we compute the spectral indices of VLASS sources with $S/N\geq15$
with their counterparts from the WSRT, VLA and LOFAR catalogs. We
find that the mean ratio between the predicted and uncorrected VLASS
fluxes is $0.91$. This correction factor is used to put the VLASS
fluxes to the SH12 flux scale . The $1\sigma$ errors on the median
stacked fluxes are estimated using a bootstrap approach for each redshift
bin. For this purpose, the sources within the redshift bin are randomly
resampled (with replacement). The stacking procedure is repeated $100$
times for each filter. The standard deviation of the distribution
of median stacked fluxes obtained in the 100 trials is the $1\sigma$
error of the stacked flux in the corresponding filter. 

\noindent Snapshot surveys with sparse UV coverage could result in
a dirty beam with high sidelobes \citep{1995ApJ...450..559B,Condon_1998}.
These could be difficult to clean and could cause an underestimation
of the real sources in the restored image \citep{Condon_1998}. Our
radio observations have good UV coverage thanks to the high integration
times, thus, they are unlikely to be affected by a snapshot bias.
The median stacked fluxes are not corrected for any CLEAN or snapshot
bias. However, to account for any potential bias in our radio fluxes
we added a $10\%$ uncertainty in quadrature to the flux uncertainties
obtained using bootstrapping. Tables \ref{fig:radio_fluxes_median_qsos}
and \ref{fig:noise_median_qsos} list the median stacked radio fluxes
and noise values, respectively, for the quasar samples. The radio
luminosities derived from these fluxes are shown in Figure \ref{fig:L150_redshift}.
The stacking procedure allows to detect the radio emission of RUQs
that are roughly an order of magnitude less luminous than RDQs at
all redshifts. The infrared fluxes are presented in Appendix \ref{sec:appendix_A}. 

\begin{table*}
\noindent \begin{centering}
\caption{Radio fluxes of the median LOFAR radio-detected quasars (RDQs) and
LOFAR radio-undetected quasars (RUQs) stacked according to redshift
between 150 MHz and 3.0 GHz. \label{fig:radio_fluxes_median_qsos} }
\begin{tabular}{ccccc}
\hline 
$z_{bin}$ & $150\ \textrm{MHz}$ & $325\ \textrm{MHz}$ & $1.4\ \textrm{GHz}$ & $3.0\ \textrm{GHz}$\tabularnewline
\hline 
RDQs & {[}$\mu$Jy{]} & {[}$\mu$Jy{]} & {[}$\mu$Jy{]} & {[}$\mu$Jy{]}\tabularnewline
\hline 
$0.0\leq z\leq1.0$ & 569$\pm$5 & 557$\pm$4 & 154$\pm$5 & 163$\pm$5\tabularnewline
$1.0<z\leq1.8$ & 648$\pm$30 & 513$\pm$15 & 136$\pm$1 & 108$\pm$1\tabularnewline
$1.8<z\leq2.7$ & 671$\pm$30 & 600$\pm$9 & 175$\pm$2 & 123$\pm$1\tabularnewline
$2.7<z\leq6.15$ & 798$\pm$190 & 551$\pm$100 & 185$\pm$3 & 144$\pm$4\tabularnewline
\hline 
RUQs & {[}$\mu$Jy{]} & {[}$\mu$Jy{]} & {[}$\mu$Jy{]} & {[}$\mu$Jy{]}\tabularnewline
\hline 
$0.0\leq z\leq1.0$ & 113.38$\pm$8.3 & 102.28$\pm$9.6 & 31.24$\pm$3.6 & $<$30.56\tabularnewline
$1.0<z\leq1.8$ & 77.30$\pm$4.2 & 76.48$\pm$8.6 & 19.04$\pm$1.3 & $<$19.26\tabularnewline
$1.8<z\leq2.7$ & 75.85$\pm$7.5 & 55.28$\pm$4.8 & 15.03$\pm$1.6 & $<$26.98\tabularnewline
{\small{}$2.7<z\leq6.15$} & 92.85$\pm$10.2 & 100.70$\pm$19.9 & 21.86$\pm$3.3 & $<$41.18\tabularnewline
\hline 
$1\sigma$ sensitivity & $55$ & $300$ & $28$ & $140$\tabularnewline
\hline 
\end{tabular}
\par\end{centering}
\begin{centering}
\centering
\par\end{centering}
$\;$

\emph{Notes:} {\tiny{}Values without error bars are }\textbf{\tiny{}$5\sigma$}{\tiny{}
upper limits derived at the center of the median stacked images. The
$1\sigma$ sensitivity is also indicated for each frequency.}{\tiny\par}
\end{table*}
 
\begin{table*}
\noindent \begin{centering}
\caption{Noise levels ($1\sigma$) of the median radio images stacked according
to redshift of LOFAR radio-detected quasars (RDQs) and LOFAR radio-undetected
quasars (RUQs). $N_{\textrm{QSO}}$ is the number of quasars within
the corresponding redshift bin. \label{fig:noise_median_qsos} }
\begin{tabular}{cccccc}
\hline 
$z_{bin}$ & $N_{\textrm{QSO}}$ & $150\ \textrm{MHz}$ & $325\ \textrm{MHz}$ & $1.4\ \textrm{GHz}$ & $3.0\ \textrm{GHz}$\tabularnewline
\hline 
RDQs &  & {[}$\mu$mJy/beam{]} & {[}$\mu$Jy/beam{]} & {[}$\mu$Jy/beam{]} & {[}$\mu$Jy/beam{]}\tabularnewline
\hline 
{\small{}$0.0\leq z\leq1.0$} & 48 & 16.63 & 48.51 & 6.51 & 22.20\tabularnewline
{\small{}$1.0<z\leq1.8$} & 79 & 16.28 & 40.28 & 5.37 & 17.14\tabularnewline
{\small{}$1.8<z\leq2.7$} & 52 & 18.80 & 49.78 & 6.78 & 20.63\tabularnewline
{\small{}$2.7<z\leq6.15$} & 14 & 28.31 & 100.41 & 111.05 & 33.02\tabularnewline
\hline 
RUQs &  & {[}$\mu$Jy/beam{]} & {[}$\mu$Jy/beam{]} & {[}$\mu$Jy/beam{]} & {[}$\mu$Jy/beam{]}\tabularnewline
\hline 
{\small{}$0.0\leq z\leq1.0$} & 203 & 6.88 & 22.99 & 2.81 & 10.79\tabularnewline
{\small{}$1.0<z\leq1.8$} & 588 & 4.07 & 17.52 & 1.61 & 6.34\tabularnewline
{\small{}$1.8<z\leq2.7$} & 405 & 4.68 & 21.04 & 1.96 & 7.65\tabularnewline
{\small{}$2.7<z\leq6.15$} & 140 & 8.16 & 39.40 & 3.21 & 12.82\tabularnewline
\hline 
\end{tabular}
\par\end{centering}
$\;$

\emph{Notes:} {\tiny{}$N_{\textrm{QSO}}$ is the number of quasars
within the corresponding redshift bin. }{\tiny\par}
\end{table*}

\noindent The median stacked infrared and radio images of RUQs are
shown in Figures \ref{fig:infrared_postages_und_1}, \ref{fig:infrared_postages_und_2},
and \ref{fig:radio_postages_und}, respectively. The RUQs are detected
in the majority of the filters, except in the PACS and VLASS bands
where they are not detected in some redshift bins. This is somehow
expected as the PACS and VLASS are relatively shallower in comparison
with the other infrared and radio bands, respectively (see Section
\ref{sec:bootes_data}). Figures \ref{fig:infrared_postages_dect_1},
\ref{fig:infrared_postages_dect}, and \ref{fig:radio_postages_dect}
display the median infrared and radio stacked images of RDQs, respectively.
In some high-z bins with a low number of RDQs stacked, the results
from the median stacking are consistent with zero in the PACS and
VLASS images.

\subsubsection{Potential systematics}

We verify that there is no potential bias in the median stacked fluxes
by performing a null test. This test consists in stacking randomly
re-positioned pairs \citep{Zhang_2013,2015ApJ...809L..22V,2019MNRAS.490.2315H}.
A total of 1000 null tests are performed for each redshift bin with
a number of random positions equal to the number of quasars in the
bin. The stacking procedure remains identical as when the real quasar
positions are used. As expected, the results of the null tests fluctuate
around zero in both the infrared and radio maps. Finally, these random
stacks centered around zero indicates that the contribution to our
stacked fluxes of source confusion due to the blending of faint sources
is negligible \citep{2017MNRAS.472.2221S}. 

\noindent Another point to consider for the radio stacking is how
uniformly was the quasar sample constructed. \citet{2007ApJ...654...99W}
pointed out how factors such as the SDSS quasar selection algorithm
\citep{2001AJ....121.2308R}, or the dependency of radio and optical
luminosity could affect the interpretations of the stacking analysis.
Similarly, \citet{2008AJ....136.1097H} found evidence for a selection
bias towards bright radio sources in their stacking analysis of SDSS
luminous red galaxies.

\begin{figure}[tp]
\begin{centering}
\includegraphics[clip,scale=0.33]{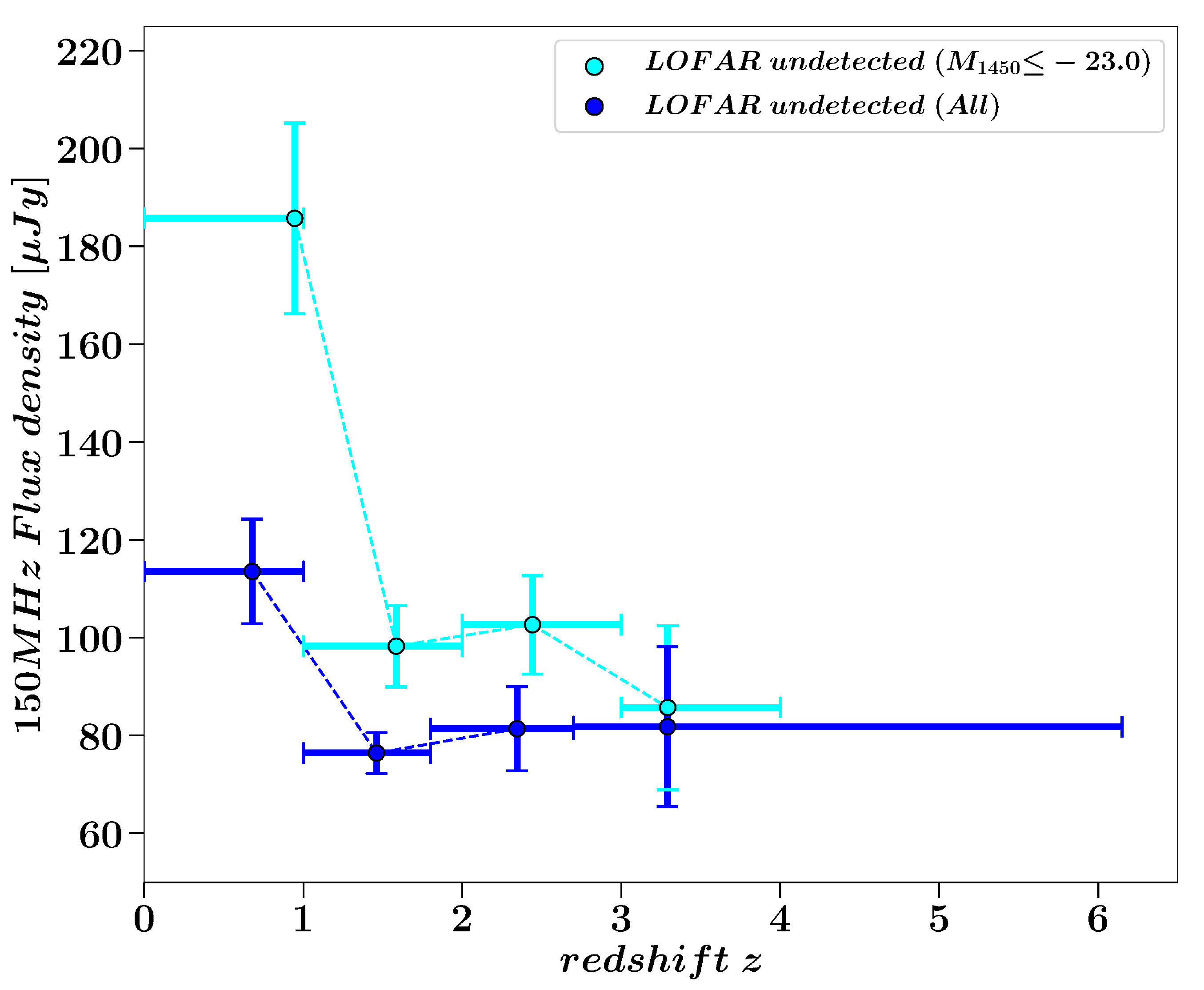}
\par\end{centering}
\centering{}\centering\caption{\label{fig:flux_z_ruqs}Median-stacked 150 MHz flux densities versus
redshift for all LOFAR radio-undetected quasars (RUQs) in our sample,
and for RUQs with $M_{1450}\protect\leq-23.0$. The samples follow
similar trends, indicating that selection effects could be present in
both samples.}
\end{figure}

\noindent Our spectroscopic quasar sample is not uniformly constructed
since it is a compilation of the observations made by way of various
surveys \citep{2006AJ....132..823C,2006ApJ...652..157M,2011ApJ...728L..26G,2012ApJS..200....8K,2018A&A...613A..51P};
with each survey having its own unique strategies to build its own
sample. Figure \ref{fig:flux_z_ruqs} shows the median 150 MHz flux
densities of RUQs as a function of redshift. The median 150 MHz fluxes
present a trend similar to that found by \citet{2007ApJ...654...99W}
in their stacking analysis of SDSS-FIRST quasars. However, we are
using redshift bins broader in comparison with \citet{2007ApJ...654...99W}.
This yields radio fluxes that have a better agreement within the error
bars except for the $z\leq1.0$ bin. Another possibility is considering
only RUQs with $M_{1450}\leq-23.0$ with $\triangle z=1.0$, where
it is expected that the median fluxes are less affected by selection
effects. However, it yields a result similar to what is obtained with
the full RUQs sample. A possible solution could be to limit our quasar
sample only to one survey with an uniform selection function, but
this in return would reduce the S/N of the stacked sources, particularly
for RUQs. This poorer S/N would makes our study of the RDQs and RUQs
more difficult as a function of redshift and luminosity, along with
its potential to investigate the origins of their radio-emission.
For these reasons, we decided to use all the known spectroscopic quasars
located in the WSRT-Bo\"otes region in our stacking analysis. However,
we did keep in mind that selection effects could affect (to some extent)
these trends in our analysis.

\noindent 

\subsection{Spectral energy distribution fitting \label{subsec:sed_modelling}}

The observed quasar SED between mid-infrared and radio wavelengths
can be fitted using three main components: AGN and SF heated thermal
dust emission, and synchrotron emission. The AGN and SF heated thermal
dust emission are represented as the combination of two terms \citep{2012MNRAS.425.3094C}:
a simple power-law with a low-frequency exponential cut-off that represents
the AGN heated dust, and a single temperature modified blackbody (BB)
that corresponds to the far-infrared emission from SF heated dust:
\noindent \begin{center}
\begin{equation}
S\left(\lambda\right)=N_{\textrm{AGN}}\lambda^{\alpha_{\textrm{AGN}}}e^{-\left(\lambda/\lambda_{\textrm{c}}\right)^{2}}+\dfrac{N_{\textrm{BB}}}{e^{hc/\lambda kT}-1}\left(\frac{c}{\lambda}\right)^{\beta+3},\label{eq:casey_model}
\end{equation}
\par\end{center}

\noindent where $\alpha$ is the power-law slope, $N_{\textrm{AGN}}$
and $N_{\textrm{BB}}$ are the normalization constants of the AGN
and BB terms, $\lambda_{\textrm{c}}$ is the fixed rest-frame cut-off
wavelength, $\beta$ is the emissivity index, and $T$ is the characteristic
temperature of the quasar host galaxy. The cut-off wavelength, and
the emissivity are fixed to the values of $\lambda_{\textrm{c}}=45\,\mu\textrm{m}$
and $\beta=2.5$, respectively \citep{2019A&A...621A..27F}. 

\noindent The synchrotron emission is fitted with a single power-law,

\begin{equation}
S_{\textrm{sync}}\left(\lambda\right)=N_{\textrm{sync}}\lambda^{\alpha},\label{eq:radio_power_law}
\end{equation}

\noindent where $\alpha$ is the synchrotron power-law slope, and
$N_{\textrm{sync}}$ is the normalization constant of the synchrotron
term. This power-law is a simple model that can be used to determine
the maximum contribution from the synchrotron emission to the radio
and sub-mm frequencies.

\begin{figure*}[t]
\centering{}\includegraphics[clip,scale=0.4]{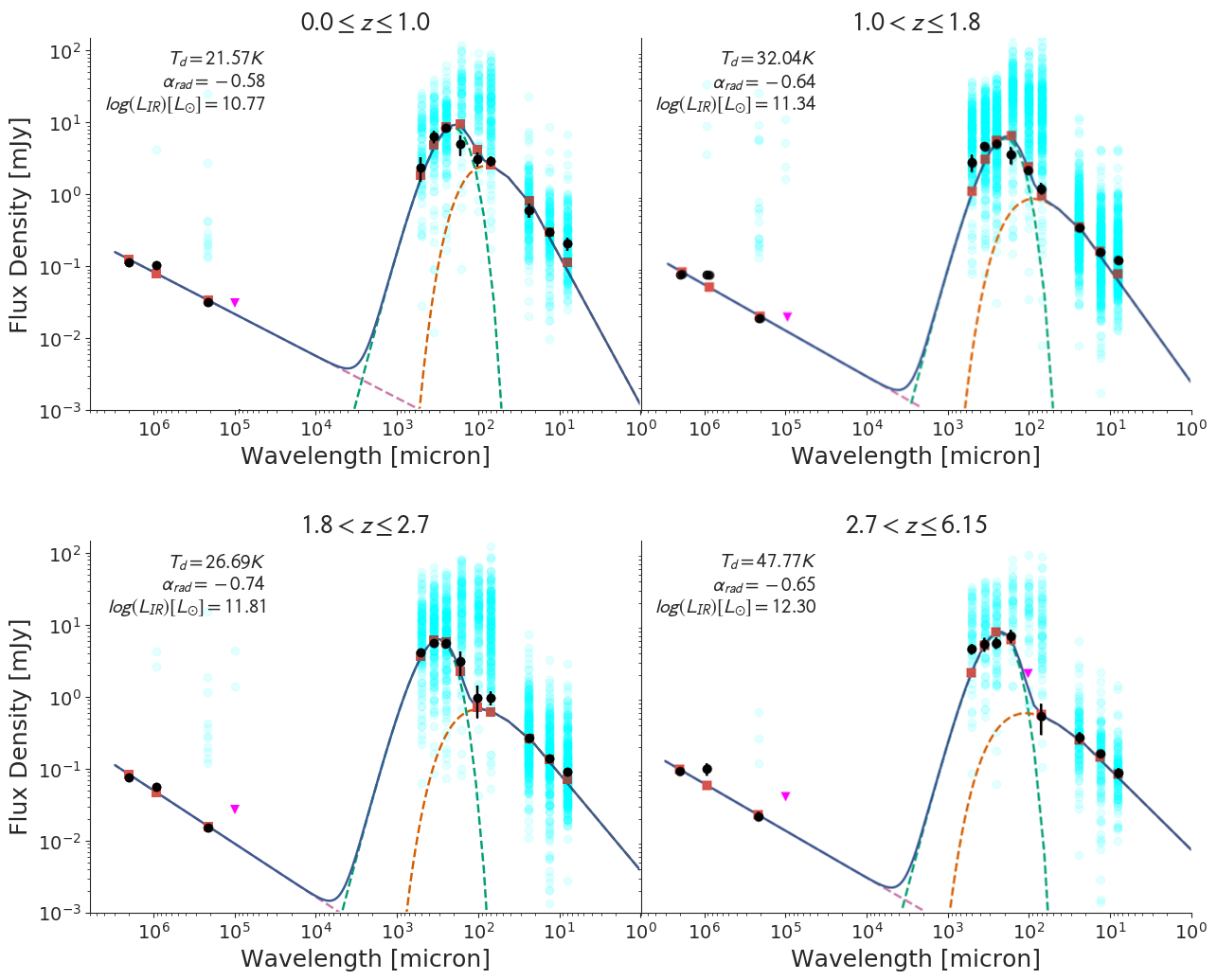}\centering\caption{\label{fig:SED_fitting_RUQs} Rest-frame spectral energy distributions
(SEDs) for median LOFAR radio-undetected quasars stacked according
to redshift. Black points plot the observed flux densities with downward-pointing
fuchsia arrows indicating $5\sigma$ upper limits. The solid blue
line shows the combined best-fit SED model, while the purple, green
and orange lines show the synchrotron, black-body, and AGN best-fit
components, respectively. The maroon squares indicate the predicted fluxes according
to the fitted SED model. The cyan points denote the fluxes of the
quasars used in the stacking. Some quasars have WSRT and VLA detections,
but remain undetected in LOFAR, thus are classified as RUQS. See Section
\ref{sec:quasar_sample} for more details. }
\end{figure*}

\begin{figure*}[t]
\centering{}\includegraphics[clip,scale=0.4]{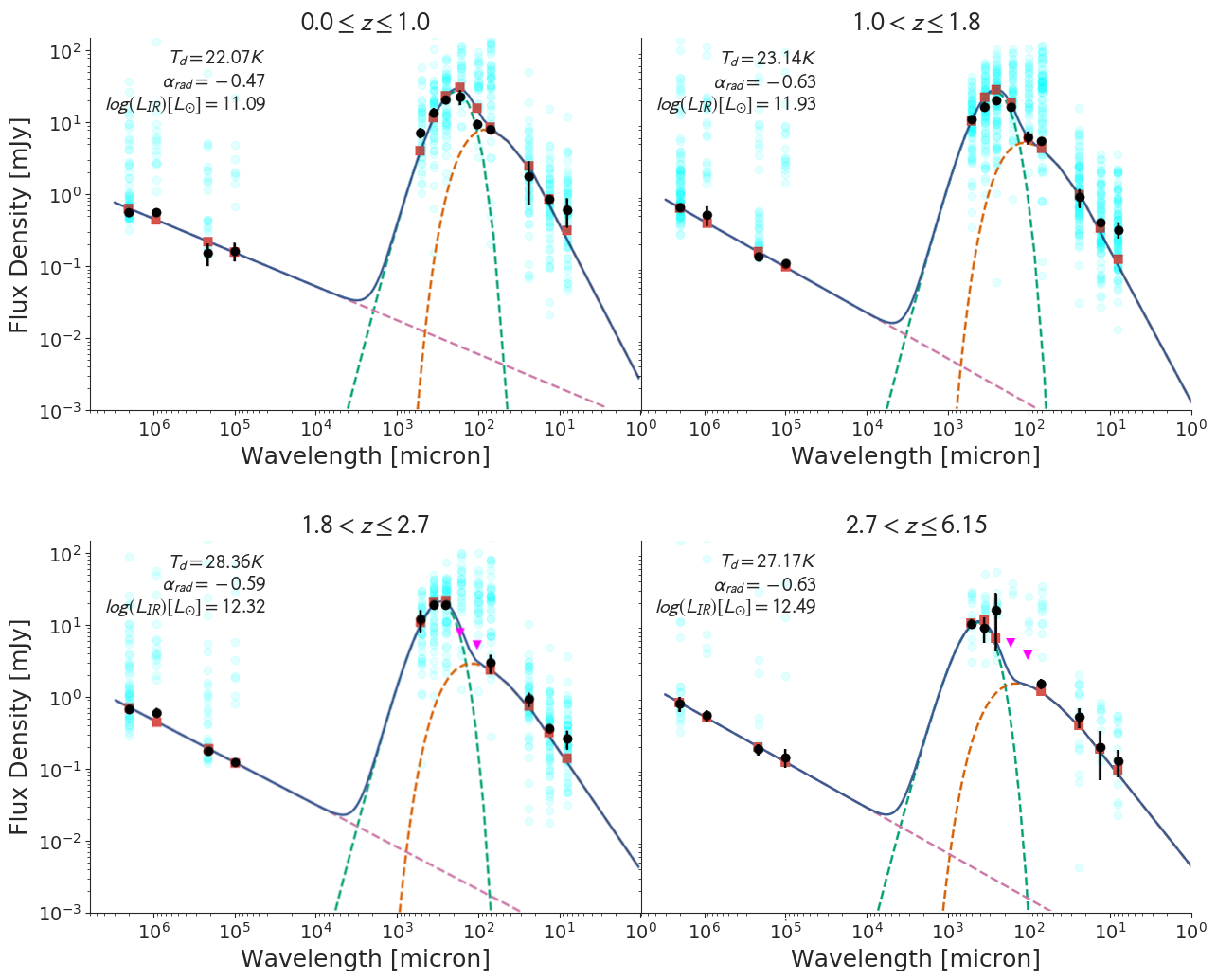}\centering\caption{\label{fig:SED_fitting_RDQs} Rest-frame spectral energy distributions
(SEDs) for median LOFAR radio-detected quasars stacked according to
redshift. Black points plot the observed flux densities with downward-pointing
fuchsia arrows indicating $5\sigma$ upper limits. The solid blue
line shows the combined best-fit SED model, while the purple, green
and orange lines show the synchrotron, black-body, and AGN best-fit
components, respectively. The maroon squares indicate the predicted fluxes according
to the fitted SED model. The cyan points denote the fluxes of the
quasars used in the stacking. }
\end{figure*}

\noindent We use Levenberg\textendash Marquardt (LM) $\chi^{2}$ minimization
to find the best-fitting parameters for the quasar SED model. The
fitting is done using the Python implementation of the LM algorithm
LmFit\footnote{https://lmfit.github.io/lmfit-py/}. Figures \ref{fig:SED_fitting_RUQs}
and \ref{fig:SED_fitting_RDQs} present the median SEDs of RUQs and
RDQs, respectively in the four redshift bins, along with the best-fit
SEDs and their different components. For all the SEDs, it is easy
to recognize the emission associated with the hotter dust heated by
the AGN, and dust heated thermally by SF, as well the synchrotron
component. We also consider the possibility that the far-infrared
flux densities are contaminated by synchrotron emission that is not
associated with star formation. By extrapolating the fitted synchrotron
power-law, we subtracted the synchrotron contribution to the far-infrared
fluxes. The fitted far-infrared fluxes are fitted again to the \citet{2012MNRAS.425.3094C}
model. We verify that our results do not change significantly by using
the far-infrared fluxes without subtracting the synchrotron contribution.
The values of the best-fit parameters are presented in Table \ref{tab:SED_fitting_parameters}.
The uncertainties are reported on the best-fit parameters correspond
to $1\sigma$ errors.

\noindent The total infrared luminosity of the SF ($L_{\textrm{SF}}$)
and AGN $(L_{\textrm{AGN}}$) components are derived by integrating
by separate each component of Eq. \eqref{eq:casey_model}. The integration
is done between the rest-frame wavelengths $8.0\,\mu\textrm{m}$ and
$\lambda_{\textrm{c}}=1000\,\mu\textrm{m}$:

\begin{equation}
L_{\textrm{IR}}=\frac{4\pi D_{\textrm{L}}^{2}}{(1+z)}\intop_{8\,\mu\textrm{m}}^{1000\,\mu\textrm{m}}S\left(\lambda\right)\textrm{d}\lambda,\label{eq:L_ir}
\end{equation}

\noindent where $z$ is the quasar redshift and $D_{\textrm{L}}$
is the luminosity distance. The total infrared luminosity, $L_{\textrm{IR}}$,
is estimated by combining the results of the SF and AGN components.
Errors in the IR luminosity components are derived from the range
of values obtained from SED fits which differ from the best fit by
$\Delta\chi_{i}=\left(\chi_{i}^{2}-\chi_{\textrm{min}}^{3}\right)<1.$
The SF rate (SFR) of the quasars host galaxies is determined using
the relation by \citet{1998ApJ...498..541K} assuming a Salpeter initial
mass function: 

\begin{equation}
SFR\:\left[M_{\odot}\:\textrm{yr}^{-1}\right]={\displaystyle \frac{L_{\textrm{IR}}}{5.8\times10^{8}}},\label{eq:kennicutt_law}
\end{equation}

\noindent where $L_{\textrm{IR}}$ is in units of solar luminosities,
$L_{\odot}$. The infrared luminosities and SFR estimates are listed
in Table \ref{tab:SED_fitting_parameters}. 
%
\begin{table*}
\noindent \begin{centering}
\caption{SED model fitting results. The best-fit models are showed in Figures
\ref{fig:SED_fitting_RDQs} and \ref{fig:SED_fitting_RUQs}. \label{tab:SED_fitting_parameters} }
\begin{tabular}{cccccccccc}
\hline 
{\small{}$z_{bin}$} & {\small{}$T$} & {\small{}$N_{\textrm{BB}}$} & {\small{}$N_{\textrm{AGN}}$} & {\small{}${\alpha_{\textrm{AGN}}}$} & {\small{}$N_{\textrm{radio}}$} & {\small{}${\alpha}$} & {\small{}${L_{\textrm{IR}}}$} & {\small{}${L_{\textrm{SF}}}$} & {\small{}${L_{\textrm{AGN}}}$}\tabularnewline
\hline 
{\small{}RDQs} & {\small{}{[}K {]}} & {\small{}{[}mJy{]}} & {\small{}{[}mJy{]}} &  & {\small{}{[}mJy{]}} &  &  &  & \tabularnewline
\hline 
{\small{}$0.0\leq z\leq1.0$} & {\small{}21.57$\pm$1.90} & {\small{}4.97$\pm$0.62} & {\small{}1.47$\pm$0.48} & {\small{}2.17$\pm$0.21} & {\small{}1.47$\pm$0.48} & {\small{}0.58$\pm$0.15} & {\small{}$10.8_{-0.11}^{+0.06}$} & {\small{}$10.7_{-0.10}^{+0.06}$} & {\small{}$10.1_{-0.24}^{+0.18}$}\tabularnewline
{\small{}$1.0<z\leq1.8$} & {\small{}32.04$\pm$2.90} & {\small{}2.57$\pm$0.56} & {\small{}0.37$\pm$0.16} & {\small{}1.64$\pm$0.27} & {\small{}0.37$\pm$0.16} & {\small{}0.64$\pm$0.17} & {\small{}$11.3_{-0.18}^{+0.15}$} & {\small{}$11.3_{-0.09}^{+0.12}$} & {\small{}$10.4_{-0.75}^{+0.27}$}\tabularnewline
{\small{}$1.8<z\leq2.7$} & {\small{}26.69$\pm$1.72} & {\small{}1.92$\pm$0.19} & {\small{}0.21$\pm$0.06} & {\small{}1.40$\pm$0.19} & {\small{}0.21$\pm$0.06} & {\small{}0.74$\pm$0.12} & {\small{}$11.8_{-0.24}^{+0.18}$} & {\small{}$11.8_{-0.16}^{+0.47}$} & {\small{}$10.8_{-0.47}^{+0.24}$}\tabularnewline
{\small{}$2.7<z\leq6.15$} & {\small{}47.77$\pm$6.73} & {\small{}1.85$\pm$0.44} & {\small{}0.14$\pm$0.02} & {\small{}1.16$\pm$0.06} & {\small{}0.14$\pm$0.02} & {\small{}0.65$\pm$0.19} & {\small{}$12.3_{-0.33}^{+0.18}$} & {\small{}$12.3_{-0.20}^{0.45}$} & {\small{}$11.2_{-0.45}^{+0.38}$}\tabularnewline
\hline 
{\small{}RUQs} &  &  &  &  &  &  &  &  & \tabularnewline
\hline 
{\small{}$0.0\leq z\leq1.0$} & {\small{}22.07$\pm$1.15} & {\small{}16.95$\pm$2.14} & {\small{}5.02$\pm$0.56} & {\small{}2.29$\pm$0.07} & {\small{}5.02$\pm$0.56} & {\small{}0.47$\pm$0.15} & {\small{}$11.1_{-0.02}^{+0.02}$} & {\small{}$11.0_{-0.02}^{+0.02}$} & {\small{}$10.4_{-0.06}^{+0.06}$}\tabularnewline
{\small{}$1.0<z\leq1.8$} & {\small{}23.14$\pm$1.58} & {\small{}11.39$\pm$1.88} & {\small{}2.22$\pm$0.55} & {\small{}2.17$\pm$0.16} & {\small{}2.22$\pm$0.55} & {\small{}0.63$\pm$0.06} & {\small{}$11.9_{-0.02}^{+0.04}$} & {\small{}$11.9_{-0.03}^{+-.03}$} & {\small{}$11.1_{-0.20}^{+0.13}$}\tabularnewline
{\small{}$1.8<z\leq2.7$} & {\small{}28.36$\pm$4.03} & {\small{}6.12$\pm$1.49} & {\small{}0.89$\pm$0.23} & {\small{}1.69$\pm$0.16} & {\small{}0.89$\pm$0.23} & {\small{}0.59$\pm$0.06} & {\small{}$12.3_{-0.06}^{+0.08}$} & {\small{}$12.3_{-0.06}^{+0.05}$} & {\small{}$11.4_{-0.34}^{+0.23}$}\tabularnewline
{\small{}$2.7<z\leq6.15$} & {\small{}27.17$\pm$3.78} & {\small{}2.53$\pm$0.63} & {\small{}0.36$\pm$0.02} & {\small{}1.46$\pm$0.04} & {\small{}0.36$\pm$0.02} & {\small{}0.63$\pm$0.06} & {\small{}$12.5_{-0.20}^{-0.16}$} & {\small{}$12.4_{-0.19}^{+0.16}$} & {\small{}$11.6_{-0.46}^{+0.23}$}\tabularnewline
\hline 
\end{tabular}
\par\end{centering}
\begin{centering}
\centering
\par\end{centering}
$\;$

\emph{Notes:} {\tiny{}See Section \ref{subsec:sed_modelling} for
more details about the models used. $L_{\textrm{IR}}$ is the total
infrared luminosity over the rest-frame wavelengths $8.0\,\mu\textrm{m}$
and $\lambda_{\textrm{c}}=1000\,\mu\textrm{m}$. $L_{\textrm{SF}}$
and $L_{\textrm{AGN}}$ are the integrated SF and AGN luminosities,
respectively. All the luminosities are in units of $\log(L_{\odot})$.}{\tiny\par}
\end{table*}

\noindent We also fit the radio-infrared photometry of the individual
quasars to Eqs. \ref{eq:casey_model} and \ref{eq:L_ir}. For quasars
that are undetected in the radio and infrared bands, we determined
the upper limits using forced photometry measurements at their optical
positions in the infrared and radio maps. We calculate the upper limits
in the radio maps considering apertures that are slightly smaller
to the corresponding beam sizes (see Section \ref{sec:bootes_data})
to prevent these measurements to be affected by imaging artifacts.
These fluxes are used as upper $5\sigma$ limits in the SED fitting.


\section{Results \label{sec:results}}

\subsection{Far-infrared radio correlation \label{sec:firc}}

The far-infrared radio correlation (FIRC) is an indicator that is
regularly used to investigate the levels of SF and AGN activity in
galaxies \citep{1985ApJ...298L...7H,1992ARA&A..30..575C,2001ApJ...554..803Y,2003ApJ...586..794B}.
Particularly, the FIRC of normal star-forming galaxies, namely, those
without significant AGN activity, is a tight relation with a scatter
of less 0.3 dex over five orders of magnitudes in luminosity \citep{2001ApJ...554..803Y}.
This is thought to be the result of the infrared and radio emission
being related to the rate of massive star formation \citep{1992ARA&A..30..575C}. 

\noindent The FIRC is parametrized using the dimensionless parameter,
$q_{\textrm{FIR}}$, that is defined as \citep{1985ApJ...298L...7H,2001ApJ...554..803Y}: 

\begin{equation}
q_{\textrm{FIR}}={\displaystyle \log\left(\frac{L_{\textrm{IR}}/3.75\times10^{12}\,\textrm{Hz}}{L_{\textrm{Radio}}}\right)},\label{eq:fir_radio_corr}
\end{equation}

\noindent where $L_{\textrm{IR}}$ is the rest-frame infrared luminosity
from $8.0\,\mu\textrm{m}$ to $1000\,\mu\textrm{m}$, and $L_{\textrm{Radio}}$
is the radio luminosity at a certain frequency.

\begin{figure*}
\centering{}\includegraphics[clip,scale=0.4]{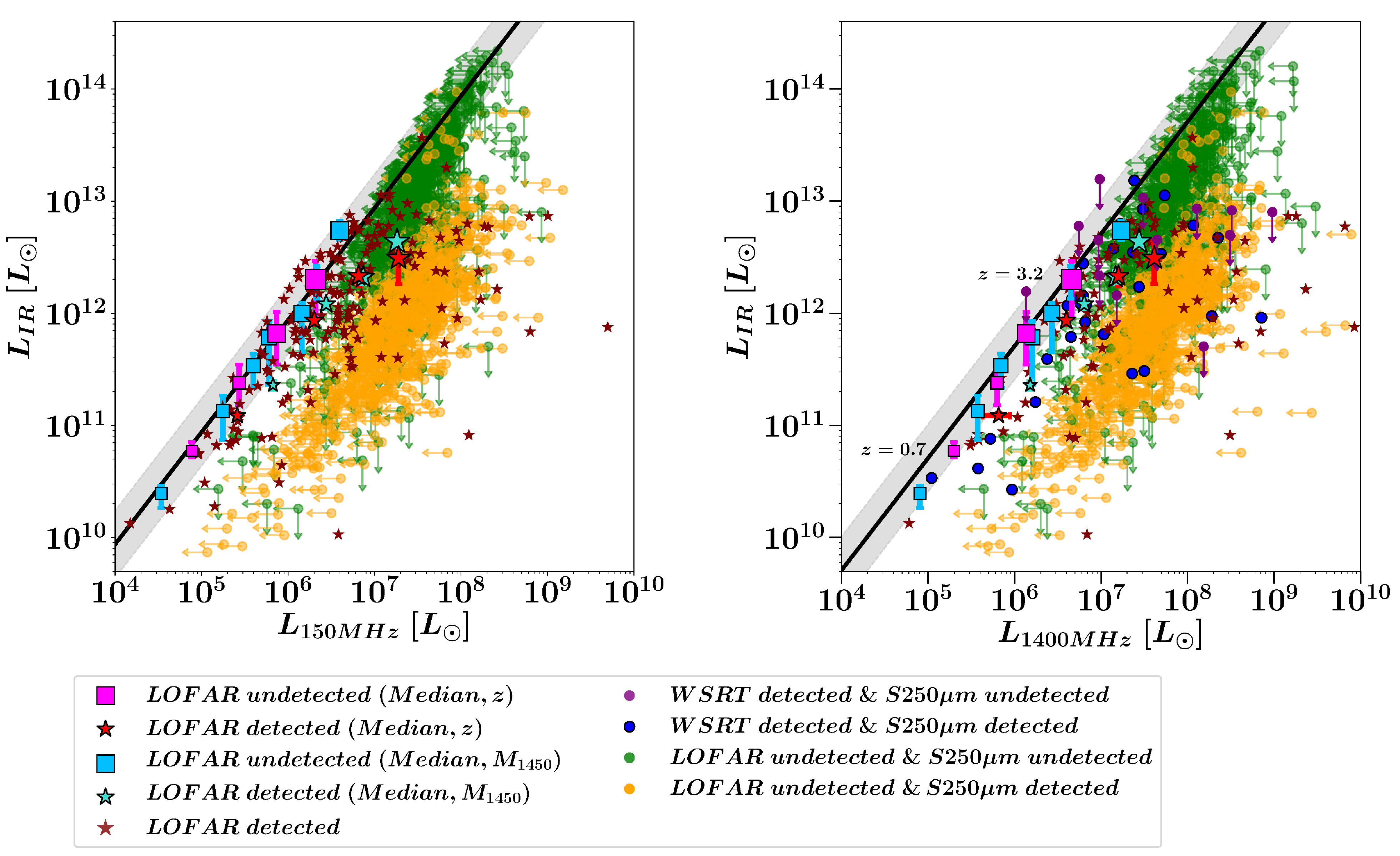}\centering\caption{\label{fig:firc_150_1400} Comparison between infrared luminosity,
$L_{\textrm{IR}},$ and radio luminosity at 150 MHz and 1400 MHz. Left:
$L_{\textrm{IR}}$ versus the radio luminosity at $150\,\textrm{MHz}$. $L_{\textrm{IR}}$ and
${\displaystyle L_{150\,MHz}}$ are in units of solar luminosity. Measurements
for median LOFAR radio-undetected quasars (RUQs, fuchsia and cyan squares), median
LOFAR radio-detected quasars (RDQs, red and turquoise stars), RDQs detected by SPIRE
at $250\,\mu\textrm{m}$ (maroon stars), RUQs detected by SPIRE at
$250\,\mu\textrm{m}$ (yellow circles), and RUQs undetected by SPIRE
at $250\,\mu\textrm{m}$ (green circles) are shown. Upper limits are
indicated by arrows in either $L_{\textrm{IR}}$ or ${\displaystyle L_{150\,MHz}}$.
For median quasars, the increasing symbol size indicates an the increment
in redshift or $M_{1450}$ absolute luminosity. The dashed line is
the far-infrared radio correlation (FIRC) at ${\displaystyle L_{150\,MHz}}$
derived by \citet{2017MNRAS.469.3468C}. The gray shaded region indicates
the spread of the FIRC scaled by a factor of $\pm2$. The text indicates the median redshift of the first and last redshift
bins of the stacking according to redshift. Right: Same
as the right panel, but for the the radio luminosity at $1400\,\textrm{MHz}$,
${\displaystyle L_{1400\,MHz}}$. Measurements for RUQs detected by
WSRT, but undetected by SPIRE at $250\,\mu\textrm{m}$ (purple circles);
and RUQs detected by both SPIRE and WSRT at $250\,\mu\textrm{m}$
and $1400\,\textrm{MHz}$ (blue circles), respectively are also displayed.
}
\end{figure*}

\noindent In Figure \ref{fig:firc_150_1400}, we plot the median infrared
and radio luminosities along with upper limits on the luminosities
for the individual quasars that are undetected by LOFAR, WSRT, and
SPIRE at $250\,\mu\textrm{m}$. Also we compare our results with the
FIRC results obtained by \citet{2017MNRAS.469.3468C} for star-forming
galaxies at $150\,\textrm{MHz}$ and $1.4\,\textrm{GHz}$ in the NDWFS-Bo\"otes
field. We define a quasar to be SF-dominated if the quasar is within
a factor of $\pm2$ of the expected value predicted by the FIRC. In
both panels of Figure \ref{fig:firc_150_1400}, the most striking
result is that median RUQs that are stacked according to redshift
closely follow the FIRC as it is expected for normal star-forming
galaxies. We did not observe any redshift trend for median quasars.
In the right panel, the median RDQ in the lowest redshift bin is located
closer to the FIRC prediction with their radio-emission likely to
have a significant contribution of SF, while the remaining three redshift
bins are located far to the right side of the FIRC. The significant
excess exhibit in their radio luminosities emission is greater than
the levels expected from SF activity alone and most likely can be
associated with AGN activity. Therefore, the radio fluxes of the mean
RDQs in the three highest redshift bins could be considered as AGN-dominated.
The upper limits on the infrared and radio luminosities of the individual
RUQs are insufficient to determine individually how closely each one
of these objects follow the FIRC. In Figure \ref{fig:firc_150_1400}
(right panel), there are a few quasars that that are undetected by
LOFAR, but are detected by WSRT at 1.4 GHz (purple and blue circles).
We classify these quasars as RUQs due to being undetected by LOFAR
at 150 MHz (see Section \ref{sec:quasar_sample}), but they show radio-excess
(i.e., radio emission associated with radio jets). Finally, in both
panels the individual RDQs also present radio-excess at 1.4 GHz which
suggests that their radio fluxes are AGN-dominated. This is consistent
with the results obtained in Section \ref{subsec:mechanism}, where
it is found that the fraction of RDQs that is AGN-dominated at 150
MHz (1.4 GHZ) is $67\%$ ($85\%$).

We also examined the behavior of the infrared and radio luminosities
as a function of the $1450 \> {\angstrom}$ absolute magnitude. For this
purpose, we stacked our sample following the steps explained in Sections
\ref{sec:stacking_method} and \ref{subsec:sed_modelling}, namely,
as a function of $1450  \> {\angstrom}$ absolute magnitude. We use bins
of size $\triangle M_{\textrm{1450}}=-1.0$ for RUQs, and $\triangle M_{\textrm{1450}}=-2.0$
for RDQs to account for the smaller size of the RDQs sample. In all
flux-limited samples, the luminosities are correlated with redshift,
which is driven by the apparent lack of bright low-z sources, and
faint sources at higher redshifts due to the detection limits. Thus,
we keep in mind that there is a possibility that the faintest bins
are affected by some incompleteness. Figure \ref{fig:firc_150_1400}
shows that median RDQs ad RUQs stacked according to their $M_{\textrm{1450}}$
luminosity follow similar trends to those found in the stacking according
to their redshift. In the same plot, we find that the main difference
is that the radio-emission of median RDQs in the faintest $M_{\textrm{1450}}$
bin is clearly AGN-dominated. In contrast with the results of the
lowest z-bin where the radio-emission is probably SF-dominated. We
did not find any $M_{\textrm{1450}}$ trend for median quasars. 

\begin{figure*}
\centering{}\includegraphics[clip,scale=0.4]{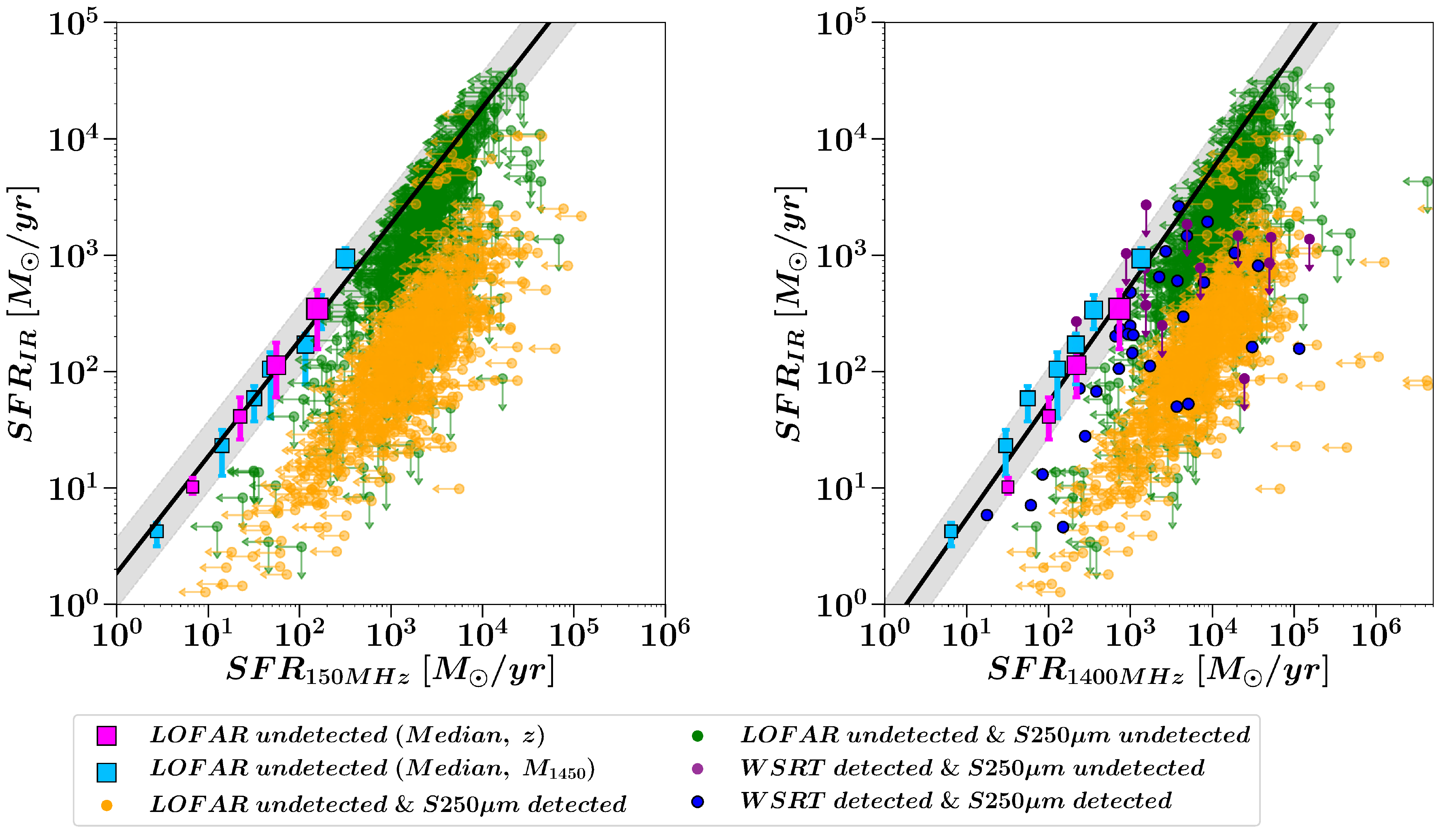}\centering\caption{\label{fig:sfr_150_1400} Comparison between star-formation rates
(SFRs) derived using infrared and radio data. Left: SFR derived from
the infrared luminosity, $SFR_{\textrm{IR}}$, versus the SFR derived
from the radio luminosity at $150\,\textrm{MHz}$. $\displaystyle SFR_{\textrm{IR}}$ and ${\displaystyle SFR_{150\,MHz}}$
are in units of solar mass per year. The measurements for median LOFAR
radio-undetected quasars (RUQs) (fuchsia and cyan squares), RUQs detected
by SPIRE at $250\,\mu\textrm{m}$ (yellow circles), and RUQs undetected
by SPIRE at $250\,\mu\textrm{m}$ (green circles) are showed. Upper
limits are indicated by arrows in either $SFR_{\textrm{IR}}$ or ${{\displaystyle SFR_{150\,MHz}}}$.
For median quasars, the increasing symbol size indicates an increment
in the redshift or $M_{1450}$ absolute luminosity. The dashed line
is the far-infrared radio correlation (FIRC) at ${\displaystyle L_{150\,MHz}}$
derived by \citet{2017MNRAS.469.3468C}. The gray shaded region indicates
the spread of the FIRC scaled by a factor of $\pm2$. The text indicates the median redshift of the first and last redshift
bins of the stacking according to redshift. Right: Same
as the right panel, but for the the SFR at $1400\,\textrm{MHz}$,
${\displaystyle SFR_{1400\,MHz}}$. The measurements for RUQs detected
by WSRT, but undetected by SPIRE at $250\,\mu\textrm{m}$ (purple
circles); and RUQs detected by both SPIRE and WSRT at $250\,\mu\textrm{m}$
and $1400\,\textrm{MHz}$ (blue circles), respectively are also displayed.}
\end{figure*}

\subsection{Star formation rates and stellar masses \label{sec:sf_rates}}

\begin{figure}[tp]
\centering{}\includegraphics[clip,scale=0.34]{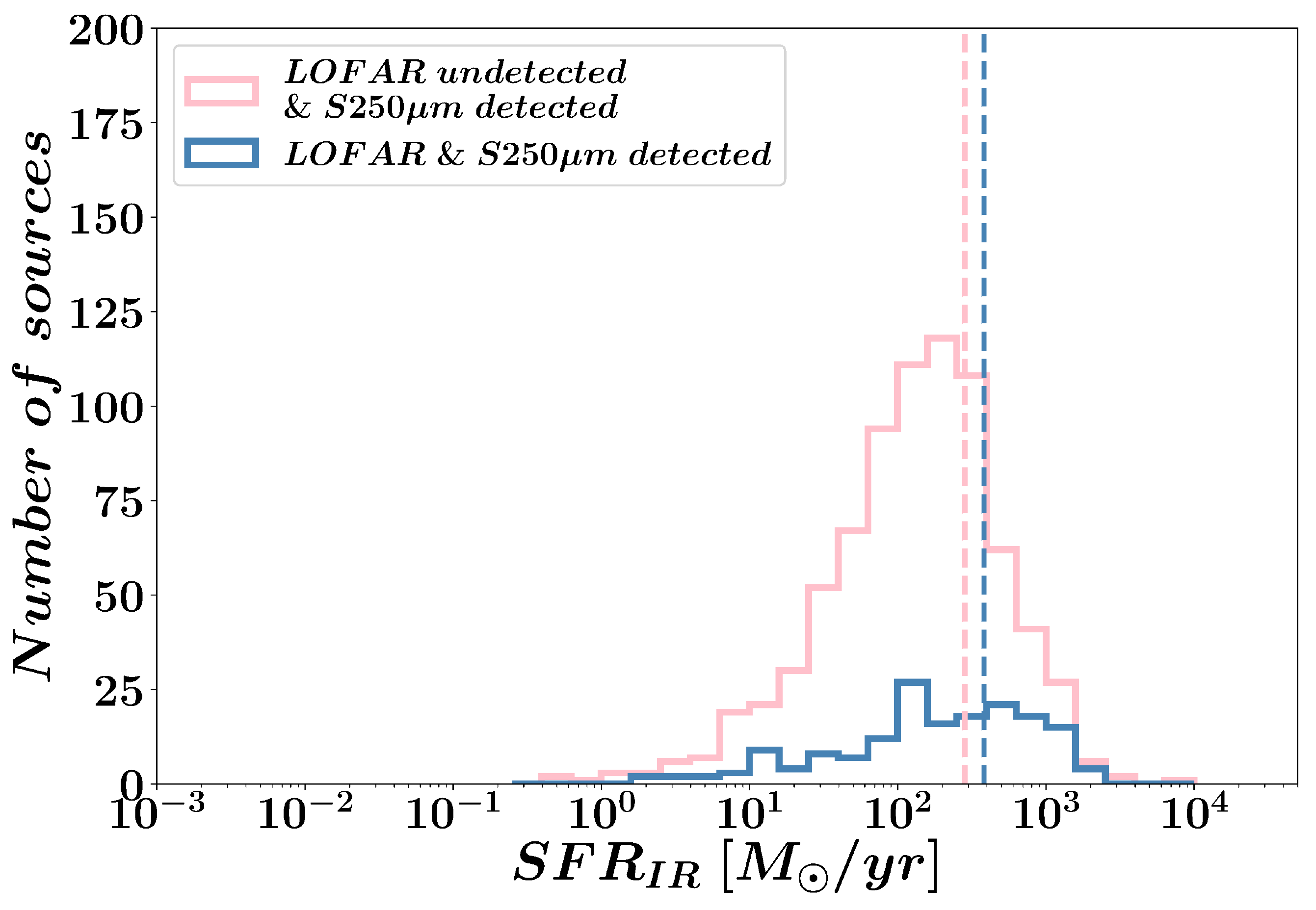}\centering\caption{\label{fig:SFR_histo} Star formation rate (SFR) distribution for
the RDQs and RUQs derived from the infrared luminosity $L_{\textrm{IR}}$
with $\protect\geq2\sigma$ SPIRE-$250\,\mu\textrm{m}$ detections.
The mean SFR of the samples are indicated by the dashed lines. The
mean SFRs of our quasar samples are consistent with those of the main
sequence star-forming galaxies. }
\end{figure}

In this section, we compare the SFRs derived using the infrared luminosity
$L_{\textrm{IR}}$ (see Section \ref{subsec:sed_modelling}) and the
radio luminosities at $150\,\textrm{MHz}$ and $1.4\,\textrm{GHz}$
to investigate the origins of the radio-emission in RUQs. Both SFR
estimates are derived independently. The SFR estimate derived from
radio observations, $\textrm{SFR}_{\textrm{Radio}}$, is obtained
assuming that the synchrotron emission produced in the host galaxies
of RUQs is associated with supernova remnants with short cooling times.
Thus, the radio emission can be considered as a measure of the instantaneous
SFR in the host galaxy \citep{1992ARA&A..30..575C}. From the $1.4\,\textrm{GHz}$
radio luminosity, ${\displaystyle L_{1.4\,GHz}}$, this is carried
out using the following relation \citep{2001ApJ...554..803Y}:

\begin{equation}
\textrm{SFR}_{\textrm{Radio}}\:\left[M_{\odot}\:\textrm{yr}^{-1}\right]=5.9\times10^{-22}\times{\displaystyle L_{1.4\,GHz}}.\label{eq:yun_law_2001}
\end{equation}

\noindent The SFR from the $150\,\textrm{MHz}$ radio luminosity,
${\displaystyle L_{150\,MHz}}$, is calculated using the relations
presented by \citet{2017MNRAS.469.3468C}. These relations were derived
using early LOFAR radio-continuum observation of the NDWFS-Bo\"otes
field \citep{2016MNRAS.460.2385W} for a sample of spectroscopic galaxies
between $0\leq z\leq3.5$. 

\noindent In Figure \ref{fig:sfr_150_1400}, we compare the SFRs derived
from the infrared and radio luminosities. We also plot the FIRC correlation
for star-forming galaxies (from \citealt{2017MNRAS.469.3468C}) to
establish whether or not SF could account for the radio-emission in
RUQs. In agreement with our results of the Section \ref{sec:firc},
we find that the data points of the median RUQs stacked according
to their redshift lie along the FIRC at 150 MHz. However, the data
point at $z=0.7$ (see text in right panel. Fig. \ref{fig:firc_150_1400})
shows a slightly larger scatter than the rest of the other data points
at $1.4\,\textrm{GHz}$. However, these results suggest that the SFRs
of the median RUQs agree with the predictions of the FIRC. Their radio-emission
could be explained by star-forming processes assuming that the FIRC
is a valid indicator to determine the origins of the radio-emission
in RUQs (see discussion in Section \ref{subsec:mechanism}). The situation
for the RUQs with upper limits on $250\,\mu\textrm{m}$ and radio
bands is more difficult to assess on a per-source basis. The upper
limits only indicate that these sources could be located close to
the FIRC but their exact location in the $\textrm{SFR}_{\textrm{Radio}}-\textrm{SFR}_{\textrm{IR}}$
plane is uncertain without deeper observations. The majority of RUQs
detected by WSRT (purple circles), but not by LOFAR show values that
exceed the predictions of the FIRC, thus, their radio-emission is
above that expected from star formation alone suggesting that it is
AGN-dominated. At $150\,\textrm{MHz}$, the median RUQs stacked using
$M_{\textrm{1450}}$ present a trend that is similar to that found
when the stacking is done according to redshift with values following
the FIRC predictions. At $1.4\,\textrm{GHz}$, the $M_{\textrm{1450}}$
-stacked RUQs have $\textrm{SFR}_{\textrm{1.4GHz}}$ values larger
than the FIRC predictions. However, these values are still consistent
with being dominated by SF. In summary, the trends found for the SFRs
are similar to those found in Section \ref{sec:firc} with the comparison
between the infrared and radio luminosities. 

The distribution of the SFRs in our quasar sample derived using the
infrared luminosity $L_{\textrm{IR}}$ is shown in Figure \ref{fig:SFR_histo}.
Only quasars detected by SPIRE at $250\,\mu\textrm{m}$ at $\geq2\sigma$
significance are plotted in order to have robust estimations of the
$\textrm{SFR}_{\textrm{IR}}$. The mean $\textrm{SFR}_{\textrm{IR}}$
of the RUQs with $\textrm{S}250\,\mu\textrm{m}$ detections is $149\:M_{\odot}\:\textrm{yr}^{-1}$,
while the corresponding $\textrm{SFR}_{\textrm{IR}}$ of RDQs is $196\:M_{\odot}\:\textrm{yr}^{-1}$.
These values are consistent with the median $\textrm{SFRs}_{\textrm{IR}}$
of the median quasars, which are $125\:M_{\odot}\:\textrm{yr}^{-1}$
and $267\:M_{\odot}\:\textrm{yr}^{-1}$ for RUQs and RDQs, respectively.
It is clear that with SFRs of only a few hundred $M_{\odot}\:\textrm{yr}^{-1}$,
the SF processes occurring in the quasars in our sample is less vigorous
in comparison with other far-infrared sources with very powerful ongoing
starbursts such as sub-millimeter galaxies \citep{2011MNRAS.415.1479W,2013Natur.498..338F,2014ApJ...784....9B},
and luminous optical quasars \citep{2016MNRAS.462.4067P,2017ApJ...836....8T,2017A&A...604A..67D}.
These sources usually present SFRs at the level of thousands of $M_{\odot}\:\textrm{yr}^{-1}$
at redshifts $z>2$. These powerful starbursts are associated with
the most massive and rarest galaxies at any redshift \citep{2012MNRAS.421..621B,2013A&A...556A..55I}
and, therefore, they have low spatial densities. Due to these low
numbers their potential contribution to our median SFR measurements
is negligible.

\begin{figure}[tp]
\centering{}\includegraphics[clip,scale=0.33]{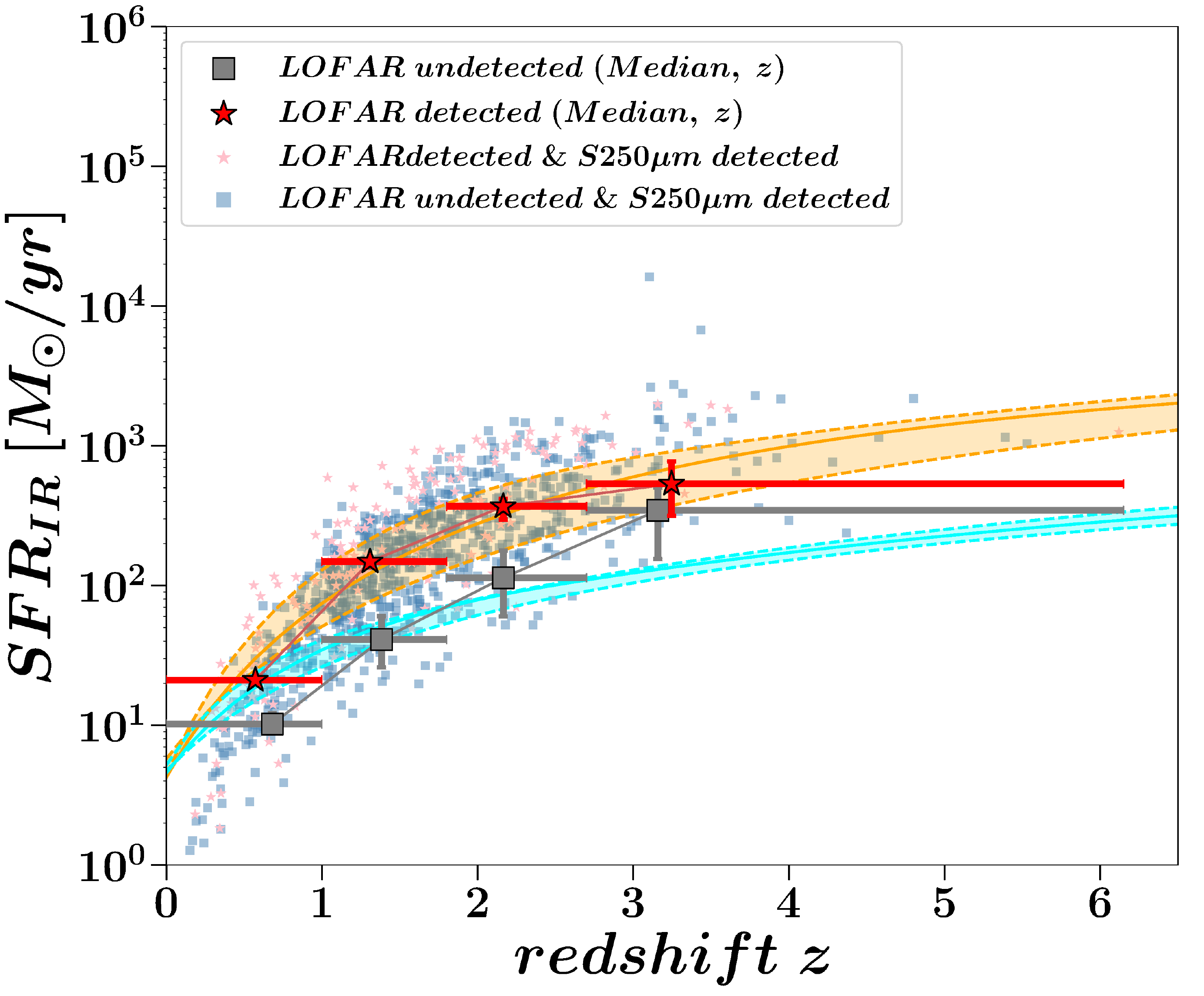}\centering\caption{\label{fig:RDQ_RUQ_comparison} {\small{}Infrared star formation }rate
($\textrm{SFR}_{\textrm{IR}}$) of LOFAR radio-detected quasars (RDQs)
and LOFAR radio-undetected quasars (RUQs) versus redshift. The filled
stars and squares and associated vertical error bars show the median value in each bin. 
The horizontal error bars indicate the extend of the redshift bins used.
The marron stars and cyan squares show the results for the RDQs and RUQs,
respectively, with $\protect\geq2\sigma$ SPIRE-$250\,\mu\textrm{m}$
detections. 
The solid lines represent the predicted SFRs as a function of redshift
derived for the median quasars using the star-formation main sequence
(SFMS) relation by \citet{2015A&A...575A..74S}. The cyan and orange
shaded regions indicate the uncertainty intervals of the fitting provided
by \citet{2015A&A...575A..74S}. The vertical errror bars indicate
the errors of the $\textrm{SFR}_{\textrm{IR}}$ estimates obtained
using the SED fitting, while the horizontal errors indicate the size
of the redshift intervals considered. There is a good agreement between
the measured median $\textrm{SFR}_{\textrm{IR}}$ for RDQs and RUQs,
and the SFMS predictions considering the width of the redshift bins
employed and the intrinsic scatter of the SFMS relation. }
\end{figure}

In Figure \ref{fig:RDQ_RUQ_comparison}, we display the $\textrm{SFRs}_{\textrm{IR}}$
derived from the infrared luminosities, as a function of redshift
for RUQs and RDQs. For the sake of clarity, only objects with SPIRE
$250\,\mu\textrm{m}$ detections at $\geq2\sigma$ are plotted. The
larger symbols represent the measurements obtained from the stacking
analysis. We see an increase in the median $\textrm{SFR}_{\textrm{IR}}$
of quasars with cosmic look-back time. At $z\sim0.6$, we find the
median $\textrm{SFR}_{\textrm{IR}}$ of RDQs (RUDs) to be $21.12\:_{-1.02}^{+0.95}M_{\odot}\:\textrm{yr}^{-1}$
$(10.13_{-2.02}^{+1.26}\:M_{\odot}\:\textrm{yr}^{-1})$, whereas at
$z\sim1.35$ the increment is approximately seven (four) times greater,
and in our high-z bins at $z\sim2.16$ and $z\sim3.2$ the $\textrm{SFR}_{\textrm{IR}}$
is thirty four (twenty five) times greater. These ranges of median
$\textrm{SFR}_{\textrm{IR}}$ are wide, however, the trend of median
$\textrm{SFR}_{\textrm{IR}}$ increasing with redshift in quasars
is clear and consistent with previous studies \citep{2011MNRAS.413.1777S,2014MNRAS.442.1181K,2017MNRAS.472.2221S}.
Finally, RDQs present higher $\textrm{SFR}_{\textrm{IR}}$ values
than RUQs at all redshifts, except at the last $z$-bin where the
$\textrm{SFR}_{\textrm{IR}}$ values between the two quasar samples
are consistent taking into account the error bars.

Studies of star-forming galaxies at a wide range of cosmic epochs
have revealed a strong correlation called the SF main sequence (SFMS,
hereafter) at fixed redshift between SFR and stellar mass $\left(M_{*}\right)$.
This relationship has been shown to hold over four to five orders
of magnitude in mass \citep{2009A&A...504..751S,2017ApJ...847...76S}
and from $z=0$ to $z=6$ \citep{2007ApJ...660L..43N,2007ApJ...670..156D,2014ApJ...791L..25S,2014ApJS..214...15S,2015A&A...575A..74S},
with only a scatter of $0.25-0.35\;\textrm{dex}$ at any redshift
\citep{2007ApJ...670..156D,2012ApJ...754L..29W}. Using the SFMS formula
derived by \citet{2015A&A...575A..74S}, we find that the median stellar
masses of the median RUQs and RDQs are $M_{*}\sim5\times10^{10}\:M_{\odot}$
and $M_{*}\sim3\times10^{11}\:M_{\odot}$, respectively. The solid
lines represent the SFRs as a function of redshift derived for the
median quasars using the \citet{2015A&A...575A..74S} relation. There
is a good agreement between the measured median $\textrm{SFR}_{\textrm{IR}}$
for RDQs and RUQs, and the SFMS predictions considering the width
of the redshift bins employed and the intrinsic scatter of the SFMS
correlation. The differences in the scatter between RDQs and RUQs
could be explained considering that the former includes almost entirely
quasars that might be located in massive galaxies due to the radio-selection
being associated with massive systems \citep{2017A&A...600A..97R,2020MNRAS.493.3838M}.
The latter includes quasar host galaxies with a wide range of stellar
masses. Thus, for galaxies hosting RUQs a larger scatter with respect
SFMS could be expected due to the wide range of stellar masses included
in the sample. We conclude the $\textrm{SFRs}_{\textrm{IR}}$ and
stellar mass measurements of galaxies hosting RUQs and RDQs are broadly
consistent with those of massive star-forming galaxies following the
main sequence. Moreover, these estimates are in good agreement with
previous works of main sequence galaxies at $0.2<z<2.5$ \citep{2013A&A...560A..72R,2017MNRAS.472.2221S,2019MNRAS.488.1180S}.
Additionally, the stellar masses of the quasar host galaxies in our
sample aligns well with those estimated for $z<0.8$ quasars obtained
using image decomposition in CFHT \citep{2018ApJ...863...21Y} and
SDDS S82 deep imaging \citep{2014ApJ...780..162M,2015MNRAS.454.4103B}.
These authors found stellar masses larger than $M_{*}>10^{10}\:M_{\odot}$
for the quasar host galaxies, but no distinction was made between
radio-detected and radio-undetected objects. 

\subsection{Median star formation rate as a function of optical and radio luminosities
\label{subsec:luminsity_correlations}}

In this section, we examine the behaviour of the SFR as a function
of the $1450 \> {\angstrom}$ absolute magnitude and 150 MHz radio luminosities.
For this purpose, we stacked our sample following the steps explained
in Sections \ref{sec:stacking_method} and \ref{sec:firc} as a function
of $1450 \> {\angstrom}$ absolute magnitude and 150 MHz radio luminosity.
We used bins of size\textbf{ }$\triangle\log\left(L_{150MHz}\right)=1.0$
for the radio luminosity, and we use the same bin sizes for $M_{\textrm{1450}}$
that was employed in the previous section.

\begin{figure}[tp]
\centering{}\includegraphics[clip,scale=0.49]{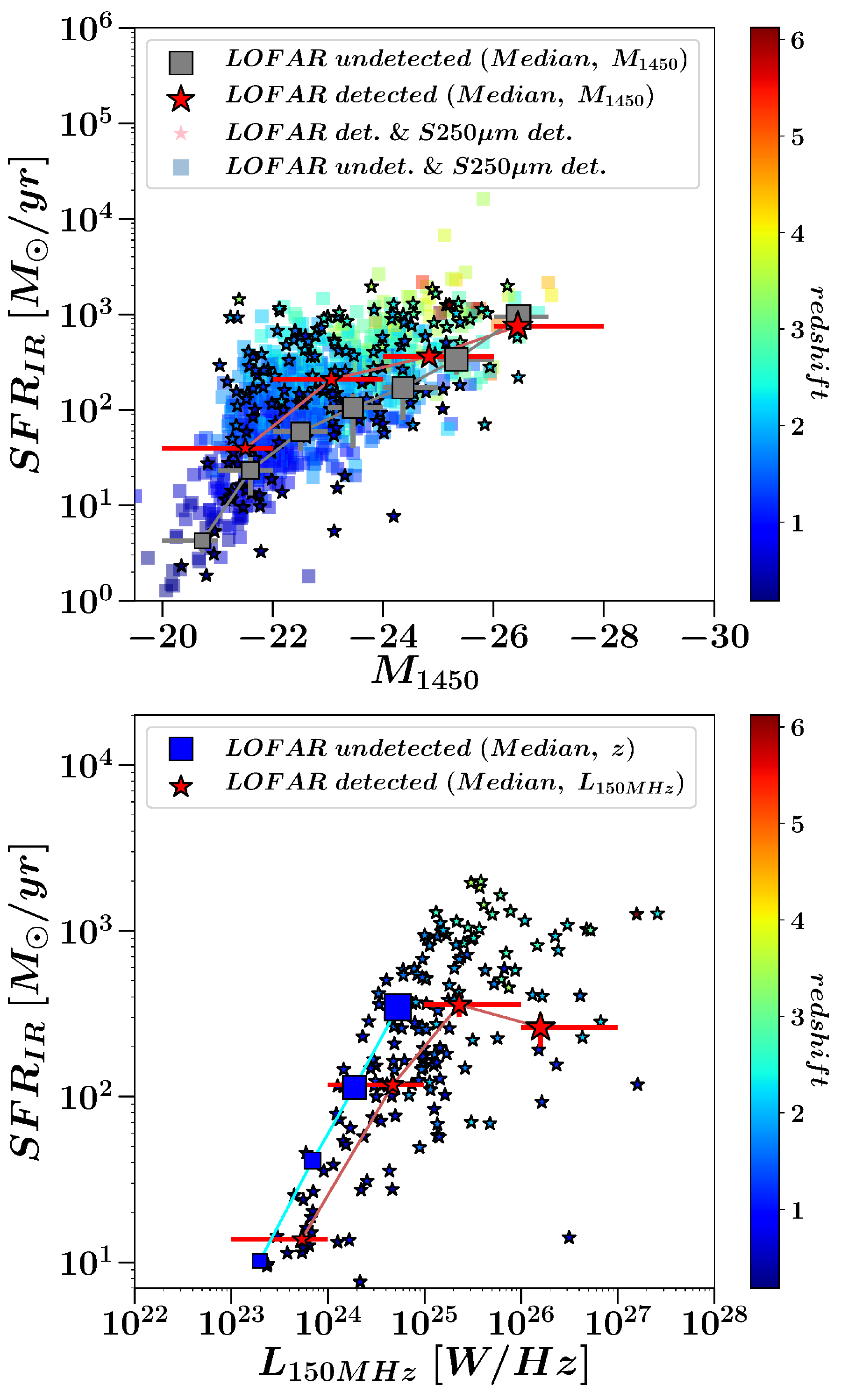}\centering\caption{\label{fig:RDQ_RUQ_SFR_luminosity} {\small{}Infrared star-formation rates (}$\textrm{SFRs}_{\textrm{IR}}${\small{})
versus absolute magnitude at $1450  \> {\angstrom} $, $M_{\textrm{1450}}$,
(top panel) and 150 MHz radio luminosity (bottom panel) for LOFAR
radio-detected quasars (RDQs ) and radio-undetected quasars (RUQs)}.
Only RDQs (stars) and RUQs (squares) with $\protect\geq2\sigma$ SPIRE-$250\,\mu\textrm{m}$
detections are shown (see Section \ref{sec:fir_ir_data}). For median
quasars, the increasing symbol size indicates an increment in the
redshift, $M_{1450}$ absolute luminosity, or 150 MHz radio luminosity.
The errors on the horizontal axis indicate the extend of the luminosity
bins used. Vertical error bars are the $1\sigma$ uncertainties for
the $\textrm{SFRs}_{\textrm{IR}}$ obtained from the SED fitting. The
positive correlation between optical and radio luminosities suggest
that there is a link between their SFR and AGN activity in median
RDQs and RUQs, with indications of star formation being suppressed
in the host galaxies of the most luminous radio quasars. }
\end{figure}

\noindent Figure \ref{fig:RDQ_RUQ_SFR_luminosity} shows the distributions
of $1450 \> {\angstrom}$ absolute magnitudes, 150 MHz radio luminosities,
and $\textrm{SFRs}_{\textrm{IR}}$. Both panels shows a correlation
between median optical and 150 MHz radio luminosites and SFR for quasars
which extends over more than two orders of magnitude in $\textrm{SFR}_{\textrm{IR}}$.
The first thing that comes to mind are the similarities and differences
between the behavior of the median RDQs and RUQs. The first similarity
is that both median quasar samples present a positive trend of $\textrm{SFR}_{\textrm{IR}}$
as function of the $1450 \> {\angstrom}$ absolute magnitude in which
$\textrm{SFR}_{\textrm{IR}}$ increases with $M_{\textrm{1450}}$.
The main difference is that for $M_{\textrm{1450}}\geq-24.0$ median
RDQs (red stars) have higher $\textrm{SFRs}_{\textrm{IR}}$ in comparison
with their radio-undetected counterparts (gray squares). For $M_{\textrm{1450}}\leqslant-24.5$
median RDQ and RUQs present similar $\textrm{SFR}_{\textrm{IR}}$
values within their corresponding uncertainties. This agrees with
previous results found by \citet{2014MNRAS.442.1181K} for RLQs and
RQQs. These authors found that RLQs present higher SFRs than RQQs
at low and medium optical luminosities but have comparable SFRs at
high optical luminosities.

Assuming that the far-infrared is a good proxy of SF, we conclude
that these trends discussed before suggest that there is a link between
their SFR and AGN activity in median RDQs and RUQs. Previous studies
found a correlation between the far-infrared and accretion luminosities,
supporting the idea that SF and AGN activity share the same reservoir
of cold gas in the host galaxy (e.g., \citealt{2009MNRAS.397..265S,2010ApJ...712.1287L,2011MNRAS.416...13B}).
It is possible that the different efficiencies in which gas fuels
SF and AGN processes or the heating of gas via feedback processes
may explain the observed behaviors in the $\textrm{SFR}_{\textrm{IR}}-M_{\textrm{1450}}$
correlation. The existence of gas reservoirs similar to those of non-active
galaxies has been confirmed in low-z \citep{2020ApJS..247...15S}
and high-z \citep{2011ApJ...739L..32R,2020MNRAS.493.3744F,2020MNRAS.496..138B}
quasars. However, CO observations are required to determine better
constraints for quasar sample.

In the bottom panel of Figure \ref{fig:RDQ_RUQ_SFR_luminosity}, we
show the distribution of median $\textrm{SFRs}_{\textrm{IR}}$ for
RDQs as a function of 150 MHz radio luminosity. We also plot for reference
the results for median RUQs (median stacked by redshift bins) and
SPIRE-$250\,\mu\textrm{m}$ detections. We find again a rise in the
$\textrm{SFR}_{\textrm{IR}}$ with increasing 150 MHz radio luminosity.
Median RUQs are shifted to the left as expected due to being less
luminous in the radio in comparison with the median RDQs. We see that
median RDQs preset a flattening, or possible decline in $\textrm{SFR}_{\textrm{IR}}$
in the highest $L_{150MHz}$ bin. This decline roughly coincides with
the spread of the individual detections. This contrasts with the top
panel of Figure \ref{fig:RDQ_RUQ_SFR_luminosity} that shows no downturn
in $\textrm{SFR}_{\textrm{IR}}$ at the highest $M_{\textrm{1450}}$
values. This perhaps could indicate that SF is suppressed in host
galaxies of luminous radio quasars due to the impact of radio-jets.
Additionally, this is consistent with the trend found in the left
panel of Figure \ref{fig:firc_150_1400}: the radio-emission of the
most luminous RDQs exceeds the FIRC predictions and it is likely AGN-dominated
(discussed in more detail in Section \ref{subsec:agn_feedback}).

\subsection{Redshift evolution of the radio spectral indices \label{subsec:spectral_indices}}

\begin{figure*}[tp]
\centering{}\includegraphics[clip,scale=0.5]{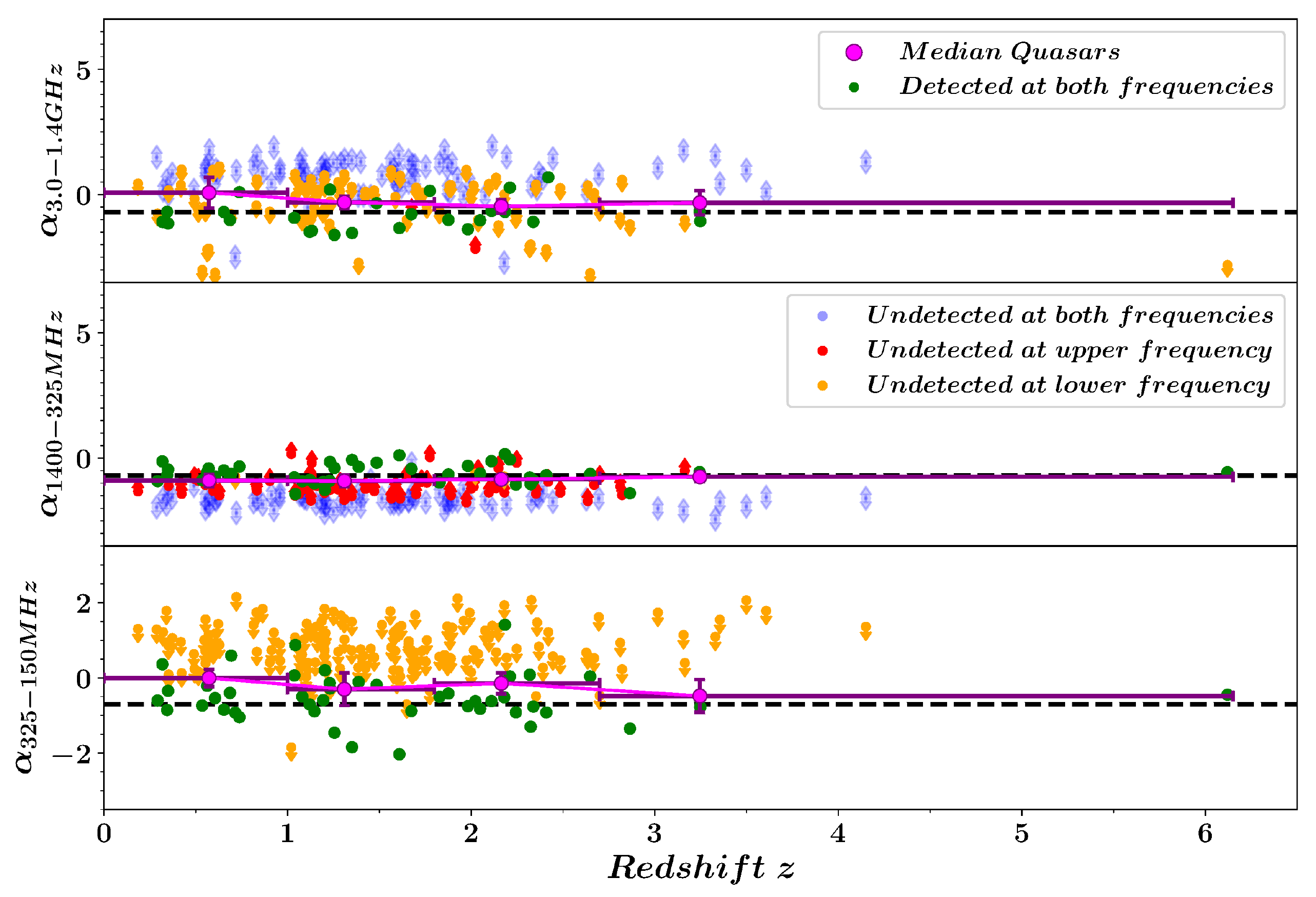}\centering\caption{\label{fig:spectral_indices_rdqs} Evolution of spectral indexes $\alpha_{3.0-1.4}$
(top), $\alpha_{1400-325}$ (middle ), and $\alpha_{325-150}$ (bottom)
as a function of redshift for LOFAR radio-detected quasars (RDQs).
The purple circles represent the median RDQs; green circles are the
RDQs detected at both frequencies; yellow or red circles indicated
RDQs that are undetected in the upper or lower frequency, respectively;
and blue circles are RDQs undetected at both frequencies. The horizontal
dashed line indicates the canonical value of $\alpha=-0.7$.}
\end{figure*}
 
\begin{figure*}[tp]
\centering{}\includegraphics[clip,scale=0.5]{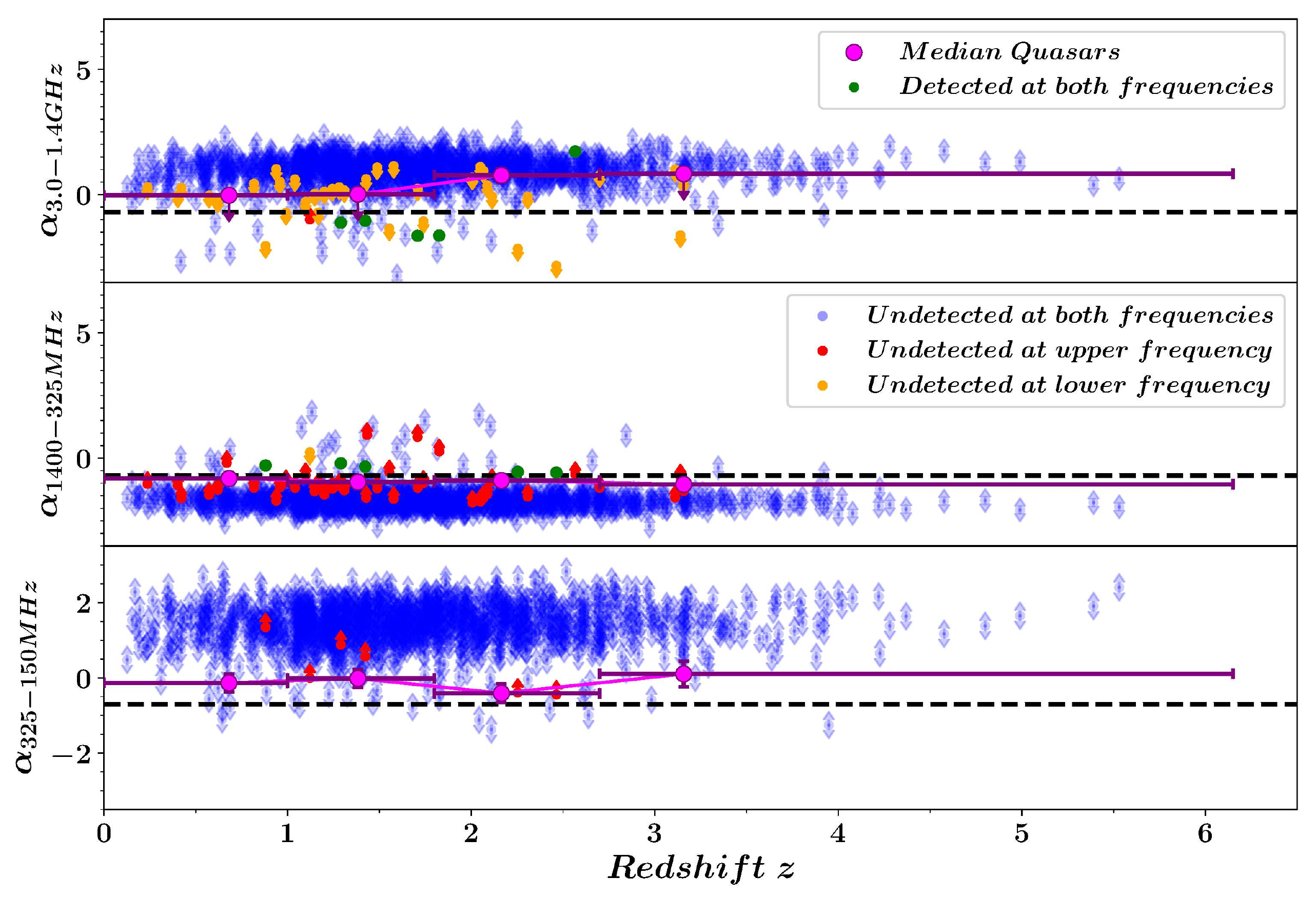}\centering\caption{\label{fig:spectral_indices_ruqs} Evolution of spectral indexes $\alpha_{3.0-1.4}$
(top), $\alpha_{1400-325}$ (middle ), and $\alpha_{325-150}$ (bottom)
as a function of redshift for LOFAR radio-undetected quasars (RUQs).
The purple circles represent the median RUQs; green circles are the
RUQs detected at both frequencies; yellow or red circles indicated
RUQs that are undetected in the upper or lower frequency, respectively;
and blue circles are RUQs undetected at both frequencies. The horizontal
dashed line indicates the canonical value of $\alpha=-0.7$.}
\end{figure*}

\noindent We calculated the spectral index between adjacent frequency
pairs at 150 MHz, 325 MHz, 1.4 GHz and 3.0 GHz. For quasars that are
detected in both frequencies, we derived an upper limit for the spectral
index when the quasar is only detected in one (two) frequency (ies)
in the pair. The spectral indices are calculated using the formula:

\begin{equation}
\alpha_{\nu_{1}}^{\nu_{2}}={\displaystyle \frac{\log\left(S_{\nu_{1}}/S_{\nu_{2}}\right)}{\log\left(\nu_{2}/\nu_{1}\right)}},\label{eq:spectral_index}
\end{equation}

\noindent where $\alpha_{\nu_{1}}^{\nu_{2}}$ is the spectral index
between the corresponding frequencies, $S_{\nu_{i}}$ denotes the
flux density at frequency $v_{i}$. In the cases of non-detections,
we use forced photometry in the corresponding frequency (ies) as described
in Section \ref{subsec:sed_modelling}. The redshift evolution the
spectral indices are presented in Figures \ref{fig:spectral_indices_rdqs}
and \ref{fig:spectral_indices_ruqs} for RDQs and RUQs, respectively.
For RDQs, it can be seen that the results for median RDQs within errors
are consisted with no evolution a function of redshift for the three
spectral indices. The robustness of this result is verified by fitting
a straight line to the data points. It is found that the slope of
the straight line is close to zero, which rules out a redshift evolution
for the spectral indices. This agrees with previous studies that found
no evidence for the redshift evolution of the radio spectral indexes
\citep{2015A&A...573A..45M,2017MNRAS.469.3468C}. The spectral indices
of the RDQs detected show scatter around the values obtained from
the stacking analysis, while the RDQs with upper limits are still
consistent with these values although with a larger dispersion.

\noindent We also note that for median RDQs the mean for the spectral
index between 325 MHz and 1.4 GHz, $\alpha_{1400}^{325}$, is steeper
than the canonical value of $\alpha=-0.7$ used in previous radio
continuum studies with a value of $\alpha_{1400}^{325}=-0.85\pm0.10$.
For comparison, our result is in good agreement with the published
mean value by \citet{2009MNRAS.395..269S} of $\alpha_{1400}^{325}=-0.83$,
but it is steeper than the mean value reported of $\alpha_{1400}^{325}=-0.71$
by \citet{2013MNRAS.435..650M} for all radio sources in their corresponding
catalogs. For median RUQs, the mean spectral index is steeper with
a value of $\alpha_{1400}^{325}=-0.97\pm0.07$. The spectral index
between 1400 MHz and 3000 MHz, $\alpha_{3000}^{1400}$, for median
RDQs has a value of $\alpha_{3000}^{1400}=-0.26\pm0.05$, while for
median RUQs the shallower flux density limit of the 3.0 GHz VLASS
mosaic makes only possible to obtain an estimate with $>3\sigma$
significance in one redshift bin. For the remaining three redshift
bins only upper limits are estimated. The mean spectral index of median
RUQs is $\alpha_{3000}^{1400}=0.39\pm0.20$, calculated using the
Kaplan-Meier estimator (e.g., \citealt{1985ApJ...293..192F}) to account
for the upper limits. Our values are flatter in comparison to the
mean value of $\alpha_{3000}^{1400}=-0.68\pm0.02$ found by \citet{2017A&A...602A...1S}.
We find that that the mean spectral index between 325 MHz and 150
MHz, $\alpha_{150}^{325}$, has a value of $\alpha_{150}^{325}=-0.23\pm0.06$
($-0.11\pm0.03$) for median RDQs (RUQs). In contrast, these values
are flatter than those reported by \citet{2017MNRAS.469.3468C} of
$\alpha_{150}^{325}=-0.87\pm0.04$ and $\alpha_{150}^{325}=-0.60\pm0.05$
for AGN and star-forming galaxies, respectively. However, our results
are consistent with the spectral flattening of radio sources towards
low-frequencies previously observed in LOFAR \citep{2014ApJ...793...82V,2016MNRAS.463.2997M}
and GMRT \citep{2011A&A...535A..38I,2013AA...549A..55W} observations.

\section{Discussion \label{subsec:discussion}}

The stacking technique is an important tool in understanding the nature
of radio emission in quasars, as basically all quasars emit in the
radio when observed down to deep limits. In this work, we find evidence
for concurrent AGN and star-forming activity in the host galaxies
of median-stacked quasars with a positive correlation between $\textrm{SFR}_{\textrm{IR}}$
and optical and radio luminosities. These pieces of evidence suggest
that the radio-emission in quasars is the result of a complex interplay
between SF and SMBH accretion processes with each process having significant
contributions to the quasar energy output. In this section, we discuss
and interpret our findings. 

\subsection{No bimodality for the radio-loudness parameter \label{subsec:radioloudness}}

Here, we investigate the radio-loudness parameter using the results
of our analysis stacking, and discuss the origins of the radio-emission
in quasars with the lowest radio-loudness values. 

\noindent We define the radio-loudness parameter ($R$ hereafter)
for our quasar samples using the ratio of radio to infrared AGN luminosity:

\begin{equation}
R={\displaystyle \log_{\textrm{10}}\left(\frac{L_{\textrm{rad}}}{L_{\textrm{AGN}}}\right)},\label{eq:radio_loudness}
\end{equation}

\noindent where $L_{\textrm{rad}}$ is the radio luminosity at 150
MHz or 1.4 GHz, and $L_{\textrm{AGN}}$ is the infrared AGN luminosity
estimated in Section \ref{subsec:sed_modelling}. We decided to use
the infrared AGN luminosity instead of an optically derived luminosity.
This is done because the mid-infrared continuum of the majority of
quasars arises from dust heated by the central AGN \citep{2006ApJS..166..470R,2007ApJ...661...30G}.
This makes the mid-infrared luminosity robust against extinction effects. 

\begin{figure*}
\centering{}\includegraphics[clip,scale=0.4]{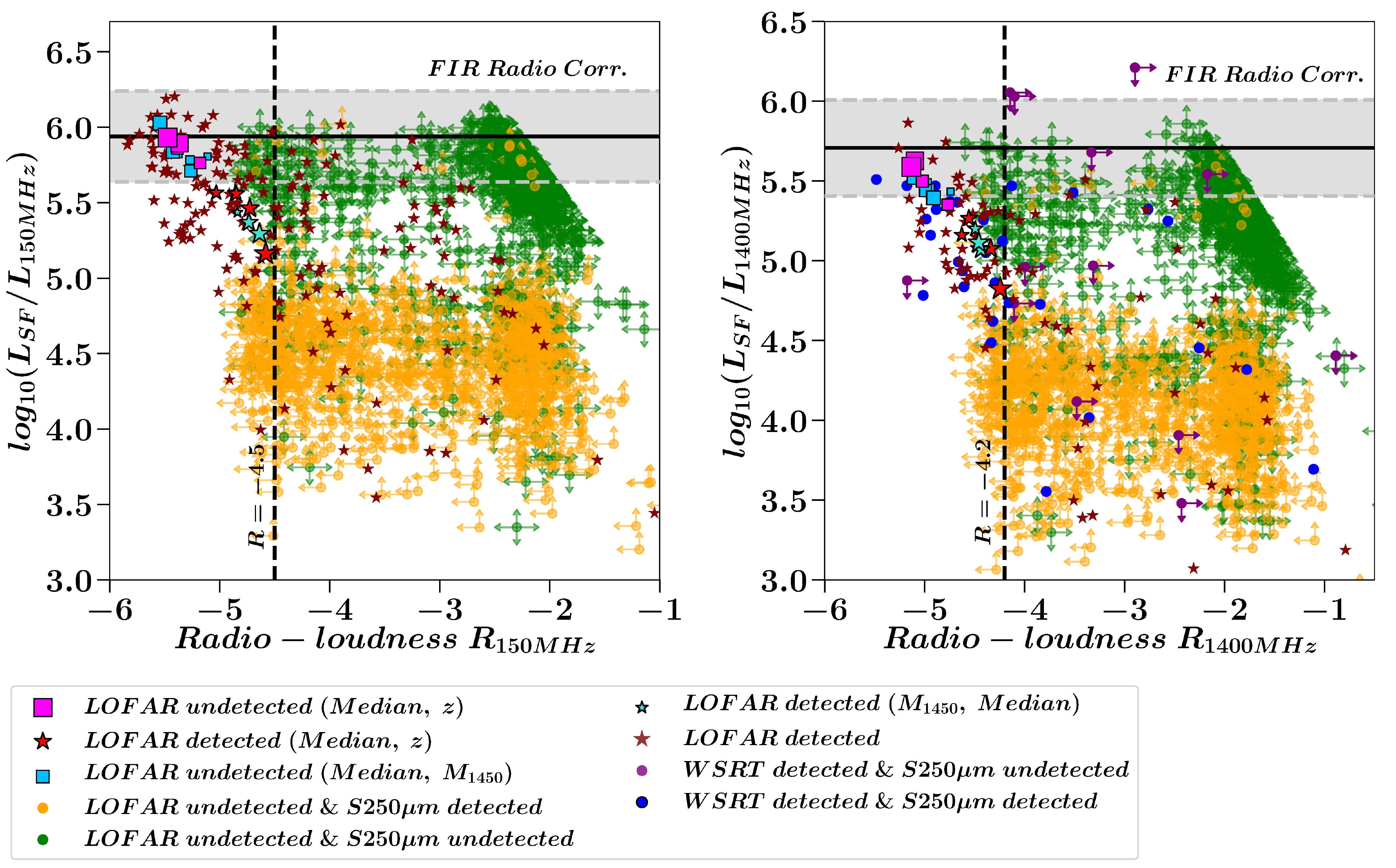}\centering\caption{\label{fig:radio_loudness} Infrared star formation luminosity, $L_{\textrm{SF}}$,
versus radio-loudness, $R$, for the quasar samples. The measurements
for median LOFAR radio-undetected quasars (RUQs, fuchsia squares),
median LOFAR radio-undetected quasars (RDQs, red stars), RDQs (maroon
stars) and RUQs detected by SPIRE at $250\,\mu\textrm{m}$ (orange
circles), and RUQs undetected by SPIRE and WSRT at $250\,\mu\textrm{m}$
and $1.4\,\textrm{GHz}$ (green circles), respectively. The measurements
for RUQs detected by WSRT, but undetected by SPIRE at $250\,\mu\textrm{m}$
(purple circles, right); and RUQs detected by both SPIRE and WSRT at $250\,\mu\textrm{m}$
and $1.4\,\textrm{GHz}$ (blue circles, right), respectively are also displayed.
Upper limits in either $L_{\textrm{SF}}$ or ${\displaystyle L_{150\,MHz}}$
are indicated by arrows. The solid line is the far-infrared radio
correlation (FIRC) at 150 MHz and 1.4 GHz. The gray shaded region
indicates the spread of the FIRC scaled by a factor of $\pm2$. The
dashed vertical lines ($R=-4.5$ at 150 MHz and $R=-4.2$ at 1.4 GHz)
denote the threshold between radio-loud and radio-quiet quasars (see
Section \ref{subsec:discussion} for more details). The radio-emission
of bright RDQs with $R\gtrsim-4.5$ is consistent with being dominated
by SMBH accretion, while for lower radio luminosity quasars with $R<-4.5$
the relative contribution of SF to their radio fluxes becomes more
important.}
\end{figure*}

In Section \ref{sec:firc}, we demonstrate that the SF could dominate
the radio-emission of RUQs and a significant fraction of RDQs. As
explained in Sections \ref{sec:firc} and \ref{sec:sf_rates}, we
define a quasar to be SF-dominated if the quasar is within a factor
of $\pm2$ of the expected value predicted by the FIRC. Figure \ref{fig:radio_loudness}
shows the ratio of $L_{\textrm{SF}}/L_{\textrm{rad}}$ as a function
of $R$ for our quasar samples. The FIRC is denoted by a solid line
in both panels, and the shaded gray horizontal regions indicate the
FIRC scaled by a factor of $\pm2$. The dashed vertical lines indicates
the thresholds of $R=-4.5$ (150 MHz) and $R=-4.2$ (1.4 GHz) utilize
to classify quasars into radio-loud/radio-quiet (e.g., \citealt{2019MNRAS.488.3109K,2020MNRAS.494.3061R}).
We see in both panels of Figure \ref{fig:radio_loudness} that median
RUQs (stacked according to their redshift and $M_{\textrm{1450}}$) and a significant fraction of RDQs have $R$ values below the thresholds
at 150 MHz and 1.4 GHz. These quasars would be classified as radio-quiet
in previous works. Moreover, many of these quasars fall within the
region delimited by $\pm2$ FIRC factors suggesting that their radio
emission dominated by SF, which is in agreement with the results obtained
in Section \ref{sec:firc}. On the other hand, the radio fluxes of
median RDQs (stacked according to their redshift and $M_{\textrm{1450}}$)
are AGN-dominated according to their location outside the predictions
of the FIRC. However, some of them have $R$ values below the classification
thresholds. It is clear that there is a trend for RDQs with high $R$
values (thus high radio-luminosities) to have their radio-emission
dominated by SMBH accretion. This should not come as a surprise, based
on their predominant FR-II-like \citep{1974MNRAS.167P..31F} radio-morphologies.
Representative examples of their radio-morphologies can be found in
\citet{2020AA}, Appendix A. On the other side, for low $R$ values
$\leq-4.5$, around $50\%$ of the RDQs have faint and compact radio
morphologies suggesting that AGN processes are not longer dominating
the radio-emission as they do at higher $R$ values. For these RDQs,
the relative contribution to the radio emission due to SF increases
as the AGN component becomes weaker towards lower $R$ values. Finally,
we also notice that WSRT RUQs (blue circles, right panel) show a similar
trend to that of the LOFAR RDQs (maroon stars, left panel). For the
majority of individual RUQs it is difficult to make an assessment
as these are non-detections in the LOFAR, WSRT, and $\textrm{S}250\,\mu\textrm{m}$
bands and only upper limits can be estimated on their $R$ and $L_{\textrm{SF}}/L_{\textrm{rad}}$
values. However, most likely these quasars follow a similar trend
and occupy a comparable region to that of the median RUQs (stacked
according to redshift and $M_{\textrm{1450}}$).

\begin{figure}[tp]
\centering{}\includegraphics[bb=0bp 0bp 788bp 771bp,clip,scale=0.33]{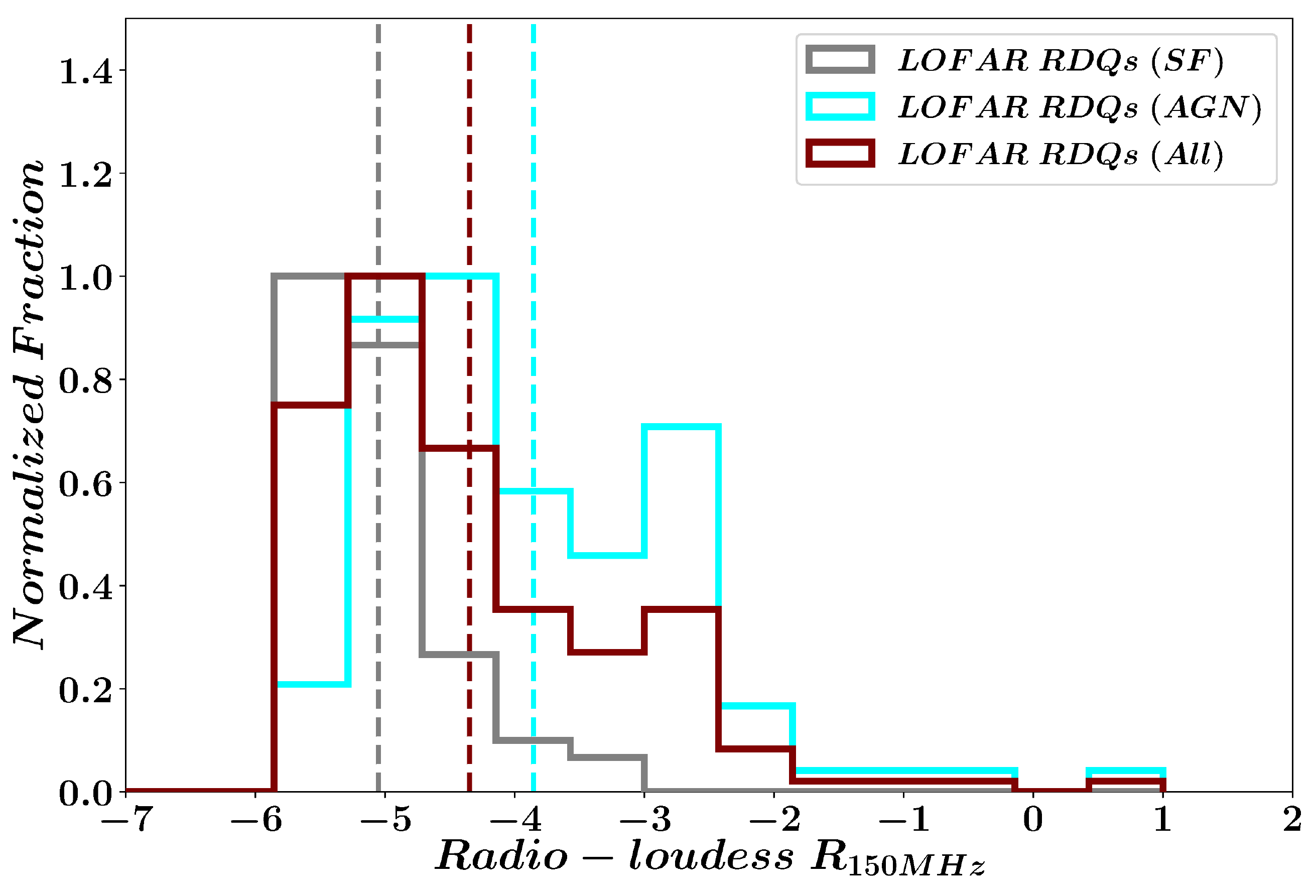}\centering\caption{\label{fig:radioloudness_histo} Distributions of the radio-loudness
parameter, $R$, derived using the 150 MHz radio luminosity for different
LOFAR radio-detected quasar (RDQs) samples. The means of each distribution
is indicated by the corresponding dashed line.}
\end{figure}

\noindent Another important conclusion that can be inferred from Figure
\ref{fig:radio_loudness} is the high fraction of RDQs with values
of $R\leq-4.5$ ($R\leq-4.2$) is $55\%$ ($53\%$) at 150 MHz (1.4
GHz), respectively. Specifically, the fraction of SF-dominated RDQs
at 150 MHz (1.4 GHz) with $R\lesssim-4.5$ ($R\lesssim-4.2$) is $28\%$
($13\%$). The higher sensitivity of LOFAR for objects with typical
synchrotron spectra makes possible to detect the radio emission of
a significant number of these quasars. Figure \ref{fig:radioloudness_histo}
shows the distribution of the $R$ parameter for the RDQs samples
in our study. It can be seen that the distributions of all RDQs and
SF-dominated RDQs with $R\leqslant-4.5$ present no obvious bimodality,
while the distribution of AGN-dominated RDQs with $R>-4.5$ exhibits
two peaks. We fit these distributions to a single Gaussian and a two-component
Gaussian mixture using Markov chain Monte Carlo. The odd ratio for
the models is $O_{\textrm{12}}\gtrsim1.0$, which suggests that neither
model is favored by the data. This suggests that the distribution
of the $R$ parameter is not bimodal, and there is not a dichotomy
as previously reported in the literature (e.g., \citealt{1989AJ.....98.1195K,1990MNRAS.244..207M,2007ApJ...654...99W}).
Our result agrees with previous works where the $R$ parameter exhibits
a continuous distribution rather than a clear bimodality (e.g., \citealt{2001ApJ...551L..17L,2019A&A...622A..11G,2020MNRAS.494.4802F}).

\subsection{Origins of the radio-emission in radio-undetected quasars: Dominant
mechanism or complex interplay \label{subsec:mechanism}}

\noindent The dominant mechanism responsible for the bulk of the radio-emission
in RQQs is a matter of ongoing debate. Many recent studies had pointed
towards synchrotron emission from SF processes as the mechanism source
of radio emission in RQQs (see \citealt{2016A&ARv..24...13P} for
a review). However, the presence of weak small-scale radio jets in
RQQs points towards a non-thermal origin for their radio emission
\citep{2006A&A...455..161L,2016A&A...589L...2H}. The comparison between
the VLBI and lower-resolution VLA images indicate that around $\sim25-50\%$
of the radio emission in these quasars could originate from star-forming
processes \citep{2016A&A...589L...2H}. Similar conclusions have been
reached by other VLBI studies of RQQs and radio-quiet AGN \citep{2005ApJ...621..123U,2016A&A...589L...3M}.

Various studies of RQQs have provided different hints on the origins
of their radio-emission. For instance, \citet{2017MNRAS.468..217W}
analyzed a sample of SDSS 70 RQQs at $z\sim1$  and found that their
RQQs showed an excess of radio emission compared to that expected
from the FIRC, which cannot be explained by SF alone. These authors
concluded that AGN processes could account for $\sim80\%$ of the
total radio luminosity in over $90\%$ of the RQQs included in their
sample. A similar conclusion was reached by \citet{2015MNRAS.448.2665W}
using the radio-stacking technique with 1.4 GHz VLA observations.
\citet{2011AJ....141..182K} constructed the 6.0 GHz radio-luminosity
function from a sample of SDSS quasars at $0.2\leq z\leq0.3$. These
authors concluded that quasars with $L_{\textrm{6.0GHz}}\leq10^{22.5}\:WHz^{-1}$
are mainly driven by SF, while at $L_{\textrm{6.0GHz}}>10^{22.5}\:WHz^{-1}$
the luminosity is related to SMBH accretion. Moreover, the advent
of deep radio surveys offers the possibility of performing studies
that are complementary to the aforementioned works. For instance,
\citet{2015MNRAS.452.1263P} and \citet{2015MNRAS.453.1079B} found
that significant numbers of radio-quiet AGN are powered mainly by
star-forming processes using deep 1.4 GHz VLA observations of the
E-CDFS region.

The analysis presented in Sections \ref{sec:results} and \ref{subsec:radioloudness}
indicates that median RUQs does not show radio excess according to
the FIRC. This suggests that radio emission in the RUQs included in
our sample is to a certain amount powered by star formation in their
host galaxies (e.g., \citealt{2015MNRAS.452.1263P,2020ApJ...901..168A,2020ApJ...903..139A}).
On the other hand, the radio fluxes of the majority of RDQs in our
sample show a radio excess which could be considered as evidence that
supports radio activity related to radio-mode feedback \citep{2007MNRAS.376.1849H,2014ARA&A..52..589H}.
However, the remaining $33\%$ of RDQs in our sample do not show an
radio excess, possibly related to the weakening of the AGN component
contributing to the their radio emission towards lower $R$ values.

We establish the level of contribution of AGN accretion and SF to
the radio-emission of median quasars by calculating the radio luminosity
associated with AGN accretion $L_{\textrm{150MHz, acc}}$:

\begin{equation}
L_{\textrm{150MHz, acc}}=L_{\textrm{150MHz}}-L_{\textrm{150MHz, SF}},\label{eq:accretion_luminosity}
\end{equation}

\noindent where $L_{\textrm{150MHz}}$ is the total radio luminosity,
and $L_{\textrm{150MHz,SF}}$ is the radio luminosity connected to
SF estimated using the FIRC. For the median RDQs stacked according
to redshift, we find that the accretion-related radio luminosity accounts
for $45.3-81.1\%$ the total radio luminosity at 150 MHz. These percentages
for median RDQs stacked according to $M_{\textrm{1450}}$ and 150
MHz radio luminosity are $49.5-72.4\%$ and $56.2-97.2\%$, respectively.
We repeat the same calculation for median RUQs, but these quasars
are SF-dominated and except for one have a negligible contribution
from AGN accretion to their radio-fluxes, resulting in unphysical
$L_{\textrm{150MHz, acc}}$ values according to the definition in
Eq. \ref{eq:accretion_luminosity}. Instead we used the 1400 MHz radio
luminosity to calculate the percentage of radio luminosity due to
AGN accretion. The resulting estimate for the percentage of accretion-related
1400 MHz radio luminosity in median RUQs stacked according to redshift
($M_{\textrm{1450}}$) is $5.4-41.6\%$ ($3.8-40.2\%$). Interestingly,
our findings for RDQs are comparable to those of \citet{2016A&A...589L...2H}
and \citet{2017MNRAS.468..217W}, who showed that AGN accretion can
account $\sim50-83\%$ of the total radio-emission in their RQQs sample.
Finally, we obtained similar percentages at 1400 MHz for RDQs to those
found at 150 MHz.

The percentages found here are within what is expected from our results.
Median RUQs exhibit lower percentages of accretion-related radio\textendash emission
in comparison with median RDQs, with some RUQs almost being dominated
by SF according to the FIRC derived by \citet{2017MNRAS.469.3468C}
at 150 MHz. Median RDQs present higher percentages of accretion-related
radio\textendash emission but still the contribution of SF to their
total radio luminosity is non-negligible. From our results, it is
clear that the radio luminosity in quasars is the result of a complex
interplay between SF and SMBH accretion with each process having a
non-negligible contribution.

Finally, an important point to be considered is the use of the FIRC
to determine the main radio-emission mechanism in quasar samples.
While, it has been shown that the FIRC holds for star-forming galaxies
fainter than $\lesssim150\,\mu\textrm{Jy}$ using both stacking \citep{2009MNRAS.394..105G}
and individual detections \citep{2020ApJ...888...61M}, this correlation
could have an intrinsic bias as the influence of AGN on the SFRs could
affect control samples of non-active star-forming galaxies \citep{2018MNRAS.475.1288S}.
For instance, recent VLBI observations revealed unambiguous evidence
for jet activity in the lensed RQQ HS 0810+2554 \citep{2019MNRAS.485.3009H}.
With a flux density of only $880\:\textrm{nJy}$ at 1.65 GHz, this
is the faintest radio quasar observed until now. Moreover, it was
found that the host galaxy of this lensed quasar follows the FIRC
as it does not display the radio-excess typically associated with
AGN activity \citep{2017MNRAS.472.2221S}. These findings suggest
that radio jet production is able to coexist alongside star-forming
activity in the faintest radio quasars despite their host galaxies
follows the FIRC. This makes clear that the FIRC must be used with
caution to determine the main radio emission mechanism in RUQs. The
question of what is the contribution of radio-jets to the energy output
of RUQs could be only answered satisfactorily with future high-resolution
radio observations. 
\begin{figure*}
\centering{}\includegraphics[clip,scale=0.4]{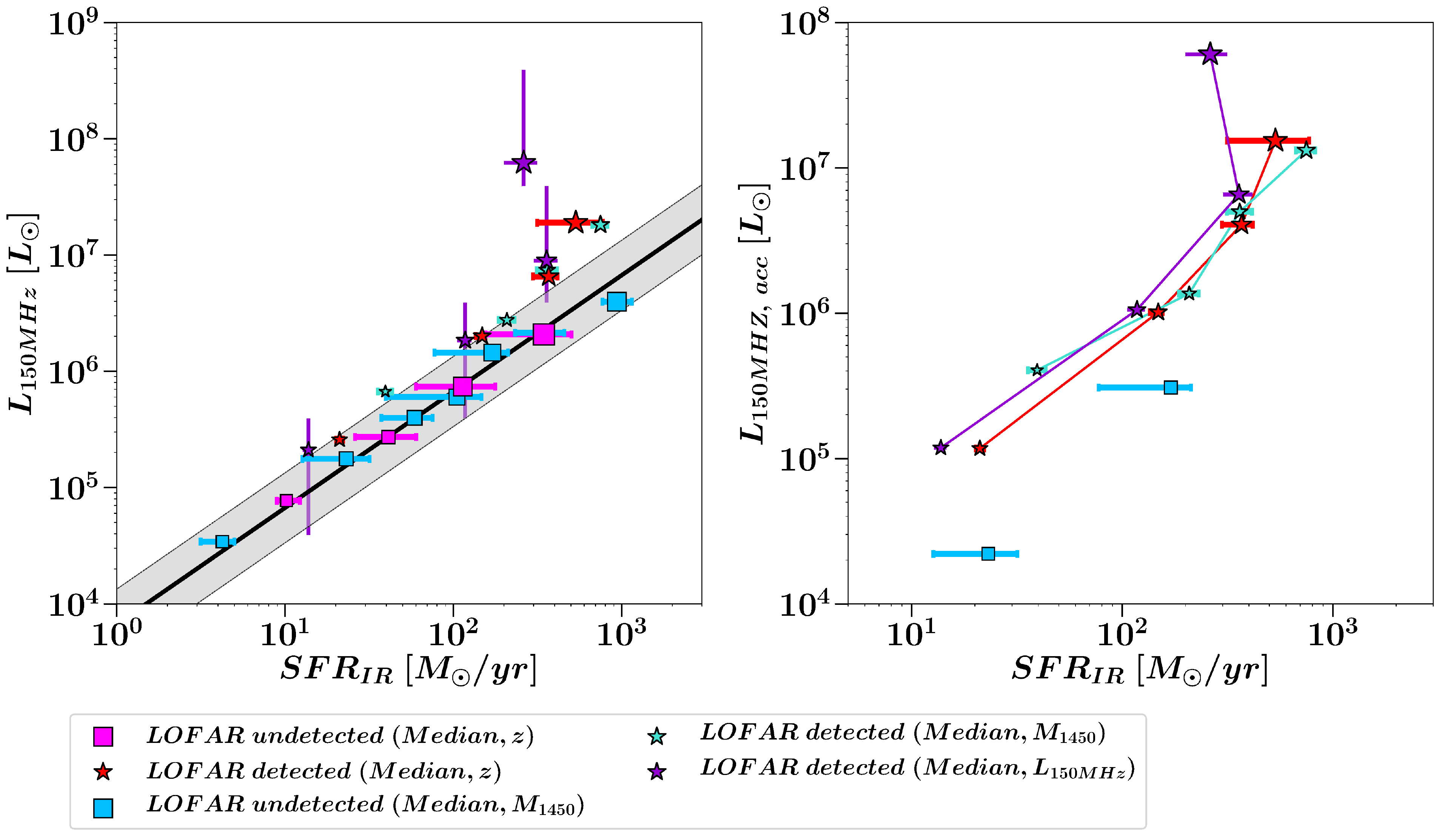}\centering\caption{\label{fig:accretion_luminosity} Comparison between the infrared
star-formation rates, $SFRs_{\textrm{IR}}$, and the total and accretion
radio luminosities at 150 MHz. Left: $SFRs_{\textrm{IR}}$ of the median
quasar samples versus their total radio luminosities at 150 MHz. Error
bars on the x-axis indicate $1\sigma$ errors on the $SFRs_{\textrm{IR}}$, while error
bars for LOFAR radio-detected quasars stacked according to their radio
luminosity denotes the bin size used. The dashed line is the far-infrared
radio correlation (FIRC) at ${\displaystyle L_{150\,MHz}}$ derived
by \citet{2017MNRAS.469.3468C}. The gray shaded region indicates
the spread of the FIRC scaled by a factor of $\pm2$. Right: $SFR_{\textrm{IR}}$
versus the radio luminosity associated with AGN accretion, $L_{\textrm{150MHz, acc}}$.
The radio luminosity connected to AGN accretion is calculated by subtracting
the contribution of SF to the total radio luminosity estimated using
the FIRC. Negative AGN feedback could be taken place in the hosts
of RDQs with the highest 150 MHz radio luminosities due to their radio-emission
being dominated by AGN accretion, and the relative lower $SFR_{\textrm{IR}}$.}
 
\end{figure*}

\subsection{Positive and negative AGN feedback \label{subsec:agn_feedback}}

We find in Sections \ref{sec:sf_rates} and \ref{subsec:luminsity_correlations}
(see Figs. \ref{fig:RDQ_RUQ_comparison} and \ref{fig:RDQ_RUQ_SFR_luminosity})
that the median host galaxies of RDQs are associated with higher $\textrm{SFRs}_{\textrm{IR}}$
than median RUQs. Excluding the highest $M_{\textrm{1450}}$ and z
bins where less sources are found, which produces larger error bars
for the $\textrm{SFR}_{\textrm{IR}}$ estimates, the $\textrm{SFRs}_{\textrm{IR}}$
of median RDQs are at least larger by a factor of $2$ than those
of median RUQs. These results demonstrate that vigorous SF activity
is coeval with the growth of the SMBHs residing in the nuclei of the
RLQs host galaxies. This agrees with previous works that found evidence
for high SFRs in radio AGN (e.g., \citealt{2011MNRAS.413.1777S,2014MNRAS.442..682M,2015A&A...575A..80P,2016MNRAS.456..431M,2020MNRAS.493.3838M}).
Moreover, bright RDQs have been associated with higher SFRs in comparison
to their radio-quiet counterparts (e.g., \citealt{2014MNRAS.442.1181K,2017MNRAS.471...28K})
The enhancement in the star formation in radio AGN could be explained
considering the formation of bow shocks caused by radio jets that
compress the interstellar medium in the host galaxy. This compression
results in higher gas clumping leading to an increased SF efficiency
\citep{2004ApJ...612L..97K,2004IAUS..217..472V,2012MNRAS.425..438G,2012MNRAS.427.2998I}.
Observational studies support the existence of jet-induced star formation
(positive feedback) in the local universe (Minkowski\textquoteright s
object, \citealt{2006ApJ...647.1040C}; Centaurus A, \citealt{2000ApJ...536..266M,2017A&A...608A..98S}),
at intermediate redshift (PKS2250-41, \citealt{2008MNRAS.386.1797I})
and high redshift (4C 41.17, \citealt{1997ApJ...490..698D}). While
our results suggests that jet-induced SF could be taking place in
our RDQs sample, it is important to remark that the conditions under
which radio jets would enhance the SF are not completely understood
yet.

Another critical aspect to understand AGN feedback in RUQs and RDQs
are the differences between their host galaxies. Particularly, radio
AGN such as radio-galaxies and bright RDQs are known to populate overdense
regions \citep{2007A&A...461..823V,2010MNRAS.405..347F,2017A&A...600A..97R}
and are usually associated with massive galaxies \citep{2019A&A...621A..27F,2005MNRAS.362...25B}.
In Section \ref{sec:sf_rates}, we show how the $\textrm{SFRs}_{\textrm{IR}}$
of the median-stacked quasars in our sample follow the SFMS predictions
with median stellar masses of $M_{*}\sim3\times10^{11}\:M_{\odot}$
and $M_{*}\sim5\times10^{10}\:M_{\odot}$ for RDQs and RUQs, respectively.
The former have the higher stellar masses comparable to those of the
most massive galaxies \citep{2011MNRAS.413..162C,2005ApJ...624L..81L},
while the latter have lower stellar masses consistent with the mass
scale associated with quasar hosts \citep{2013A&A...560A..72R,2014ApJ...780..162M}.
Additionally, this difference is in line with what has been found
by previous studies of the host galaxies of radio sources (e.g., \citealt{2009ApJ...699L..43S,2012MNRAS.421.1569B,2017A&A...602A...3D}).
Finally, as can be seen in Figure \ref{fig:RDQ_RUQ_comparison}, the
star-forming properties of the median RDQs hosts are similar to those
of the general population of massive normal (i.e., non-quasar) galaxies
at comparable redshifts. This indicates that there is no quenching
of star formation in the median RDQs hosts with median stacking carried
out according to redshift and $M_{\textrm{1450}}$ in all the redshift
bins considered.

The apparent lack of signature for negative AGN feedback (considering
only stacking to redshift and $M_{\textrm{1450}}$) is puzzling. The
most luminous quasars, such as the RDQs in our sample, should be effective
in ionizing the interstellar medium in their hosts and hence impact
their star formation, namely, according to \citet{2008ApJ...686..219M,2012ApJ...745L..34Z,2016MNRAS.458..816H}.
Therefore, luminous quasars should be expected to have the clearest
signatures of AGN feedback. This implies that a significant fraction
of them should be on their way to being quenched or be already quenched.
However, the outcome is different as can be seen in the bottom panel
of Figure \ref{fig:RDQ_RUQ_SFR_luminosity} and right panel of Figure
\ref{fig:accretion_luminosity} when the median stacking is done using
the 150 MHz radio luminosity, $L_{150\textrm{MHz}}$, instead of redshift
and $M_{\textrm{1450}}$. We find that the median hosts of RLQs with
$L_{\textrm{150MHz}}\gtrsim8\times10^{26}\,\textrm{W}\,\textrm{Hz}^{-1}$
present a flattening, or possible decline in SF at the highest $L_{150\textrm{MHz}}$
bin. This implies that the $\textrm{SFRs}_{\textrm{IR}}$ of the most
radio luminous RDQs are lower than those of RDQs with lower radio
luminosities. Similar signatures of negative AGN feedback have been
found previously \citep{2014MNRAS.442.1181K,2017MNRAS.471...28K}.
Moreover, the most luminous median RDQ at 150 MHz exhibits the highest
level of accretion-dominated radio emission with $97.2\%$ as found
in Section \ref{subsec:mechanism}. This high level of AGN-dominated
radio emission along with the relative lower $\textrm{SFR}_{\textrm{IR}}$
provides evidence supporting negative AGN feedback taking place in
the most luminous RDQs in our sample. 

While jet-induced star formation could be taking place in the host
galaxies of some RDQs as mentioned earlier, it has also been suggested
that radio jets could be associated with negative AGN feedback. This
either by suppressing gas cooling or by expelling gas from the host
galaxy, hence quenching SF in host galaxies. Evidence of gas being
expelled by an AGN outflow has been observed in bright radio-galaxies
(e.g., \citealt{2006ApJ...650..693N}). Additionally, \citet{2017MNRAS.471...28K}
proposed a model with a jet power threshold that controls radio-jet
feedback, where sources with low and intermediate jet powers present
SF enhancement, while sources with the highest jet powers above the
threshold at which radio-jet feedback changes from positive to negative
suppressing SF. Although this model could explain our results for
RDQs with $L_{\textrm{150MHz}}\gtrsim8\times10^{26}\,\textrm{W}\,\textrm{Hz}^{-1}$,
other factors such as merger history and environment might influence
star formation in the host galaxies of these RDQs.

Finally, it is important to acknowledge that despite the fact that
median stacking enables statistical fluxes of undetected sources,
it also leads to the lost of information about individual detections.
This in the context of the implication that based on AGN feedback,
only median trends could be recovered in some cases (i.e., RUQs),
while trend outliers where positive or negative AGN feedback could
take place have a limited weight in the trends obtained through stacking.
Future studies will utilize deeper infrared and radio continuum observations
to study a number of open questions regarding AGN feedback without
the need to rely on stacking.

\section{Conclusions \label{subsec:conclusions}}

In this work, we construct the infrared-radio median spectral energy
distributions for LOFAR RDQs and RUQs in the NDWFS-Bo\"otes field.
To obtain the infrared and radio photometry, we used a stacking analysis
of Spitzer, WISE, Herschel, VLA, WSRT, and LOFAR maps. The stacking
analysis allows us to probe the radio-emission for quasars that are
up to one order of magnitude fainter than the ones detected individually.
These results help us to address the questions posed in Section \ref{sec:intro}.
We arrive at the following conclusions:
\begin{enumerate}
\item The radio-emission of RUQs is consistent to be linked to SF processes,
while the radio activity in RDQs is likely associated with accretion
onto SMBHs. This is supported by the radio-excess showed by median
RDQs with respect to the predictions of the FIRC, while RUQs follow
tightly this correlation.
\item The observed behaviour of the radio-loudness parameter, $R$, is as
follows: for bright RDQs with $R\gtrsim-4.5$ the radio-emission is
consistent with being dominated by SMBH accretion, while for lower-radio
luminosity quasars with $R<-4.5$ the relative contribution of SF
to their radio fluxes increases as the SMBH component becomes weaker. 
\item For median RDQs, the SMBH accretion can account for $\sim50-84\%$
of the total 150 MHz radio luminosity, while for median RUQs the contribution
is $\sim5-41\%$ . This strongly suggests that vigorous SF activity
is coeval with the growth of the SMBHs residing in the nuclei of quasar
host galaxies.
\item There is no evidence for the suppression of SF in the host galaxies
of median RDQs stacked according to redshift and $M_{\textrm{1450}}$
(not considering the highest $M_{\textrm{1450}}$ and z bins where
less sources are found). However, when the 150 MHz radio luminosity
is considered for the stacking, the highest radio luminosity median
RDQs have lower SFRs than median the RDQs with lower radio luminosities.
In particular, for the brightest median RDQs at 150 MHz the accretion-related
radio luminosity accounts for $97.2\%$ of the total 150 MHz radio
luminosity, and has a relative lower $\textrm{SFR}_{\textrm{IR}}$.
This suggests that it is likely that negative AGN feedback has taken
place in the hosts of RDQs with the highest 150 MHz radio luminosities
in our sample.
\item The spectral indices of median quasars (undetected and detected by
LOFAR) do not evolve significantly with redshift, but become flatter
towards lower frequencies. Overall, the spectral curvature of RUQs
is similar to that of RDQs. 
\end{enumerate}
Finally, our results highlight the importance of using deep low- and
high- frequency radio surveys to study the properties of RDQs and
RUQs, and particularly the radio properties of quasars that generally
remain radio silent and hidden in shallow flux-density-limited surveys.

\bibliographystyle{aa}
\addcontentsline{toc}{section}{\refname}\bibliography{mybib}

\begin{thebibliography}{171}
\expandafter\ifx\csname natexlab\endcsname\relax\def\natexlab#1{#1}\fi

\bibitem[{{Alberts} {et~al.}(2020){Alberts}, {Rujopakarn}, {Rieke},
  {Jagannathan}, \& {Nyland}}]{2020ApJ...901..168A}
{Alberts}, S., {Rujopakarn}, W., {Rieke}, G.~H., {Jagannathan}, P., \&
  {Nyland}, K. 2020, \apj, 901, 168

\bibitem[{{Algera} {et~al.}(2020){Algera}, {van der Vlugt}, {Hodge}, {Smail},
  {Novak}, {Radcliffe}, {Riechers}, {R{\"o}ttgering}, {Smol{\v{c}}i{\'c}}, \&
  {Walter}}]{2020ApJ...903..139A}
{Algera}, H.~S.~B., {van der Vlugt}, D., {Hodge}, J.~A., {et~al.} 2020, \apj,
  903, 139

\bibitem[{{Ashby} {et~al.}(2009){Ashby}, {Stern}, {Brodwin}, {Griffith},
  {Eisenhardt}, {Koz{\l}owski}, {Kochanek}, {Bock}, {Borys}, {Brand}, {Brown},
  {Cool}, {Cooray}, {Croft}, {Dey}, {Eisenstein}, {Gonzalez}, {Gorjian},
  {Grogin}, {Ivison}, {Jacob}, {Jannuzi}, {Mainzer}, {Moustakas},
  {R{\"o}ttgering}, {Seymour}, {Smith}, {Stanford}, {Stauffer}, {Sullivan},
  {van Breugel}, {Willner}, \& {Wright}}]{2009ApJ...701..428A}
{Ashby}, M.~L.~N., {Stern}, D., {Brodwin}, M., {et~al.} 2009, \apj, 701, 428

\bibitem[{{Azadi} {et~al.}(2017){Azadi}, {Coil}, {Aird}, {Reddy}, {Shapley},
  {Freeman}, {Kriek}, {Leung}, {Mobasher}, {Price}, {Sanders}, {Shivaei}, \&
  {Siana}}]{2017ApJ...835...27A}
{Azadi}, M., {Coil}, A.~L., {Aird}, J., {et~al.} 2017, \apj, 835, 27

\bibitem[{{Badole} {et~al.}(2020){Badole}, {Jackson}, {Hartley}, {Sluse},
  {Stacey}, \& {Vives-Arias}}]{2020MNRAS.496..138B}
{Badole}, S., {Jackson}, N., {Hartley}, P., {et~al.} 2020, \mnras, 496, 138

\bibitem[{{Baldry} {et~al.}(2012){Baldry}, {Driver}, {Loveday}, {Taylor},
  {Kelvin}, {Liske}, {Norberg}, {Robotham}, {Brough}, {Hopkins}, {Bamford},
  {Peacock}, {Bland-Hawthorn}, {Conselice}, {Croom}, {Jones}, {Parkinson},
  {Popescu}, {Prescott}, {Sharp}, \& {Tuffs}}]{2012MNRAS.421..621B}
{Baldry}, I.~K., {Driver}, S.~P., {Loveday}, J., {et~al.} 2012, \mnras, 421,
  621

\bibitem[{{Balokovi{\'c}} {et~al.}(2012){Balokovi{\'c}}, {Smol{\v{c}}i{\'c}},
  {Ivezi{\'c}}, {Zamorani}, {Schinnerer}, \& {Kelly}}]{2012ApJ...759...30B}
{Balokovi{\'c}}, M., {Smol{\v{c}}i{\'c}}, V., {Ivezi{\'c}}, {\v{Z}}., {et~al.}
  2012, \apj, 759, 30

\bibitem[{{Barger} {et~al.}(2014){Barger}, {Cowie}, {Chen}, {Owen}, {Wang},
  {Casey}, {Lee}, {Sanders}, \& {Williams}}]{2014ApJ...784....9B}
{Barger}, A.~J., {Cowie}, L.~L., {Chen}, C.~C., {et~al.} 2014, \apj, 784, 9

\bibitem[{{Becker} {et~al.}(1995){Becker}, {White}, \&
  {Helfand}}]{1995ApJ...450..559B}
{Becker}, R.~H., {White}, R.~L., \& {Helfand}, D.~J. 1995, \apj, 450, 559

\bibitem[{{Bell}(2003)}]{2003ApJ...586..794B}
{Bell}, E.~F. 2003, \apj, 586, 794

\bibitem[{{Best} \& {Heckman}(2012)}]{2012MNRAS.421.1569B}
{Best}, P.~N. \& {Heckman}, T.~M. 2012, \mnras, 421, 1569

\bibitem[{{Best} {et~al.}(2005){Best}, {Kauffmann}, {Heckman}, {Brinchmann},
  {Charlot}, {Ivezi{\'c}}, \& {White}}]{2005MNRAS.362...25B}
{Best}, P.~N., {Kauffmann}, G., {Heckman}, T.~M., {et~al.} 2005, \mnras, 362,
  25

\bibitem[{{Bettoni} {et~al.}(2015){Bettoni}, {Falomo}, {Kotilainen},
  {Karhunen}, \& {Uslenghi}}]{2015MNRAS.454.4103B}
{Bettoni}, D., {Falomo}, R., {Kotilainen}, J.~K., {Karhunen}, K., \&
  {Uslenghi}, M. 2015, \mnras, 454, 4103

\bibitem[{{Bian} {et~al.}(2013){Bian}, {Fan}, {Jiang}, {McGreer}, {Dey},
  {Green}, {Maiolino}, {Walter}, {Lee}, \& {Dav{\'e}}}]{2013ApJ...774...28B}
{Bian}, F., {Fan}, X., {Jiang}, L., {et~al.} 2013, \apj, 774, 28

\bibitem[{Blundell {et~al.}(1996)Blundell, Beasley, Lacy, \&
  Garrington}]{Blundell_1996}
Blundell, K.~M., Beasley, A.~J., Lacy, M., \& Garrington, S.~T. 1996, The
  Astrophysical Journal, 468, L91

\bibitem[{{Bonfield} {et~al.}(2011){Bonfield}, {Jarvis}, {Hardcastle},
  {Cooray}, {Hatziminaoglou}, {Ivison}, {Page}, {Stevens}, {de Zotti}, {Auld},
  {Baes}, {Buttiglione}, {Cava}, {Dariush}, {Dunlop}, {Dunne}, {Dye}, {Eales},
  {Fritz}, {Hopwood}, {Ibar}, {Maddox}, {Micha{\l}owski}, {Pascale}, {Pohlen},
  {Rigby}, {Rodighiero}, {Serjeant}, {Smith}, {Temi}, \& {van der
  Werf}}]{2011MNRAS.416...13B}
{Bonfield}, D.~G., {Jarvis}, M.~J., {Hardcastle}, M.~J., {et~al.} 2011, \mnras,
  416, 13

\bibitem[{{Bonzini} {et~al.}(2015){Bonzini}, {Mainieri}, {Padovani},
  {Andreani}, {Berta}, {Bethermin}, {Lutz}, {Rodighiero}, {Rosario}, {Tozzi},
  \& {Vattakunnel}}]{2015MNRAS.453.1079B}
{Bonzini}, M., {Mainieri}, V., {Padovani}, P., {et~al.} 2015, \mnras, 453, 1079

\bibitem[{{Brown} {et~al.}(2007){Brown}, {Dey}, {Jannuzi}, {Brand}, {Benson},
  {Brodwin}, {Croton}, \& {Eisenhardt}}]{2007ApJ...654..858B}
{Brown}, M.~J.~I., {Dey}, A., {Jannuzi}, B.~T., {et~al.} 2007, \apj, 654, 858

\bibitem[{{Calistro-Rivera} {et~al.}(2017){Calistro-Rivera}, {Williams},
  {Hardcastle}, {Duncan}, {R{\"o}ttgering}, {Best}, {Br{\"u}ggen}, {Chy{\.z}y},
  {Conselice}, {de Gasperin}, {Engels}, {G{\"u}rkan}, {Intema}, {Jarvis},
  {Mahony}, {Miley}, {Morabito}, {Prandoni}, {Sabater}, {Smith}, {Tasse}, {van
  der Werf}, \& {White}}]{2017MNRAS.469.3468C}
{Calistro-Rivera}, G., {Williams}, W.~L., {Hardcastle}, M.~J., {et~al.} 2017,
  \mnras, 469, 3468

\bibitem[{{Caputi} {et~al.}(2011){Caputi}, {Cirasuolo}, {Dunlop}, {McLure},
  {Farrah}, \& {Almaini}}]{2011MNRAS.413..162C}
{Caputi}, K.~I., {Cirasuolo}, M., {Dunlop}, J.~S., {et~al.} 2011, \mnras, 413,
  162

\bibitem[{{Casey}(2012)}]{2012MNRAS.425.3094C}
{Casey}, C.~M. 2012, \mnras, 425, 3094

\bibitem[{{Chambers} {et~al.}(2016){Chambers}, {Magnier}, {Metcalfe},
  {Flewelling}, {Huber}, {Waters}, {Denneau}, {Draper}, {Farrow}, {Finkbeiner},
  {Holmberg}, {Koppenhoefer}, {Price}, {Rest}, {Saglia}, {Schlafly}, {Smartt},
  {Sweeney}, {Wainscoat}, {Burgett}, {Chastel}, {Grav}, {Heasley}, {Hodapp},
  {Jedicke}, {Kaiser}, {Kudritzki}, {Luppino}, {Lupton}, {Monet}, {Morgan},
  {Onaka}, {Shiao}, {Stubbs}, {Tonry}, {White}, {Ba{\~n}ados}, {Bell},
  {Bender}, {Bernard}, {Boegner}, {Boffi}, {Botticella}, {Calamida},
  {Casertano}, {Chen}, {Chen}, {Cole}, {Deacon}, {Frenk}, {Fitzsimmons},
  {Gezari}, {Gibbs}, {Goessl}, {Goggia}, {Gourgue}, {Goldman}, {Grant},
  {Grebel}, {Hambly}, {Hasinger}, {Heavens}, {Heckman}, {Henderson}, {Henning},
  {Holman}, {Hopp}, {Ip}, {Isani}, {Jackson}, {Keyes}, {Koekemoer}, {Kotak},
  {Le}, {Liska}, {Long}, {Lucey}, {Liu}, {Martin}, {Masci}, {McLean}, {Mindel},
  {Misra}, {Morganson}, {Murphy}, {Obaika}, {Narayan}, {Nieto-Santisteban},
  {Norberg}, {Peacock}, {Pier}, {Postman}, {Primak}, {Rae}, {Rai}, {Riess},
  {Riffeser}, {Rix}, {R{\"o}ser}, {Russel}, {Rutz}, {Schilbach}, {Schultz},
  {Scolnic}, {Strolger}, {Szalay}, {Seitz}, {Small}, {Smith}, {Soderblom},
  {Taylor}, {Thomson}, {Taylor}, {Thakar}, {Thiel}, {Thilker}, {Unger},
  {Urata}, {Valenti}, {Wagner}, {Walder}, {Walter}, {Watters}, {Werner},
  {Wood-Vasey}, \& {Wyse}}]{2016arXiv161205560C}
{Chambers}, K.~C., {Magnier}, E.~A., {Metcalfe}, N., {et~al.} 2016, arXiv
  e-prints, arXiv:1612.05560

\bibitem[{{Cirasuolo} {et~al.}(2003){Cirasuolo}, {Magliocchetti}, {Celotti}, \&
  {Danese}}]{2003MNRAS.341..993C}
{Cirasuolo}, M., {Magliocchetti}, M., {Celotti}, A., \& {Danese}, L. 2003,
  \mnras, 341, 993

\bibitem[{{Condon}(1992)}]{1992ARA&A..30..575C}
{Condon}, J.~J. 1992, \araa, 30, 575

\bibitem[{Condon {et~al.}(1998)Condon, Cotton, Greisen, Yin, Perley, Taylor, \&
  Broderick}]{Condon_1998}
Condon, J.~J., Cotton, W.~D., Greisen, E.~W., {et~al.} 1998, The Astronomical
  Journal, 115, 1693

\bibitem[{{Cool} {et~al.}(2006){Cool}, {Kochanek}, {Eisenstein}, {Stern},
  {Brand}, {Brown}, {Dey}, {Eisenhardt}, {Fan}, {Gonzalez}, {Green}, {Jannuzi},
  {McKenzie}, {Rieke}, {Rieke}, {Soifer}, {Spinrad}, \&
  {Elston}}]{2006AJ....132..823C}
{Cool}, R.~J., {Kochanek}, C.~S., {Eisenstein}, D.~J., {et~al.} 2006, \aj, 132,
  823

\bibitem[{{Coppejans} {et~al.}(2015){Coppejans}, {Cseh}, {Williams}, {van
  Velzen}, \& {Falcke}}]{2015MNRAS.450.1477C}
{Coppejans}, R., {Cseh}, D., {Williams}, W.~L., {van Velzen}, S., \& {Falcke},
  H. 2015, \mnras, 450, 1477

\bibitem[{{Croft} {et~al.}(2006){Croft}, {van Breugel}, {de Vries}, {Dopita},
  {Martin}, {Morganti}, {Neff}, {Oosterloo}, {Schiminovich}, {Stanford}, \&
  {van Gorkom}}]{2006ApJ...647.1040C}
{Croft}, S., {van Breugel}, W., {de Vries}, W., {et~al.} 2006, \apj, 647, 1040

\bibitem[{{Croom} {et~al.}(2004){Croom}, {Smith}, {Boyle}, {Shanks}, {Miller},
  {Outram}, \& {Loaring}}]{2004MNRAS.349.1397C}
{Croom}, S.~M., {Smith}, R.~J., {Boyle}, B.~J., {et~al.} 2004, \mnras, 349,
  1397

\bibitem[{{Daddi} {et~al.}(2007){Daddi}, {Dickinson}, {Morrison}, {Chary},
  {Cimatti}, {Elbaz}, {Frayer}, {Renzini}, {Pope}, {Alexander}, {Bauer},
  {Giavalisco}, {Huynh}, {Kurk}, \& {Mignoli}}]{2007ApJ...670..156D}
{Daddi}, E., {Dickinson}, M., {Morrison}, G., {et~al.} 2007, \apj, 670, 156

\bibitem[{{de Vries} {et~al.}(2002){de Vries}, {Morganti}, {R{\"o}ttgering},
  {Vermeulen}, {van Breugel}, {Rengelink}, \& {Jarvis}}]{2002AJ....123.1784D}
{de Vries}, W.~H., {Morganti}, R., {R{\"o}ttgering}, H.~J.~A., {et~al.} 2002,
  \aj, 123, 1784

\bibitem[{{Delvecchio} {et~al.}(2017){Delvecchio}, {Smol{\v{c}}i{\'c}},
  {Zamorani}, {Lagos}, {Berta}, {Delhaize}, {Baran}, {Alexander}, {Rosario},
  {Gonzalez-Perez}, {Ilbert}, {Lacey}, {Le F{\`e}vre}, {Miettinen}, {Aravena},
  {Bondi}, {Carilli}, {Ciliegi}, {Mooley}, {Novak}, {Schinnerer}, {Capak},
  {Civano}, {Fanidakis}, {Herrera Ruiz}, {Karim}, {Laigle}, {Marchesi},
  {McCracken}, {Middleberg}, {Salvato}, \& {Tasca}}]{2017A&A...602A...3D}
{Delvecchio}, I., {Smol{\v{c}}i{\'c}}, V., {Zamorani}, G., {et~al.} 2017, \aap,
  602, A3

\bibitem[{{Dey} {et~al.}(1997){Dey}, {van Breugel}, {Vacca}, \&
  {Antonucci}}]{1997ApJ...490..698D}
{Dey}, A., {van Breugel}, W., {Vacca}, W.~D., \& {Antonucci}, R. 1997, \apj,
  490, 698

\bibitem[{{Dong} \& {Wu}(2016)}]{2016ApJ...824...70D}
{Dong}, X.~Y. \& {Wu}, X.-B. 2016, \apj, 824, 70

\bibitem[{{Duras} {et~al.}(2017){Duras}, {Bongiorno}, {Piconcelli}, {Bianchi},
  {Pappalardo}, {Valiante}, {Bischetti}, {Feruglio}, {Martocchia}, {Schneider},
  {Vietri}, {Vignali}, {Zappacosta}, {La Franca}, \&
  {Fiore}}]{2017A&A...604A..67D}
{Duras}, F., {Bongiorno}, A., {Piconcelli}, E., {et~al.} 2017, \aap, 604, A67

\bibitem[{{Elbaz} {et~al.}(2011){Elbaz}, {Dickinson}, {Hwang},
  {D{\'\i}az-Santos}, {Magdis}, {Magnelli}, {Le Borgne}, {Galliano},
  {Pannella}, {Chanial}, {Armus}, {Charmandaris}, {Daddi}, {Aussel}, {Popesso},
  {Kartaltepe}, {Altieri}, {Valtchanov}, {Coia}, {Dannerbauer}, {Dasyra},
  {Leiton}, {Mazzarella}, {Alexander}, {Buat}, {Burgarella}, {Chary}, {Gilli},
  {Ivison}, {Juneau}, {Le Floc'h}, {Lutz}, {Morrison}, {Mullaney}, {Murphy},
  {Pope}, {Scott}, {Brodwin}, {Calzetti}, {Cesarsky}, {Charlot}, {Dole},
  {Eisenhardt}, {Ferguson}, {F{\"o}rster Schreiber}, {Frayer}, {Giavalisco},
  {Huynh}, {Koekemoer}, {Papovich}, {Reddy}, {Surace}, {Teplitz}, {Yun}, \&
  {Wilson}}]{2011A&A...533A.119E}
{Elbaz}, D., {Dickinson}, M., {Hwang}, H.~S., {et~al.} 2011, \aap, 533, A119

\bibitem[{{Falder} {et~al.}(2010){Falder}, {Stevens}, {Jarvis}, {Hardcastle},
  {Lacy}, {McLure}, {Hatziminaoglou}, {Page}, \&
  {Richards}}]{2010MNRAS.405..347F}
{Falder}, J.~T., {Stevens}, J.~A., {Jarvis}, M.~J., {et~al.} 2010, \mnras, 405,
  347

\bibitem[{{Falkendal} {et~al.}(2019){Falkendal}, {De Breuck}, {Lehnert},
  {Drouart}, {Vernet}, {Emonts}, {Lee}, {Nesvadba}, {Seymour}, {B{\'e}thermin},
  {Kolwa}, {Gullberg}, \& {Wylezalek}}]{2019A&A...621A..27F}
{Falkendal}, T., {De Breuck}, C., {Lehnert}, M.~D., {et~al.} 2019, \aap, 621,
  A27

\bibitem[{{Fanaroff} \& {Riley}(1974)}]{1974MNRAS.167P..31F}
{Fanaroff}, B.~L. \& {Riley}, J.~M. 1974, \mnras, 167, 31P

\bibitem[{{Fawcett} {et~al.}(2020){Fawcett}, {Alexander}, {Rosario}, {Klindt},
  {Fotopoulou}, {Lusso}, {Morabito}, \& {Calistro
  Rivera}}]{2020MNRAS.494.4802F}
{Fawcett}, V.~A., {Alexander}, D.~M., {Rosario}, D.~J., {et~al.} 2020, \mnras,
  494, 4802

\bibitem[{{Fazio} {et~al.}(2004){Fazio}, {Hora}, {Allen}, {Ashby}, {Barmby},
  {Deutsch}, {Huang}, {Kleiner}, {Marengo}, {Megeath}, {Melnick}, {Pahre},
  {Patten}, {Polizotti}, {Smith}, {Taylor}, {Wang}, {Willner}, {Hoffmann},
  {Pipher}, {Forrest}, {McMurty}, {McCreight}, {McKelvey}, {McMurray}, {Koch},
  {Moseley}, {Arendt}, {Mentzell}, {Marx}, {Losch}, {Mayman}, {Eichhorn},
  {Krebs}, {Jhabvala}, {Gezari}, {Fixsen}, {Flores}, {Shakoorzadeh}, {Jungo},
  {Hakun}, {Workman}, {Karpati}, {Kichak}, {Whitley}, {Mann}, {Tollestrup},
  {Eisenhardt}, {Stern}, {Gorjian}, {Bhattacharya}, {Carey}, {Nelson},
  {Glaccum}, {Lacy}, {Lowrance}, {Laine}, {Reach}, {Stauffer}, {Surace},
  {Wilson}, {Wright}, {Hoffman}, {Domingo}, \& {Cohen}}]{2004ApJS..154...10F}
{Fazio}, G.~G., {Hora}, J.~L., {Allen}, L.~E., {et~al.} 2004, \apjs, 154, 10

\bibitem[{{Feigelson} \& {Nelson}(1985)}]{1985ApJ...293..192F}
{Feigelson}, E.~D. \& {Nelson}, P.~I. 1985, \apj, 293, 192

\bibitem[{{Flesch}(2015)}]{2015PASA...32...10F}
{Flesch}, E.~W. 2015, \pasa, 32, e010

\bibitem[{{Fogasy} {et~al.}(2020){Fogasy}, {Knudsen}, {Drouart}, {Lagos}, \&
  {Fan}}]{2020MNRAS.493.3744F}
{Fogasy}, J., {Knudsen}, K.~K., {Drouart}, G., {Lagos}, C.~D.~P., \& {Fan}, L.
  2020, \mnras, 493, 3744

\bibitem[{{Fu} {et~al.}(2013){Fu}, {Cooray}, {Feruglio}, {Ivison}, {Riechers},
  {Gurwell}, {Bussmann}, {Harris}, {Altieri}, {Aussel}, {Baker}, {Bock},
  {Boylan-Kolchin}, {Bridge}, {Calanog}, {Casey}, {Cava}, {Chapman},
  {Clements}, {Conley}, {Cox}, {Farrah}, {Frayer}, {Hopwood}, {Jia}, {Magdis},
  {Marsden}, {Mart{\'\i}nez-Navajas}, {Negrello}, {Neri}, {Oliver}, {Omont},
  {Page}, {P{\'e}rez-Fournon}, {Schulz}, {Scott}, {Smith}, {Vaccari},
  {Valtchanov}, {Vieira}, {Viero}, {Wang}, {Wardlow}, \&
  {Zemcov}}]{2013Natur.498..338F}
{Fu}, H., {Cooray}, A., {Feruglio}, C., {et~al.} 2013, \nat, 498, 338

\bibitem[{{Gaibler} {et~al.}(2012){Gaibler}, {Khochfar}, {Krause}, \&
  {Silk}}]{2012MNRAS.425..438G}
{Gaibler}, V., {Khochfar}, S., {Krause}, M., \& {Silk}, J. 2012, \mnras, 425,
  438

\bibitem[{{Gallagher} {et~al.}(2007){Gallagher}, {Richards}, {Lacy}, {Hines},
  {Elitzur}, \& {Storrie-Lombardi}}]{2007ApJ...661...30G}
{Gallagher}, S.~C., {Richards}, G.~T., {Lacy}, M., {et~al.} 2007, \apj, 661, 30

\bibitem[{{Garn} \& {Alexander}(2009)}]{2009MNRAS.394..105G}
{Garn}, T. \& {Alexander}, P. 2009, \mnras, 394, 105

\bibitem[{{Glikman} {et~al.}(2011){Glikman}, {Djorgovski}, {Stern}, {Dey},
  {Jannuzi}, \& {Lee}}]{2011ApJ...728L..26G}
{Glikman}, E., {Djorgovski}, S.~G., {Stern}, D., {et~al.} 2011, \apjl, 728, L26

\bibitem[{{Griffin} {et~al.}(2010){Griffin}, {Abergel}, {Abreu}, {Ade},
  {Andr{\'e}}, {Augueres}, {Babbedge}, {Bae}, {Baillie}, {Baluteau}, {Barlow},
  {Bendo}, {Benielli}, {Bock}, {Bonhomme}, {Brisbin}, {Brockley-Blatt},
  {Caldwell}, {Cara}, {Castro-Rodriguez}, {Cerulli}, {Chanial}, {Chen},
  {Clark}, {Clements}, {Clerc}, {Coker}, {Communal}, {Conversi}, {Cox},
  {Crumb}, {Cunningham}, {Daly}, {Davis}, {de Antoni}, {Delderfield}, {Devin},
  {di Giorgio}, {Didschuns}, {Dohlen}, {Donati}, {Dowell}, {Dowell}, {Duband},
  {Dumaye}, {Emery}, {Ferlet}, {Ferrand}, {Fontignie}, {Fox}, {Franceschini},
  {Frerking}, {Fulton}, {Garcia}, {Gastaud}, {Gear}, {Glenn}, {Goizel},
  {Griffin}, {Grundy}, {Guest}, {Guillemet}, {Hargrave}, {Harwit}, {Hastings},
  {Hatziminaoglou}, {Herman}, {Hinde}, {Hristov}, {Huang}, {Imhof}, {Isaak},
  {Israelsson}, {Ivison}, {Jennings}, {Kiernan}, {King}, {Lange}, {Latter},
  {Laurent}, {Laurent}, {Leeks}, {Lellouch}, {Levenson}, {Li}, {Li},
  {Lilienthal}, {Lim}, {Liu}, {Lu}, {Madden}, {Mainetti}, {Marliani}, {McKay},
  {Mercier}, {Molinari}, {Morris}, {Moseley}, {Mulder}, {Mur}, {Naylor},
  {Nguyen}, {O'Halloran}, {Oliver}, {Olofsson}, {Olofsson}, {Orfei}, {Page},
  {Pain}, {Panuzzo}, {Papageorgiou}, {Parks}, {Parr-Burman}, {Pearce},
  {Pearson}, {P{\'e}rez-Fournon}, {Pinsard}, {Pisano}, {Podosek}, {Pohlen},
  {Polehampton}, {Pouliquen}, {Rigopoulou}, {Rizzo}, {Roseboom}, {Roussel},
  {Rowan-Robinson}, {Rownd}, {Saraceno}, {Sauvage}, {Savage}, {Savini},
  {Sawyer}, {Scharmberg}, {Schmitt}, {Schneider}, {Schulz}, {Schwartz},
  {Shafer}, {Shupe}, {Sibthorpe}, {Sidher}, {Smith}, {Smith}, {Smith},
  {Spencer}, {Stobie}, {Sudiwala}, {Sukhatme}, {Surace}, {Stevens}, {Swinyard},
  {Trichas}, {Tourette}, {Triou}, {Tseng}, {Tucker}, {Turner}, {Vaccari},
  {Valtchanov}, {Vigroux}, {Virique}, {Voellmer}, {Walker}, {Ward}, {Waskett},
  {Weilert}, {Wesson}, {White}, {Whitehouse}, {Wilson}, {Winter}, {Woodcraft},
  {Wright}, {Xu}, {Zavagno}, {Zemcov}, {Zhang}, \&
  {Zonca}}]{2010A&A...518L...3G}
{Griffin}, M.~J., {Abergel}, A., {Abreu}, A., {et~al.} 2010, \aap, 518, L3

\bibitem[{{G{\"u}rkan} {et~al.}(2019){G{\"u}rkan}, {Hardcastle}, {Best},
  {Morabito}, {Prandoni}, {Jarvis}, {Duncan}, {Calistro Rivera}, {Callingham},
  {Cochrane}, {Croston}, {Heald}, {Mingo}, {Mooney}, {Sabater},
  {R{\"o}ttgering}, {Shimwell}, {Smith}, {Tasse}, \&
  {Williams}}]{2019A&A...622A..11G}
{G{\"u}rkan}, G., {Hardcastle}, M.~J., {Best}, P.~N., {et~al.} 2019, \aap, 622,
  A11

\bibitem[{{Hall} {et~al.}(2019){Hall}, {Zakamska}, {Addison}, {Battaglia},
  {Crichton}, {Devlin}, {Dunkley}, {Gralla}, {Hill}, {Hilton}, {Hubmayr},
  {Hughes}, {Huffenberger}, {Kosowsky}, {Marriage}, {Maurin}, {Moodley},
  {Niemack}, {Page}, {Partridge}, {D{\"u}nner Planella}, {Schillaci},
  {Sif{\'o}n}, {Staggs}, {Wollack}, \& {Xu}}]{2019MNRAS.490.2315H}
{Hall}, K.~R., {Zakamska}, N.~L., {Addison}, G.~E., {et~al.} 2019, \mnras, 490,
  2315

\bibitem[{{Hardcastle} {et~al.}(2007){Hardcastle}, {Evans}, \&
  {Croston}}]{2007MNRAS.376.1849H}
{Hardcastle}, M.~J., {Evans}, D.~A., \& {Croston}, J.~H. 2007, \mnras, 376,
  1849

\bibitem[{{Harris} {et~al.}(2016){Harris}, {Farrah}, {Schulz},
  {Hatziminaoglou}, {Viero}, {Anderson}, {B{\'e}thermin}, {Chapman},
  {Clements}, {Cooray}, {Efstathiou}, {Feltre}, {Hurley}, {Ibar}, {Lacy},
  {Oliver}, {Page}, {P{\'e}rez-Fournon}, {Petty}, {Pitchford}, {Rigopoulou},
  {Scott}, {Symeonidis}, {Vieira}, \& {Wang}}]{2016MNRAS.457.4179H}
{Harris}, K., {Farrah}, D., {Schulz}, B., {et~al.} 2016, \mnras, 457, 4179

\bibitem[{{Hartley} {et~al.}(2019){Hartley}, {Jackson}, {Sluse}, {Stacey}, \&
  {Vives-Arias}}]{2019MNRAS.485.3009H}
{Hartley}, P., {Jackson}, N., {Sluse}, D., {Stacey}, H.~R., \& {Vives-Arias},
  H. 2019, \mnras, 485, 3009

\bibitem[{{Heckman} \& {Best}(2014)}]{2014ARA&A..52..589H}
{Heckman}, T.~M. \& {Best}, P.~N. 2014, \araa, 52, 589

\bibitem[{{Helou} {et~al.}(1985){Helou}, {Soifer}, \&
  {Rowan-Robinson}}]{1985ApJ...298L...7H}
{Helou}, G., {Soifer}, B.~T., \& {Rowan-Robinson}, M. 1985, \apjl, 298, L7

\bibitem[{{Herrera Ruiz} {et~al.}(2016){Herrera Ruiz}, {Middelberg}, {Norris},
  \& {Maini}}]{2016A&A...589L...2H}
{Herrera Ruiz}, N., {Middelberg}, E., {Norris}, R.~P., \& {Maini}, A. 2016,
  \aap, 589, L2

\bibitem[{{Hodapp} {et~al.}(2004){Hodapp}, {Siegmund}, {Kaiser}, {Chambers},
  {Laux}, {Morgan}, \& {Mannery}}]{2004SPIE.5489..667H}
{Hodapp}, K.~W., {Siegmund}, W.~A., {Kaiser}, N., {et~al.} 2004, in Society of
  Photo-Optical Instrumentation Engineers (SPIE) Conference Series, Vol. 5489,
  Ground-based Telescopes, ed. J.~{Oschmann}, Jacobus~M., 667--678

\bibitem[{{Hodge} {et~al.}(2008){Hodge}, {Becker}, {White}, \& {de
  Vries}}]{2008AJ....136.1097H}
{Hodge}, J.~A., {Becker}, R.~H., {White}, R.~L., \& {de Vries}, W.~H. 2008,
  \aj, 136, 1097

\bibitem[{{Hopkins} {et~al.}(2016){Hopkins}, {Torrey}, {Faucher-Gigu{\`e}re},
  {Quataert}, \& {Murray}}]{2016MNRAS.458..816H}
{Hopkins}, P.~F., {Torrey}, P., {Faucher-Gigu{\`e}re}, C.-A., {Quataert}, E.,
  \& {Murray}, N. 2016, \mnras, 458, 816

\bibitem[{{Hwang} {et~al.}(2018){Hwang}, {Zakamska}, {Alexand roff}, {Hamann},
  {Greene}, {Perrotta}, \& {Richards}}]{2018MNRAS.477..830H}
{Hwang}, H.-C., {Zakamska}, N.~L., {Alexand roff}, R.~M., {et~al.} 2018,
  \mnras, 477, 830

\bibitem[{{Ilbert} {et~al.}(2013){Ilbert}, {McCracken}, {Le F{\`e}vre},
  {Capak}, {Dunlop}, {Karim}, {Renzini}, {Caputi}, {Boissier}, {Arnouts},
  {Aussel}, {Comparat}, {Guo}, {Hudelot}, {Kartaltepe}, {Kneib}, {Krogager},
  {Le Floc'h}, {Lilly}, {Mellier}, {Milvang-Jensen}, {Moutard}, {Onodera},
  {Richard}, {Salvato}, {Sanders}, {Scoville}, {Silverman}, {Taniguchi},
  {Tasca}, {Thomas}, {Toft}, {Tresse}, {Vergani}, {Wolk}, \&
  {Zirm}}]{2013A&A...556A..55I}
{Ilbert}, O., {McCracken}, H.~J., {Le F{\`e}vre}, O., {et~al.} 2013, \aap, 556,
  A55

\bibitem[{{Inskip} {et~al.}(2008){Inskip}, {Villar-Mart{\'\i}n}, {Tadhunter},
  {Morganti}, {Holt}, \& {Dicken}}]{2008MNRAS.386.1797I}
{Inskip}, K.~J., {Villar-Mart{\'\i}n}, M., {Tadhunter}, C.~N., {et~al.} 2008,
  \mnras, 386, 1797

\bibitem[{{Intema} {et~al.}(2011){Intema}, {van Weeren}, {R{\"o}ttgering}, \&
  {Lal}}]{2011A&A...535A..38I}
{Intema}, H.~T., {van Weeren}, R.~J., {R{\"o}ttgering}, H.~J.~A., \& {Lal},
  D.~V. 2011, \aap, 535, A38

\bibitem[{{Ishibashi} \& {Fabian}(2012)}]{2012MNRAS.427.2998I}
{Ishibashi}, W. \& {Fabian}, A.~C. 2012, \mnras, 427, 2998

\bibitem[{{Ivezi{\'c}} {et~al.}(2002){Ivezi{\'c}}, {Menou}, {Knapp}, {Strauss},
  {Lupton}, {Vand en Berk}, {Richards}, {Tremonti}, {Weinstein}, {Anderson},
  {Bahcall}, {Becker}, {Bernardi}, {Blanton}, {Eisenstein}, {Fan},
  {Finkbeiner}, {Finlator}, {Frieman}, {Gunn}, {Hall}, {Kim}, {Kinkhabwala},
  {Narayanan}, {Rockosi}, {Schlegel}, {Schneider}, {Strateva}, {SubbaRao},
  {Thakar}, {Voges}, {White}, {Yanny}, {Brinkmann}, {Doi}, {Fukugita},
  {Hennessy}, {Munn}, {Nichol}, \& {York}}]{2002AJ....124.2364I}
{Ivezi{\'c}}, {\v{Z}}., {Menou}, K., {Knapp}, G.~R., {et~al.} 2002, \aj, 124,
  2364

\bibitem[{{Jannuzi} \& {Dey}(1999)}]{1999ASPC..191..111J}
{Jannuzi}, B.~T. \& {Dey}, A. 1999, in Astronomical Society of the Pacific
  Conference Series, Vol. 191, Photometric Redshifts and the Detection of High
  Redshift Galaxies, ed. R.~{Weymann}, L.~{Storrie-Lombardi}, M.~{Sawicki}, \&
  R.~{Brunner}, 111

\bibitem[{{Jarvis} {et~al.}(2013){Jarvis}, {Bonfield}, {Bruce}, {Geach},
  {McAlpine}, {McLure}, {Gonz{\'a}lez-Solares}, {Irwin}, {Lewis}, {Yoldas},
  {Andreon}, {Cross}, {Emerson}, {Dalton}, {Dunlop}, {Hodgkin}, {Le},
  {Karouzos}, {Meisenheimer}, {Oliver}, {Rawlings}, {Simpson}, {Smail},
  {Smith}, {Sullivan}, {Sutherland}, {White}, \& {Zwart}}]{2013MNRAS.428.1281J}
{Jarvis}, M.~J., {Bonfield}, D.~G., {Bruce}, V.~A., {et~al.} 2013, \mnras, 428,
  1281

\bibitem[{{Jiang} {et~al.}(2007){Jiang}, {Fan}, {Ivezi{\'c}}, {Richards},
  {Schneider}, {Strauss}, \& {Kelly}}]{2007ApJ...656..680J}
{Jiang}, L., {Fan}, X., {Ivezi{\'c}}, {\v Z}., {et~al.} 2007, \apj, 656, 680

\bibitem[{{Kalfountzou} {et~al.}(2014){Kalfountzou}, {Stevens}, {Jarvis},
  {Hardcastle}, {Smith}, {Bourne}, {Dunne}, {Ibar}, {Eales}, {Ivison},
  {Maddox}, {Smith}, {Valiante}, \& {de Zotti}}]{2014MNRAS.442.1181K}
{Kalfountzou}, E., {Stevens}, J.~A., {Jarvis}, M.~J., {et~al.} 2014, \mnras,
  442, 1181

\bibitem[{{Kalfountzou} {et~al.}(2017){Kalfountzou}, {Stevens}, {Jarvis},
  {Hardcastle}, {Wilner}, {Elvis}, {Page}, {Trichas}, \&
  {Smith}}]{2017MNRAS.471...28K}
{Kalfountzou}, E., {Stevens}, J.~A., {Jarvis}, M.~J., {et~al.} 2017, \mnras,
  471, 28

\bibitem[{{Kellermann} {et~al.}(1989){Kellermann}, {Sramek}, {Schmidt},
  {Shaffer}, \& {Green}}]{1989AJ.....98.1195K}
{Kellermann}, K.~I., {Sramek}, R., {Schmidt}, M., {Shaffer}, D.~B., \& {Green},
  R. 1989, \aj, 98, 1195

\bibitem[{{Kennicutt}(1998)}]{1998ApJ...498..541K}
{Kennicutt}, Robert~C., J. 1998, \apj, 498, 541

\bibitem[{{Kimball} {et~al.}(2011){Kimball}, {Ivezi{\'c}}, {Wiita}, \&
  {Schneider}}]{2011AJ....141..182K}
{Kimball}, A.~E., {Ivezi{\'c}}, {\v{Z}}., {Wiita}, P.~J., \& {Schneider}, D.~P.
  2011, \aj, 141, 182

\bibitem[{{Klamer} {et~al.}(2004){Klamer}, {Ekers}, {Sadler}, \&
  {Hunstead}}]{2004ApJ...612L..97K}
{Klamer}, I.~J., {Ekers}, R.~D., {Sadler}, E.~M., \& {Hunstead}, R.~W. 2004,
  \apjl, 612, L97

\bibitem[{{Klindt} {et~al.}(2019){Klindt}, {Alexander}, {Rosario}, {Lusso}, \&
  {Fotopoulou}}]{2019MNRAS.488.3109K}
{Klindt}, L., {Alexander}, D.~M., {Rosario}, D.~J., {Lusso}, E., \&
  {Fotopoulou}, S. 2019, \mnras, 488, 3109

\bibitem[{{Kochanek} {et~al.}(2012){Kochanek}, {Eisenstein}, {Cool},
  {Caldwell}, {Assef}, {Jannuzi}, {Jones}, {Murray}, {Forman}, {Dey}, {Brown},
  {Eisenhardt}, {Gonzalez}, {Green}, \& {Stern}}]{2012ApJS..200....8K}
{Kochanek}, C.~S., {Eisenstein}, D.~J., {Cool}, R.~J., {et~al.} 2012, \apjs,
  200, 8

\bibitem[{{Labb{\'e}} {et~al.}(2005){Labb{\'e}}, {Huang}, {Franx}, {Rudnick},
  {Barmby}, {Daddi}, {van Dokkum}, {Fazio}, {Schreiber}, {Moorwood}, {Rix},
  {R{\"o}ttgering}, {Trujillo}, \& {van der Werf}}]{2005ApJ...624L..81L}
{Labb{\'e}}, I., {Huang}, J., {Franx}, M., {et~al.} 2005, \apjl, 624, L81

\bibitem[{{Lacy} {et~al.}(2020){Lacy}, {Baum}, {Chandler}, {Chatterjee},
  {Clarke}, {Deustua}, {English}, {Farnes}, {Gaensler}, {Gugliucci},
  {Hallinan}, {Kent}, {Kimball}, {Law}, {Lazio}, {Marvil}, {Mao}, {Medlin},
  {Mooley}, {Murphy}, {Myers}, {Osten}, {Richards}, {Rosolowsky}, {Rudnick},
  {Schinzel}, {Sivakoff}, {Sjouwerman}, {Taylor}, {White}, {Wrobel},
  {Andernach}, {Beasley}, {Berger}, {Bhatnager}, {Birkinshaw}, {Bower},
  {Brandt}, {Brown}, {Burke-Spolaor}, {Butler}, {Comerford}, {Demorest}, {Fu},
  {Giacintucci}, {Golap}, {G{\"u}th}, {Hales}, {Hiriart}, {Hodge}, {Horesh},
  {Ivezi{\'c}}, {Jarvis}, {Kamble}, {Kassim}, {Liu}, {Loinard}, {Lyons},
  {Masters}, {Mezcua}, {Moellenbrock}, {Mroczkowski}, {Nyland},
  {O{\textquoteright}Dea}, {O{\textquoteright}Sullivan}, {Peters}, {Radford},
  {Rao}, {Robnett}, {Salcido}, {Shen}, {Sobotka}, {Witz}, {Vaccari}, {van
  Weeren}, {Vargas}, {Williams}, \& {Yoon}}]{2020PASP..132c5001L}
{Lacy}, M., {Baum}, S.~A., {Chandler}, C.~J., {et~al.} 2020, \pasp, 132, 035001

\bibitem[{{Lacy} {et~al.}(2001){Lacy}, {Laurent-Muehleisen}, {Ridgway},
  {Becker}, \& {White}}]{2001ApJ...551L..17L}
{Lacy}, M., {Laurent-Muehleisen}, S.~A., {Ridgway}, S.~E., {Becker}, R.~H., \&
  {White}, R.~L. 2001, \apjl, 551, L17

\bibitem[{{Laor} {et~al.}(2019){Laor}, {Baldi}, \&
  {Behar}}]{2019MNRAS.482.5513L}
{Laor}, A., {Baldi}, R.~D., \& {Behar}, E. 2019, \mnras, 482, 5513

\bibitem[{{Laor} \& {Behar}(2008)}]{2008MNRAS.390..847L}
{Laor}, A. \& {Behar}, E. 2008, \mnras, 390, 847

\bibitem[{{Leipski} {et~al.}(2006){Leipski}, {Falcke}, {Bennert}, \&
  {H{\"u}ttemeister}}]{2006A&A...455..161L}
{Leipski}, C., {Falcke}, H., {Bennert}, N., \& {H{\"u}ttemeister}, S. 2006,
  \aap, 455, 161

\bibitem[{{Lutz} {et~al.}(2010){Lutz}, {Mainieri}, {Rafferty}, {Shao},
  {Hasinger}, {Wei{\ss}}, {Walter}, {Smail}, {Alexander}, {Brandt}, {Chapman},
  {Coppin}, {F{\"o}rster Schreiber}, {Gawiser}, {Genzel}, {Greve}, {Ivison},
  {Koekemoer}, {Kurczynski}, {Menten}, {Nordon}, {Popesso}, {Schinnerer},
  {Silverman}, {Wardlow}, \& {Xue}}]{2010ApJ...712.1287L}
{Lutz}, D., {Mainieri}, V., {Rafferty}, D., {et~al.} 2010, \apj, 712, 1287

\bibitem[{{Magliocchetti} {et~al.}(2014){Magliocchetti}, {Lutz}, {Rosario},
  {Berta}, {Le Floc'h}, {Magnelli}, {Pozzi}, {Riguccini}, \&
  {Santini}}]{2014MNRAS.442..682M}
{Magliocchetti}, M., {Lutz}, D., {Rosario}, D., {et~al.} 2014, \mnras, 442, 682

\bibitem[{{Magliocchetti} {et~al.}(2016){Magliocchetti}, {Lutz}, {Santini},
  {Salvato}, {Popesso}, {Berta}, \& {Pozzi}}]{2016MNRAS.456..431M}
{Magliocchetti}, M., {Lutz}, D., {Santini}, P., {et~al.} 2016, \mnras, 456, 431

\bibitem[{{Magliocchetti} {et~al.}(2020){Magliocchetti}, {Pentericci},
  {Cirasuolo}, {Zamorani}, {Amorin}, {Bongiorno}, {Cimatti}, {Fontana},
  {Garilli}, {Gargiulo}, {Hathi}, {McLeod}, {McLure}, {Brusa}, {Saxena}, \&
  {Talia}}]{2020MNRAS.493.3838M}
{Magliocchetti}, M., {Pentericci}, L., {Cirasuolo}, M., {et~al.} 2020, \mnras,
  493, 3838

\bibitem[{{Magnelli} {et~al.}(2015){Magnelli}, {Ivison}, {Lutz}, {Valtchanov},
  {Farrah}, {Berta}, {Bertoldi}, {Bock}, {Cooray}, {Ibar}, {Karim}, {Le
  Floc'h}, {Nordon}, {Oliver}, {Page}, {Popesso}, {Pozzi}, {Rigopoulou},
  {Riguccini}, {Rodighiero}, {Rosario}, {Roseboom}, {Wang}, \&
  {Wuyts}}]{2015A&A...573A..45M}
{Magnelli}, B., {Ivison}, R.~J., {Lutz}, D., {et~al.} 2015, \aap, 573, A45

\bibitem[{{Mahony} {et~al.}(2016){Mahony}, {Morganti}, {Prandoni}, {van
  Bemmel}, {Shimwell}, {Brienza}, {Best}, {Br{\"u}ggen}, {Calistro Rivera}, {de
  Gasperin}, {Hardcastle}, {Harwood}, {Heald}, {Jarvis}, {Mandal}, {Miley},
  {Retana-Montenegro}, {R{\"o}ttgering}, {Sabater}, {Tasse}, {van Velzen}, {van
  Weeren}, {Williams}, \& {White}}]{2016MNRAS.463.2997M}
{Mahony}, E.~K., {Morganti}, R., {Prandoni}, I., {et~al.} 2016, \mnras, 463,
  2997

\bibitem[{{Maini} {et~al.}(2016){Maini}, {Prandoni}, {Norris}, {Giovannini}, \&
  {Spitler}}]{2016A&A...589L...3M}
{Maini}, A., {Prandoni}, I., {Norris}, R.~P., {Giovannini}, G., \& {Spitler},
  L.~R. 2016, \aap, 589, L3

\bibitem[{{Makovoz} {et~al.}(2006){Makovoz}, {Roby}, {Khan}, \&
  {Booth}}]{2006SPIE.6274E..0CM}
{Makovoz}, D., {Roby}, T., {Khan}, I., \& {Booth}, H. 2006, in Society of
  Photo-Optical Instrumentation Engineers (SPIE) Conference Series, Vol. 6274,
  Society of Photo-Optical Instrumentation Engineers (SPIE) Conference Series,
  ed. H.~{Lewis} \& A.~{Bridger}, 62740C

\bibitem[{{Matsuoka} {et~al.}(2014){Matsuoka}, {Strauss}, {Price}, \&
  {DiDonato}}]{2014ApJ...780..162M}
{Matsuoka}, Y., {Strauss}, M.~A., {Price}, Ted~N., I., \& {DiDonato}, M.~S.
  2014, \apj, 780, 162

\bibitem[{{Mauch} {et~al.}(2020){Mauch}, {Cotton}, {Condon}, {Matthews},
  {Abbott}, {Adam}, {Aldera}, {Asad}, {Bauermeister}, {Bennett}, {Bester},
  {Botha}, {Brederode}, {Brits}, {Buchner}, {Burger}, {Camilo}, {Chalmers},
  {Cheetham}, {de Villiers}, {de Villiers}, {Dikgale-Mahlakoana}, {du Toit},
  {Esterhuyse}, {Fadana}, {Fanaroff}, {Fataar}, {February}, {Frank},
  {Gamatham}, {Geyer}, {Goedhart}, {Gounden}, {Gumede}, {Heywood}, {Hlakola},
  {Horrell}, {Hugo}, {Isaacson}, {J{\'o}zsa}, {Jonas}, {Julie}, {Kapp},
  {Kasper}, {Kenyon}, {Kotz{\'e}}, {Kriek}, {Kriel}, {Kusel}, {Lehmensiek},
  {Loots}, {Lord}, {Lunsky}, {Madisa}, {Magnus}, {Main}, {Malan}, {Manley},
  {Marais}, {Martens}, {Merry}, {Millenaar}, {Mnyandu}, {Moeng}, {Mokone},
  {Monama}, {Mphego}, {New}, {Ngcebetsha}, {Ngoasheng}, {Ockards}, {Oozeer},
  {Otto}, {Patel}, {Peens-Hough}, {Perkins}, {Ramaila}, {Ramudzuli}, {Renil},
  {Richter}, {Robyntjies}, {Salie}, {Schollar}, {Schwardt}, {Serylak},
  {Siebrits}, {Sirothia}, {Smirnov}, {Sofeya}, {Stone}, {Taljaard}, {Tasse},
  {Theron}, {Tiplady}, {Toruvanda}, {Twum}, {van Balla}, {van der Byl}, {van
  der Merwe}, {Van Tonder}, {Wallace}, {Welz}, {Williams}, \&
  {Xaia}}]{2020ApJ...888...61M}
{Mauch}, T., {Cotton}, W.~D., {Condon}, J.~J., {et~al.} 2020, \apj, 888, 61

\bibitem[{{Mauch} {et~al.}(2013){Mauch}, {Kl{\"o}ckner}, {Rawlings}, {Jarvis},
  {Hardcastle}, {Obreschkow}, {Saikia}, \& {Thompson}}]{2013MNRAS.435..650M}
{Mauch}, T., {Kl{\"o}ckner}, H.-R., {Rawlings}, S., {et~al.} 2013, \mnras, 435,
  650

\bibitem[{{McGreer} {et~al.}(2006){McGreer}, {Becker}, {Helfand}, \&
  {White}}]{2006ApJ...652..157M}
{McGreer}, I.~D., {Becker}, R.~H., {Helfand}, D.~J., \& {White}, R.~L. 2006,
  \apj, 652, 157

\bibitem[{{Menci} {et~al.}(2008){Menci}, {Fiore}, {Puccetti}, \&
  {Cavaliere}}]{2008ApJ...686..219M}
{Menci}, N., {Fiore}, F., {Puccetti}, S., \& {Cavaliere}, A. 2008, \apj, 686,
  219

\bibitem[{{Miller} {et~al.}(1990){Miller}, {Peacock}, \&
  {Mead}}]{1990MNRAS.244..207M}
{Miller}, L., {Peacock}, J.~A., \& {Mead}, A.~R.~G. 1990, \mnras, 244, 207

\bibitem[{{Miller} {et~al.}(1993){Miller}, {Rawlings}, \&
  {Saunders}}]{1993MNRAS.263..425M}
{Miller}, P., {Rawlings}, S., \& {Saunders}, R. 1993, \mnras, 263, 425

\bibitem[{{Morabito} {et~al.}(2019){Morabito}, {Matthews}, {Best},
  {G{\"u}rkan}, {Jarvis}, {Prandoni}, {Duncan}, {Hardcastle},
  {Kunert-Bajraszewska}, {Mechev}, {Mooney}, {Sabater}, {R{\"o}ttgering},
  {Shimwell}, {Smith}, {Tasse}, \& {Williams}}]{2019A&A...622A..15M}
{Morabito}, L.~K., {Matthews}, J.~H., {Best}, P.~N., {et~al.} 2019, \aap, 622,
  A15

\bibitem[{{Mould} {et~al.}(2000){Mould}, {Ridgewell}, {Gallagher}, {Bessell},
  {Keller}, {Calzetti}, {Clarke}, {Trauger}, {Grillmair}, {Ballester},
  {Burrows}, {Krist}, {Crisp}, {Evans}, {Griffiths}, {Hester}, {Hoessel},
  {Holtzman}, {Scowen}, {Stapelfeldt}, {Sahai}, {Watson}, \&
  {Meadows}}]{2000ApJ...536..266M}
{Mould}, J.~R., {Ridgewell}, A., {Gallagher}, John~S., I., {et~al.} 2000, \apj,
  536, 266

\bibitem[{{Nesvadba} {et~al.}(2006){Nesvadba}, {Lehnert}, {Eisenhauer},
  {Gilbert}, {Tecza}, \& {Abuter}}]{2006ApJ...650..693N}
{Nesvadba}, N.~P.~H., {Lehnert}, M.~D., {Eisenhauer}, F., {et~al.} 2006, \apj,
  650, 693

\bibitem[{{Noeske} {et~al.}(2007){Noeske}, {Weiner}, {Faber}, {Papovich},
  {Koo}, {Somerville}, {Bundy}, {Conselice}, {Newman}, {Schiminovich}, {Le
  Floc'h}, {Coil}, {Rieke}, {Lotz}, {Primack}, {Barmby}, {Cooper}, {Davis},
  {Ellis}, {Fazio}, {Guhathakurta}, {Huang}, {Kassin}, {Martin}, {Phillips},
  {Rich}, {Small}, {Willmer}, \& {Wilson}}]{2007ApJ...660L..43N}
{Noeske}, K.~G., {Weiner}, B.~J., {Faber}, S.~M., {et~al.} 2007, \apjl, 660,
  L43

\bibitem[{{Oke} \& {Gunn}(1983)}]{1983ApJ...266..713O}
{Oke}, J.~B. \& {Gunn}, J.~E. 1983, \apj, 266, 713

\bibitem[{{Oliver} {et~al.}(2012){Oliver}, {Bock}, {Altieri}, {Amblard},
  {Arumugam}, {Aussel}, {Babbedge}, {Beelen}, {B{\'e}thermin}, {Blain},
  {Boselli}, {Bridge}, {Brisbin}, {Buat}, {Burgarella},
  {Castro-Rodr{\'\i}guez}, {Cava}, {Chanial}, {Cirasuolo}, {Clements},
  {Conley}, {Conversi}, {Cooray}, {Dowell}, {Dubois}, {Dwek}, {Dye}, {Eales},
  {Elbaz}, {Farrah}, {Feltre}, {Ferrero}, {Fiolet}, {Fox}, {Franceschini},
  {Gear}, {Giovannoli}, {Glenn}, {Gong}, {Gonz{\'a}lez Solares}, {Griffin},
  {Halpern}, {Harwit}, {Hatziminaoglou}, {Heinis}, {Hurley}, {Hwang}, {Hyde},
  {Ibar}, {Ilbert}, {Isaak}, {Ivison}, {Lagache}, {Le Floc'h}, {Levenson},
  {Faro}, {Lu}, {Madden}, {Maffei}, {Magdis}, {Mainetti}, {Marchetti},
  {Marsden}, {Marshall}, {Mortier}, {Nguyen}, {O'Halloran}, {Omont}, {Page},
  {Panuzzo}, {Papageorgiou}, {Patel}, {Pearson}, {P{\'e}rez-Fournon}, {Pohlen},
  {Rawlings}, {Raymond}, {Rigopoulou}, {Riguccini}, {Rizzo}, {Rodighiero},
  {Roseboom}, {Rowan-Robinson}, {S{\'a}nchez Portal}, {Schulz}, {Scott},
  {Seymour}, {Shupe}, {Smith}, {Stevens}, {Symeonidis}, {Trichas}, {Tugwell},
  {Vaccari}, {Valtchanov}, {Vieira}, {Viero}, {Vigroux}, {Wang}, {Ward},
  {Wardlow}, {Wright}, {Xu}, \& {Zemcov}}]{2012MNRAS.424.1614O}
{Oliver}, S.~J., {Bock}, J., {Altieri}, B., {et~al.} 2012, \mnras, 424, 1614

\bibitem[{{Padovani}(2016)}]{2016A&ARv..24...13P}
{Padovani}, P. 2016, \aapr, 24, 13

\bibitem[{{Padovani} {et~al.}(2015){Padovani}, {Bonzini}, {Kellermann},
  {Miller}, {Mainieri}, \& {Tozzi}}]{2015MNRAS.452.1263P}
{Padovani}, P., {Bonzini}, M., {Kellermann}, K.~I., {et~al.} 2015, \mnras, 452,
  1263

\bibitem[{{P{\^a}ris} {et~al.}(2018){P{\^a}ris}, {Petitjean}, {Aubourg},
  {Myers}, {Streblyanska}, {Lyke}, {Anderson}, {Armengaud}, {Bautista},
  {Blanton}, {Blomqvist}, {Brinkmann}, {Brownstein}, {Brandt}, {Burtin},
  {Dawson}, {de la Torre}, {Georgakakis}, {Gil-Mar{\'{\i}}n}, {Green}, {Hall},
  {Kneib}, {LaMassa}, {Le Goff}, {MacLeod}, {Mariappan}, {McGreer}, {Merloni},
  {Noterdaeme}, {Palanque-Delabrouille}, {Percival}, {Ross}, {Rossi},
  {Schneider}, {Seo}, {Tojeiro}, {Weaver}, {Weijmans}, {Y{\`e}che}, {Zarrouk},
  \& {Zhao}}]{2018A&A...613A..51P}
{P{\^a}ris}, I., {Petitjean}, P., {Aubourg}, {\'E}., {et~al.} 2018, \aap, 613,
  A51

\bibitem[{{Perger} {et~al.}(2019){Perger}, {Frey}, {Gab{\'a}nyi}, \&
  {T{\'o}th}}]{2019MNRAS.490.2542P}
{Perger}, K., {Frey}, S., {Gab{\'a}nyi}, K.~{\'E}., \& {T{\'o}th}, L.~V. 2019,
  \mnras, 490, 2542

\bibitem[{{Pitchford} {et~al.}(2016){Pitchford}, {Hatziminaoglou}, {Feltre},
  {Farrah}, {Clarke}, {Harris}, {Hurley}, {Oliver}, {Page}, \&
  {Wang}}]{2016MNRAS.462.4067P}
{Pitchford}, L.~K., {Hatziminaoglou}, E., {Feltre}, A., {et~al.} 2016, \mnras,
  462, 4067

\bibitem[{{Podigachoski} {et~al.}(2015){Podigachoski}, {Barthel}, {Haas},
  {Leipski}, {Wilkes}, {Kuraszkiewicz}, {Westhues}, {Willner}, {Ashby},
  {Chini}, {Clements}, {Fazio}, {Labiano}, {Lawrence}, {Meisenheimer},
  {Peletier}, {Siebenmorgen}, \& {Verdoes Kleijn}}]{2015A&A...575A..80P}
{Podigachoski}, P., {Barthel}, P.~D., {Haas}, M., {et~al.} 2015, \aap, 575, A80

\bibitem[{{Poglitsch} {et~al.}(2010){Poglitsch}, {Waelkens}, {Geis},
  {Feuchtgruber}, {Vandenbussche}, {Rodriguez}, {Krause}, {Renotte}, {van
  Hoof}, {Saraceno}, {Cepa}, {Kerschbaum}, {Agn{\`e}se}, {Ali}, {Altieri},
  {Andreani}, {Augueres}, {Balog}, {Barl}, {Bauer}, {Belbachir}, {Benedettini},
  {Billot}, {Boulade}, {Bischof}, {Blommaert}, {Callut}, {Cara}, {Cerulli},
  {Cesarsky}, {Contursi}, {Creten}, {De Meester}, {Doublier}, {Doumayrou},
  {Duband }, {Exter}, {Genzel}, {Gillis}, {Gr{\"o}zinger}, {Henning},
  {Herreros}, {Huygen}, {Inguscio}, {Jakob}, {Jamar}, {Jean}, {de Jong},
  {Katterloher}, {Kiss}, {Klaas}, {Lemke}, {Lutz}, {Madden}, {Marquet},
  {Martignac}, {Mazy}, {Merken}, {Montfort}, {Morbidelli}, {M{\"u}ller},
  {Nielbock}, {Okumura}, {Orfei}, {Ottensamer}, {Pezzuto}, {Popesso},
  {Putzeys}, {Regibo}, {Reveret}, {Royer}, {Sauvage}, {Schreiber}, {Stegmaier},
  {Schmitt}, {Schubert}, {Sturm}, {Thiel}, {Tofani}, {Vavrek}, {Wetzstein},
  {Wieprecht}, \& {Wiezorrek}}]{2010A&A...518L...2P}
{Poglitsch}, A., {Waelkens}, C., {Geis}, N., {et~al.} 2010, \aap, 518, L2

\bibitem[{{Retana-Montenegro} \& {R{\"o}ttgering}(2018)}]{2018FrASS...5....5R}
{Retana-Montenegro}, E. \& {R{\"o}ttgering}, H. 2018, Frontiers in Astronomy
  and Space Sciences, 5, 5

\bibitem[{{Retana-Montenegro} \& {R\"ottgering}(2020)}]{2020AA}
{Retana-Montenegro}, E. \& {R\"ottgering}, H. 2020, \aap

\bibitem[{{Retana-Montenegro} \& {R{\"o}ttgering}(2017)}]{2017A&A...600A..97R}
{Retana-Montenegro}, E. \& {R{\"o}ttgering}, H.~J.~A. 2017, \aap, 600, A97

\bibitem[{{Retana-Montenegro} {et~al.}(2018){Retana-Montenegro},
  {R{\"o}ttgering}, {Shimwell}, {van Weeren}, {Prandoni}, {Brunetti}, {Best},
  \& {Br{\"u}ggen}}]{2018A&A...620A..74R}
{Retana-Montenegro}, E., {R{\"o}ttgering}, H.~J.~A., {Shimwell}, T.~W.,
  {et~al.} 2018, \aap, 620, A74

\bibitem[{{Richards} {et~al.}(2001){Richards}, {Fan}, {Schneider}, {Vanden
  Berk}, {Strauss}, {York}, {Anderson}, {Anderson}, {Annis}, {Bahcall},
  {Bernardi}, {Briggs}, {Brinkmann}, {Brunner}, {Burles}, {Carey}, {Castand
  er}, {Connolly}, {Crocker}, {Csabai}, {Doi}, {Finkbeiner}, {Friedman},
  {Frieman}, {Fukugita}, {Gunn}, {Hindsley}, {Ivezi{\'c}}, {Kent}, {Knapp},
  {Lamb}, {Leger}, {Long}, {Loveday}, {Lupton}, {McKay}, {Meiksin}, {Merrelli},
  {Munn}, {Newberg}, {Newcomb}, {Nichol}, {Owen}, {Pier}, {Pope}, {Richmond},
  {Rockosi}, {Schlegel}, {Siegmund}, {Smee}, {Snir}, {Stoughton}, {Stubbs},
  {SubbaRao}, {Szalay}, {Szokoly}, {Tremonti}, {Uomoto}, {Waddell}, {Yanny}, \&
  {Zheng}}]{2001AJ....121.2308R}
{Richards}, G.~T., {Fan}, X., {Schneider}, D.~P., {et~al.} 2001, \aj, 121, 2308

\bibitem[{{Richards} {et~al.}(2006){Richards}, {Lacy}, {Storrie-Lombardi},
  {Hall}, {Gallagher}, {Hines}, {Fan}, {Papovich}, {Vanden Berk}, {Trammell},
  {Schneider}, {Vestergaard}, {York}, {Jester}, {Anderson}, {Budav{\'a}ri}, \&
  {Szalay}}]{2006ApJS..166..470R}
{Richards}, G.~T., {Lacy}, M., {Storrie-Lombardi}, L.~J., {et~al.} 2006, \apjs,
  166, 470

\bibitem[{{Riechers} {et~al.}(2011){Riechers}, {Carilli}, {Maddalena}, {Hodge},
  {Harris}, {Baker}, {Walter}, {Wagg}, {Vand en Bout}, {Wei{\ss}}, \&
  {Sharon}}]{2011ApJ...739L..32R}
{Riechers}, D.~A., {Carilli}, C.~L., {Maddalena}, R.~J., {et~al.} 2011, \apjl,
  739, L32

\bibitem[{{Rieke} {et~al.}(2004){Rieke}, {Young}, {Engelbracht}, {Kelly},
  {Low}, {Haller}, {Beeman}, {Gordon}, {Stansberry}, {Misselt}, {Cadien},
  {Morrison}, {Rivlis}, {Latter}, {Noriega-Crespo}, {Padgett}, {Stapelfeldt},
  {Hines}, {Egami}, {Muzerolle}, {Alonso-Herrero}, {Blaylock}, {Dole}, {Hinz},
  {Le Floc'h}, {Papovich}, {P{\'e}rez-Gonz{\'a}lez}, {Smith}, {Su}, {Bennett},
  {Frayer}, {Henderson}, {Lu}, {Masci}, {Pesenson}, {Rebull}, {Rho}, {Keene},
  {Stolovy}, {Wachter}, {Wheaton}, {Werner}, \&
  {Richards}}]{2004ApJS..154...25R}
{Rieke}, G.~H., {Young}, E.~T., {Engelbracht}, C.~W., {et~al.} 2004, \apjs,
  154, 25

\bibitem[{{Rosario} {et~al.}(2020){Rosario}, {Fawcett}, {Klindt}, {Alexand er},
  {Morabito}, {Fotopoulou}, {Lusso}, \& {Calistro
  Rivera}}]{2020MNRAS.494.3061R}
{Rosario}, D.~J., {Fawcett}, V.~A., {Klindt}, L., {et~al.} 2020, \mnras, 494,
  3061

\bibitem[{{Rosario} {et~al.}(2013){Rosario}, {Trakhtenbrot}, {Lutz}, {Netzer},
  {Trump}, {Silverman}, {Schramm}, {Lusso}, {Berta}, {Bongiorno}, {Brusa},
  {F{\"o}rster-Schreiber}, {Genzel}, {Lilly}, {Magnelli}, {Mainieri},
  {Maiolino}, {Merloni}, {Mignoli}, {Nordon}, {Popesso}, {Salvato}, {Santini},
  {Tacconi}, \& {Zamorani}}]{2013A&A...560A..72R}
{Rosario}, D.~J., {Trakhtenbrot}, B., {Lutz}, D., {et~al.} 2013, \aap, 560, A72

\bibitem[{{Roseboom} {et~al.}(2010){Roseboom}, {Oliver}, {Kunz}, {Altieri},
  {Amblard}, {Arumugam}, {Auld}, {Aussel}, {Babbedge}, {B{\'e}thermin},
  {Blain}, {Bock}, {Boselli}, {Brisbin}, {Buat}, {Burgarella},
  {Castro-Rodr{\'\i}guez}, {Cava}, {Chanial}, {Chapin}, {Clements}, {Conley},
  {Conversi}, {Cooray}, {Dowell}, {Dwek}, {Dye}, {Eales}, {Elbaz}, {Farrah},
  {Fox}, {Franceschini}, {Gear}, {Glenn}, {Solares}, {Griffin}, {Halpern},
  {Harwit}, {Hatziminaoglou}, {Huang}, {Ibar}, {Isaak}, {Ivison}, {Lagache},
  {Levenson}, {Lu}, {Madden}, {Maffei}, {Mainetti}, {Marchetti}, {Marsden},
  {Mortier}, {Nguyen}, {O'Halloran}, {Omont}, {Page}, {Panuzzo},
  {Papageorgiou}, {Patel}, {Pearson}, {P{\'e}rez-Fournon}, {Pohlen},
  {Rawlings}, {Raymond}, {Rigopoulou}, {Rizzo}, {Rowan-Robinson}, {Portal},
  {Schulz}, {Scott}, {Seymour}, {Shupe}, {Smith}, {Stevens}, {Symeonidis},
  {Trichas}, {Tugwell}, {Vaccari}, {Valtchanov}, {Vieira}, {Vigroux}, {Wang},
  {Ward}, {Wright}, {Xu}, \& {Zemcov}}]{2010MNRAS.409...48R}
{Roseboom}, I.~G., {Oliver}, S.~J., {Kunz}, M., {et~al.} 2010, \mnras, 409, 48

\bibitem[{{Rovilos} {et~al.}(2012){Rovilos}, {Comastri}, {Gilli},
  {Georgantopoulos}, {Ranalli}, {Vignali}, {Lusso}, {Cappelluti}, {Zamorani},
  {Elbaz}, {Dickinson}, {Hwang}, {Charmandaris}, {Ivison}, {Merloni}, {Daddi},
  {Carrera}, {Brandt}, {Mullaney}, {Scott}, {Alexand er}, {Del Moro},
  {Morrison}, {Murphy}, {Altieri}, {Aussel}, {Dannerbauer}, {Kartaltepe},
  {Leiton}, {Magdis}, {Magnelli}, {Popesso}, \&
  {Valtchanov}}]{2012A&A...546A..58R}
{Rovilos}, E., {Comastri}, A., {Gilli}, R., {et~al.} 2012, \aap, 546, A58

\bibitem[{{Salom{\'e}} {et~al.}(2017){Salom{\'e}}, {Salom{\'e}},
  {Miville-Desch{\^e}nes}, {Combes}, \& {Hamer}}]{2017A&A...608A..98S}
{Salom{\'e}}, Q., {Salom{\'e}}, P., {Miville-Desch{\^e}nes}, M.~A., {Combes},
  F., \& {Hamer}, S. 2017, \aap, 608, A98

\bibitem[{{Sanders} {et~al.}(2007){Sanders}, {Salvato}, {Aussel}, {Ilbert},
  {Scoville}, {Surace}, {Frayer}, {Sheth}, {Helou}, {Brooke}, {Bhattacharya},
  {Yan}, {Kartaltepe}, {Barnes}, {Blain}, {Calzetti}, {Capak}, {Carilli},
  {Carollo}, {Comastri}, {Daddi}, {Ellis}, {Elvis}, {Fall}, {Franceschini},
  {Giavalisco}, {Hasinger}, {Impey}, {Koekemoer}, {Le F{\`e}vre}, {Lilly},
  {Liu}, {McCracken}, {Mobasher}, {Renzini}, {Rich}, {Schinnerer}, {Shopbell},
  {Taniguchi}, {Thompson}, {Urry}, \& {Williams}}]{2007ApJS..172...86S}
{Sanders}, D.~B., {Salvato}, M., {Aussel}, H., {et~al.} 2007, \apjs, 172, 86

\bibitem[{{Santini} {et~al.}(2017){Santini}, {Fontana}, {Castellano}, {Di
  Criscienzo}, {Merlin}, {Amorin}, {Cullen}, {Daddi}, {Dickinson}, {Dunlop},
  {Grazian}, {Lamastra}, {McLure}, {Micha{\l}owski}, {Pentericci}, \&
  {Shu}}]{2017ApJ...847...76S}
{Santini}, P., {Fontana}, A., {Castellano}, M., {et~al.} 2017, \apj, 847, 76

\bibitem[{{Santini} {et~al.}(2009){Santini}, {Fontana}, {Grazian}, {Salimbeni},
  {Fiore}, {Fontanot}, {Boutsia}, {Castellano}, {Cristiani}, {de Santis},
  {Gallozzi}, {Giallongo}, {Menci}, {Nonino}, {Paris}, {Pentericci}, \&
  {Vanzella}}]{2009A&A...504..751S}
{Santini}, P., {Fontana}, A., {Grazian}, A., {et~al.} 2009, \aap, 504, 751

\bibitem[{{Scaife} \& {Heald}(2012)}]{2012MNRAS.423L..30S}
{Scaife}, A.~M.~M. \& {Heald}, G.~H. 2012, \mnras, 423, L30

\bibitem[{{Schlafly} \& {Finkbeiner}(2011)}]{2011ApJ...737..103S}
{Schlafly}, E.~F. \& {Finkbeiner}, D.~P. 2011, \apj, 737, 103

\bibitem[{{Schmidt}(1963)}]{1963Natur.197.1040S}
{Schmidt}, M. 1963, \nat, 197, 1040

\bibitem[{{Schneider} {et~al.}(2005){Schneider}, {Hall}, {Richards}, {Vanden
  Berk}, {Anderson}, {Fan}, {Jester}, {Stoughton}, {Strauss}, {SubbaRao},
  {Brandt}, {Gunn}, {Yanny}, {Bahcall}, {Barentine}, {Blanton}, {Boroski},
  {Brewington}, {Brinkmann}, {Brunner}, {Csabai}, {Doi}, {Eisenstein},
  {Frieman}, {Fukugita}, {Gray}, {Harvanek}, {Heckman}, {Ivezi{\'c}}, {Kent},
  {Kleinman}, {Knapp}, {Kron}, {Krzesinski}, {Long}, {Loveday}, {Lupton},
  {Margon}, {Munn}, {Neilsen}, {Newberg}, {Newman}, {Nichol}, {Nitta}, {Pier},
  {Rockosi}, {Saxe}, {Schlegel}, {Snedden}, {Szalay}, {Thakar}, {Uomoto},
  {Voges}, \& {York}}]{2005AJ....130..367S}
{Schneider}, D.~P., {Hall}, P.~B., {Richards}, G.~T., {et~al.} 2005, \aj, 130,
  367

\bibitem[{{Scholtz} {et~al.}(2018){Scholtz}, {Alexander}, {Harrison},
  {Rosario}, {McAlpine}, {Mullaney}, {Stanley}, {Simpson}, {Theuns}, {Bower},
  {Hickox}, {Santini}, \& {Swinbank}}]{2018MNRAS.475.1288S}
{Scholtz}, J., {Alexander}, D.~M., {Harrison}, C.~M., {et~al.} 2018, \mnras,
  475, 1288

\bibitem[{{Schreiber} {et~al.}(2015){Schreiber}, {Pannella}, {Elbaz},
  {B{\'e}thermin}, {Inami}, {Dickinson}, {Magnelli}, {Wang}, {Aussel}, {Daddi},
  {Juneau}, {Shu}, {Sargent}, {Buat}, {Faber}, {Ferguson}, {Giavalisco},
  {Koekemoer}, {Magdis}, {Morrison}, {Papovich}, {Santini}, \&
  {Scott}}]{2015A&A...575A..74S}
{Schreiber}, C., {Pannella}, M., {Elbaz}, D., {et~al.} 2015, \aap, 575, A74

\bibitem[{{Schulze} {et~al.}(2019){Schulze}, {Silverman}, {Daddi},
  {Rujopakarn}, {Liu}, {Schramm}, {Mainieri}, {Imanishi}, {Hirschmann}, \&
  {Jahnke}}]{2019MNRAS.488.1180S}
{Schulze}, A., {Silverman}, J.~D., {Daddi}, E., {et~al.} 2019, \mnras, 488,
  1180

\bibitem[{{Serjeant} \& {Hatziminaoglou}(2009)}]{2009MNRAS.397..265S}
{Serjeant}, S. \& {Hatziminaoglou}, E. 2009, \mnras, 397, 265

\bibitem[{{Seymour} {et~al.}(2011){Seymour}, {Symeonidis}, {Page}, {Amblard},
  {Arumugam}, {Aussel}, {Blain}, {Bock}, {Boselli}, {Buat},
  {Castro-Rodr{\'\i}guez}, {Cava}, {Chanial}, {Clements}, {Conley}, {Conversi},
  {Cooray}, {Dowell}, {Dwek}, {Eales}, {Elbaz}, {Franceschini}, {Glenn},
  {Solares}, {Griffin}, {Hatziminaoglou}, {Ibar}, {Isaak}, {Ivison}, {Lagache},
  {Levenson}, {Lu}, {Madden}, {Maffei}, {Mainetti}, {Marchetti}, {Nguyen},
  {O'Halloran}, {Oliver}, {Omont}, {Panuzzo}, {Papageorgiou}, {Pearson},
  {P{\'e}rez-Fournon}, {Pohlen}, {Rawlings}, {Rizzo}, {Roseboom},
  {Rowan-Robinson}, {Schulz}, {Scott}, {Shupe}, {Smith}, {Stevens}, {Trichas},
  {Tugwell}, {Vaccari}, {Valtchanov}, {Vigroux}, {Wang}, {Wright}, {Xu}, \&
  {Zemcov}}]{2011MNRAS.413.1777S}
{Seymour}, N., {Symeonidis}, M., {Page}, M.~J., {et~al.} 2011, \mnras, 413,
  1777

\bibitem[{{Shangguan} {et~al.}(2020){Shangguan}, {Ho}, {Bauer}, {Wang}, \&
  {Treister}}]{2020ApJS..247...15S}
{Shangguan}, J., {Ho}, L.~C., {Bauer}, F.~E., {Wang}, R., \& {Treister}, E.
  2020, \apjs, 247, 15

\bibitem[{{Singal} {et~al.}(2011){Singal}, {Petrosian}, {Lawrence}, \&
  {Stawarz}}]{2011ApJ...743..104S}
{Singal}, J., {Petrosian}, V., {Lawrence}, A., \& {Stawarz}, L. 2011, \apj,
  743, 104

\bibitem[{{Sirothia} {et~al.}(2009){Sirothia}, {Dennefeld}, {Saikia}, {Dole},
  {Ricquebourg}, \& {Roland}}]{2009MNRAS.395..269S}
{Sirothia}, S.~K., {Dennefeld}, M., {Saikia}, D.~J., {et~al.} 2009, \mnras,
  395, 269

\bibitem[{{Smol{\v c}i{\'c}} {et~al.}(2017){Smol{\v c}i{\'c}}, {Novak},
  {Bondi}, {Ciliegi}, {Mooley}, {Schinnerer}, {Zamorani}, {Navarrete},
  {Bourke}, {Karim}, {Vardoulaki}, {Leslie}, {Delhaize}, {Carilli}, {Myers},
  {Baran}, {Delvecchio}, {Miettinen}, {Banfield}, {Balokovi{\'c}}, {Bertoldi},
  {Capak}, {Frail}, {Hallinan}, {Hao}, {Herrera Ruiz}, {Horesh}, {Ilbert},
  {Intema}, {Jeli{\'c}}, {Kl{\"o}ckner}, {Krpan}, {Kulkarni}, {McCracken},
  {Laigle}, {Middleberg}, {Murphy}, {Sargent}, {Scoville}, \&
  {Sheth}}]{2017A&A...602A...1S}
{Smol{\v c}i{\'c}}, V., {Novak}, M., {Bondi}, M., {et~al.} 2017, \aap, 602, A1

\bibitem[{{Smol{\v{c}}i{\'c}}(2009)}]{2009ApJ...699L..43S}
{Smol{\v{c}}i{\'c}}, V. 2009, \apjl, 699, L43

\bibitem[{{Sopp} \& {Alexander}(1991)}]{1991MNRAS.251..112S}
{Sopp}, H.~M. \& {Alexander}, P. 1991, \mnras, 251, 112

\bibitem[{{Speagle} {et~al.}(2014){Speagle}, {Steinhardt}, {Capak}, \&
  {Silverman}}]{2014ApJS..214...15S}
{Speagle}, J.~S., {Steinhardt}, C.~L., {Capak}, P.~L., \& {Silverman}, J.~D.
  2014, \apjs, 214, 15

\bibitem[{{Stanley} {et~al.}(2017){Stanley}, {Alexander}, {Harrison},
  {Rosario}, {Wang}, {Aird}, {Bourne}, {Dunne}, {Dye}, {Eales}, {Knudsen},
  {Micha{\l}owski}, {Valiante}, {De Zotti}, {Furlanetto}, {Ivison}, {Maddox},
  \& {Smith}}]{2017MNRAS.472.2221S}
{Stanley}, F., {Alexander}, D.~M., {Harrison}, C.~M., {et~al.} 2017, \mnras,
  472, 2221

\bibitem[{{Stanley} {et~al.}(2015){Stanley}, {Harrison}, {Alexander},
  {Swinbank}, {Aird}, {Del Moro}, {Hickox}, \&
  {Mullaney}}]{2015MNRAS.453..591S}
{Stanley}, F., {Harrison}, C.~M., {Alexander}, D.~M., {et~al.} 2015, \mnras,
  453, 591

\bibitem[{{Steinhardt} {et~al.}(2014){Steinhardt}, {Speagle}, {Capak},
  {Silverman}, {Carollo}, {Dunlop}, {Hashimoto}, {Hsieh}, {Ilbert}, {Le Fevre},
  {Le Floc'h}, {Lee}, {Lin}, {Lin}, {Masters}, {McCracken}, {Nagao}, {Petric},
  {Salvato}, {Sanders}, {Scoville}, {Sheth}, {Strauss}, \&
  {Taniguchi}}]{2014ApJ...791L..25S}
{Steinhardt}, C.~L., {Speagle}, J.~S., {Capak}, P., {et~al.} 2014, \apjl, 791,
  L25

\bibitem[{{Terlevich} \& {Boyle}(1993)}]{1993MNRAS.262..491T}
{Terlevich}, R.~J. \& {Boyle}, B.~J. 1993, \mnras, 262, 491

\bibitem[{{Trakhtenbrot} {et~al.}(2017){Trakhtenbrot}, {Lira}, {Netzer},
  {Cicone}, {Maiolino}, \& {Shemmer}}]{2017ApJ...836....8T}
{Trakhtenbrot}, B., {Lira}, P., {Netzer}, H., {et~al.} 2017, \apj, 836, 8

\bibitem[{{Ulvestad} {et~al.}(2005){Ulvestad}, {Antonucci}, \&
  {Barvainis}}]{2005ApJ...621..123U}
{Ulvestad}, J.~S., {Antonucci}, R. R.~J., \& {Barvainis}, R. 2005, \apj, 621,
  123

\bibitem[{{van Breugel} {et~al.}(2004){van Breugel}, {Fragile}, {Anninos}, \&
  {Murray}}]{2004IAUS..217..472V}
{van Breugel}, W., {Fragile}, C., {Anninos}, P., \& {Murray}, S. 2004, in IAU
  Symposium, Vol. 217, Recycling Intergalactic and Interstellar Matter, ed.
  P.-A. {Duc}, J.~{Braine}, \& E.~{Brinks}, 472

\bibitem[{{van Haarlem} {et~al.}(2013){van Haarlem}, {Wise}, {Gunst}, {Heald},
  {McKean}, {Hessels}, {de Bruyn}, {Nijboer}, {Swinbank}, {Fallows},
  {Brentjens}, {Nelles}, {Beck}, {Falcke}, {Fender}, {H{\"o}randel},
  {Koopmans}, {Mann}, {Miley}, {R{\"o}ttgering}, {Stappers}, {Wijers},
  {Zaroubi}, {van den Akker}, {Alexov}, {Anderson}, {Anderson}, {van Ardenne},
  {Arts}, {Asgekar}, {Avruch}, {Batejat}, {B{\"a}hren}, {Bell}, {Bell}, {van
  Bemmel}, {Bennema}, {Bentum}, {Bernardi}, {Best}, {B{\^i}rzan}, {Bonafede},
  {Boonstra}, {Braun}, {Bregman}, {Breitling}, {van de Brink}, {Broderick},
  {Broekema}, {Brouw}, {Br{\"u}ggen}, {Butcher}, {van Cappellen}, {Ciardi},
  {Coenen}, {Conway}, {Coolen}, {Corstanje}, {Damstra}, {Davies}, {Deller},
  {Dettmar}, {van Diepen}, {Dijkstra}, {Donker}, {Doorduin}, {Dromer}, {Drost},
  {van Duin}, {Eisl{\"o}ffel}, {van Enst}, {Ferrari}, {Frieswijk}, {Gankema},
  {Garrett}, {de Gasperin}, {Gerbers}, {de Geus}, {Grie{\ss}meier}, {Grit},
  {Gruppen}, {Hamaker}, {Hassall}, {Hoeft}, {Holties}, {Horneffer}, {van der
  Horst}, {van Houwelingen}, {Huijgen}, {Iacobelli}, {Intema}, {Jackson},
  {Jelic}, {de Jong}, {Juette}, {Kant}, {Karastergiou}, {Koers}, {Kollen},
  {Kondratiev}, {Kooistra}, {Koopman}, {Koster}, {Kuniyoshi}, {Kramer},
  {Kuper}, {Lambropoulos}, {Law}, {van Leeuwen}, {Lemaitre}, {Loose}, {Maat},
  {Macario}, {Markoff}, {Masters}, {McFadden}, {McKay-Bukowski}, {Meijering},
  {Meulman}, {Mevius}, {Middelberg}, {Millenaar}, {Miller-Jones}, {Mohan},
  {Mol}, {Morawietz}, {Morganti}, {Mulcahy}, {Mulder}, {Munk}, {Nieuwenhuis},
  {van Nieuwpoort}, {Noordam}, {Norden}, {Noutsos}, {Offringa}, {Olofsson},
  {Omar}, {Orr{\'u}}, {Overeem}, {Paas}, {Pandey-Pommier}, {Pandey}, {Pizzo},
  {Polatidis}, {Rafferty}, {Rawlings}, {Reich}, {de Reijer}, {Reitsma},
  {Renting}, {Riemers}, {Rol}, {Romein}, {Roosjen}, {Ruiter}, {Scaife}, {van
  der Schaaf}, {Scheers}, {Schellart}, {Schoenmakers}, {Schoonderbeek},
  {Serylak}, {Shulevski}, {Sluman}, {Smirnov}, {Sobey}, {Spreeuw}, {Steinmetz},
  {Sterks}, {Stiepel}, {Stuurwold}, {Tagger}, {Tang}, {Tasse}, {Thomas},
  {Thoudam}, {Toribio}, {van der Tol}, {Usov}, {van Veelen}, {van der Veen},
  {ter Veen}, {Verbiest}, {Vermeulen}, {Vermaas}, {Vocks}, {Vogt}, {de Vos},
  {van der Wal}, {van Weeren}, {Weggemans}, {Weltevrede}, {White}, {Wijnholds},
  {Wilhelmsson}, {Wucknitz}, {Yatawatta}, {Zarka}, {Zensus}, \& {van
  Zwieten}}]{2013A&A...556A...2V}
{van Haarlem}, M.~P., {Wise}, M.~W., {Gunst}, A.~W., {et~al.} 2013, \aap, 556,
  A2

\bibitem[{{van Weeren} {et~al.}(2014){van Weeren}, {Williams}, {Tasse},
  {R{\"o}ttgering}, {Rafferty}, {van der Tol}, {Heald}, {White}, {Shulevski},
  {Best}, {Intema}, {Bhatnagar}, {Reich}, {Steinmetz}, {van Velzen},
  {En{\ss}lin}, {Prandoni}, {de Gasperin}, {Jamrozy}, {Brunetti}, {Jarvis},
  {McKean}, {Wise}, {Ferrari}, {Harwood}, {Oonk}, {Hoeft},
  {Kunert-Bajraszewska}, {Horellou}, {Wucknitz}, {Bonafede}, {Mohan}, {Scaife},
  {Kl{\"o}ckner}, {van Bemmel}, {Merloni}, {Chyzy}, {Engels}, {Falcke},
  {Pandey-Pommier}, {Alexov}, {Anderson}, {Avruch}, {Beck}, {Bell}, {Bentum},
  {Bernardi}, {Breitling}, {Broderick}, {Brouw}, {Br{\"u}ggen}, {Butcher},
  {Ciardi}, {de Geus}, {de Vos}, {Deller}, {Duscha}, {Eisl{\"o}ffel},
  {Fallows}, {Frieswijk}, {Garrett}, {Grie{\ss}meier}, {Gunst}, {Hamaker},
  {Hassall}, {H{\"o}randel}, {van der Horst}, {Iacobelli}, {Jackson}, {Juette},
  {Kondratiev}, {Kuniyoshi}, {Maat}, {Mann}, {McKay-Bukowski}, {Mevius},
  {Morganti}, {Munk}, {Offringa}, {Orr{\`u}}, {Paas}, {Pandey}, {Pietka},
  {Pizzo}, {Polatidis}, {Renting}, {Rowlinson}, {Schwarz}, {Serylak}, {Sluman},
  {Smirnov}, {Stappers}, {Stewart}, {Swinbank}, {Tagger}, {Tang}, {Thoudam},
  {Toribio}, {Vermeulen}, {Vocks}, \& {Zarka}}]{2014ApJ...793...82V}
{van Weeren}, R.~J., {Williams}, W.~L., {Tasse}, C., {et~al.} 2014, \apj, 793,
  82

\bibitem[{{Venemans} {et~al.}(2007){Venemans}, {R{\"o}ttgering}, {Miley}, {van
  Breugel}, {de Breuck}, {Kurk}, {Pentericci}, {Stanford}, {Overzier}, {Croft},
  \& {Ford}}]{2007A&A...461..823V}
{Venemans}, B.~P., {R{\"o}ttgering}, H.~J.~A., {Miley}, G.~K., {et~al.} 2007,
  \aap, 461, 823

\bibitem[{{Viero} {et~al.}(2015){Viero}, {Moncelsi}, {Quadri}, {B{\'e}thermin},
  {Bock}, {Burgarella}, {Chapman}, {Clements}, {Conley}, {Conversi},
  {Duivenvoorden}, {Dunlop}, {Farrah}, {Franceschini}, {Halpern}, {Ivison},
  {Lagache}, {Magdis}, {Marchetti}, {{\'A}lvarez-M{\'a}rquez}, {Marsden},
  {Oliver}, {Page}, {P{\'e}rez-Fournon}, {Schulz}, {Scott}, {Valtchanov},
  {Vieira}, {Wang}, {Wardlow}, \& {Zemcov}}]{2015ApJ...809L..22V}
{Viero}, M.~P., {Moncelsi}, L., {Quadri}, R.~F., {et~al.} 2015, \apjl, 809, L22

\bibitem[{{Wals} {et~al.}(2005){Wals}, {Boyle}, {Croom}, {Miller}, {Smith},
  {Shanks}, \& {Outram}}]{2005MNRAS.360..453W}
{Wals}, M., {Boyle}, B.~J., {Croom}, S.~M., {et~al.} 2005, \mnras, 360, 453

\bibitem[{{Wardlow} {et~al.}(2011){Wardlow}, {Smail}, {Coppin}, {Alexand er},
  {Brandt}, {Danielson}, {Luo}, {Swinbank}, {Walter}, {Wei{\ss}}, {Xue},
  {Zibetti}, {Bertoldi}, {Biggs}, {Chapman}, {Dannerbauer}, {Dunlop},
  {Gawiser}, {Ivison}, {Knudsen}, {Kov{\'a}cs}, {Lacey}, {Menten}, {Padilla},
  {Rix}, \& {van der Werf}}]{2011MNRAS.415.1479W}
{Wardlow}, J.~L., {Smail}, I., {Coppin}, K.~E.~K., {et~al.} 2011, \mnras, 415,
  1479

\bibitem[{{Whitaker} {et~al.}(2012){Whitaker}, {van Dokkum}, {Brammer}, \&
  {Franx}}]{2012ApJ...754L..29W}
{Whitaker}, K.~E., {van Dokkum}, P.~G., {Brammer}, G., \& {Franx}, M. 2012,
  \apjl, 754, L29

\bibitem[{{White} {et~al.}(2007){White}, {Helfand}, {Becker}, {Glikman}, \& {de
  Vries}}]{2007ApJ...654...99W}
{White}, R.~L., {Helfand}, D.~J., {Becker}, R.~H., {Glikman}, E., \& {de
  Vries}, W. 2007, \apj, 654, 99

\bibitem[{{White} {et~al.}(2015){White}, {Jarvis}, {H{\"a}u{\ss}ler}, \&
  {Maddox}}]{2015MNRAS.448.2665W}
{White}, S.~V., {Jarvis}, M.~J., {H{\"a}u{\ss}ler}, B., \& {Maddox}, N. 2015,
  \mnras, 448, 2665

\bibitem[{{White} {et~al.}(2017){White}, {Jarvis}, {Kalfountzou}, {Hardcastle},
  {Verma}, {Cao Orjales}, \& {Stevens}}]{2017MNRAS.468..217W}
{White}, S.~V., {Jarvis}, M.~J., {Kalfountzou}, E., {et~al.} 2017, \mnras, 468,
  217

\bibitem[{{Williams} {et~al.}(2013){Williams}, {Intema}, \&
  {R{\"o}ttgering}}]{2013AA...549A..55W}
{Williams}, W.~L., {Intema}, H.~T., \& {R{\"o}ttgering}, H.~J.~A. 2013, \aap,
  549, A55

\bibitem[{{Williams} {et~al.}(2016){Williams}, {van Weeren}, {R{\"o}ttgering},
  {Best}, {Dijkema}, {de Gasperin}, {Hardcastle}, {Heald}, {Prandoni},
  {Sabater}, {Shimwell}, {Tasse}, {van Bemmel}, {Br{\"u}ggen}, {Brunetti},
  {Conway}, {En{\ss}lin}, {Engels}, {Falcke}, {Ferrari}, {Haverkorn},
  {Jackson}, {Jarvis}, {Kapi{\'n}ska}, {Mahony}, {Miley}, {Morabito},
  {Morganti}, {Orr{\'u}}, {Retana-Montenegro}, {Sridhar}, {Toribio}, {White},
  {Wise}, \& {Zwart}}]{2016MNRAS.460.2385W}
{Williams}, W.~L., {van Weeren}, R.~J., {R{\"o}ttgering}, H.~J.~A., {et~al.}
  2016, \mnras, 460, 2385

\bibitem[{{Wright} {et~al.}(2010){Wright}, {Eisenhardt}, {Mainzer}, {Ressler},
  {Cutri}, {Jarrett}, {Kirkpatrick}, {Padgett}, {McMillan}, {Skrutskie},
  {Stanford}, {Cohen}, {Walker}, {Mather}, {Leisawitz}, {Gautier}, {McLean},
  {Benford}, {Lonsdale}, {Blain}, {Mendez}, {Irace}, {Duval}, {Liu}, {Royer},
  {Heinrichsen}, {Howard}, {Shannon}, {Kendall}, {Walsh}, {Larsen}, {Cardon},
  {Schick}, {Schwalm}, {Abid}, {Fabinsky}, {Naes}, \&
  {Tsai}}]{2010AJ....140.1868W}
{Wright}, E.~L., {Eisenhardt}, P.~R.~M., {Mainzer}, A.~K., {et~al.} 2010, \aj,
  140, 1868

\bibitem[{{Yao} {et~al.}(2019){Yao}, {Wu}, {Ai}, {Yang}, {Yang}, {Dong},
  {Joshi}, {Wang}, {Feng}, {Fu}, {Hou}, {Luo}, {Kong}, {Liu}, {Zhao}, {Zhang},
  {Yuan}, \& {Shen}}]{2019ApJS..240....6Y}
{Yao}, S., {Wu}, X.-B., {Ai}, Y.~L., {et~al.} 2019, \apjs, 240, 6

\bibitem[{{Yue} {et~al.}(2018){Yue}, {Jiang}, {Shen}, {Hall}, {Yu},
  {Schneider}, {Ho}, {Horne}, {Petitjean}, \& {Trump}}]{2018ApJ...863...21Y}
{Yue}, M., {Jiang}, L., {Shen}, Y., {et~al.} 2018, \apj, 863, 21

\bibitem[{{Yun} {et~al.}(2001){Yun}, {Reddy}, \&
  {Condon}}]{2001ApJ...554..803Y}
{Yun}, M.~S., {Reddy}, N.~A., \& {Condon}, J.~J. 2001, \apj, 554, 803

\bibitem[{{Zakamska} \& {Greene}(2014)}]{2014MNRAS.442..784Z}
{Zakamska}, N.~L. \& {Greene}, J.~E. 2014, \mnras, 442, 784

\bibitem[{{Zakamska} {et~al.}(2016){Zakamska}, {Lampayan}, {Petric}, {Dicken},
  {Greene}, {Heckman}, {Hickox}, {Ho}, {Krolik}, {Nesvadba}, {Strauss},
  {Geach}, {Oguri}, \& {Strateva}}]{2016MNRAS.455.4191Z}
{Zakamska}, N.~L., {Lampayan}, K., {Petric}, A., {et~al.} 2016, \mnras, 455,
  4191

\bibitem[{Zhang {et~al.}(2013)Zhang, Dietrich, McKay, Sheldon, \&
  Nguyen}]{Zhang_2013}
Zhang, Y., Dietrich, J.~P., McKay, T.~A., Sheldon, E.~S., \& Nguyen, A. T.~Q.
  2013, The Astrophysical Journal, 773, 115

\bibitem[{{Zubovas} \& {King}(2012)}]{2012ApJ...745L..34Z}
{Zubovas}, K. \& {King}, A. 2012, \apjl, 745, L34

\end{thebibliography}


\begin{acknowledgements}

ERM acknowledges financial support from NWO Top project, No. 614.001.006. \\

We thank the anonymous referee for the helpful comments that improved this work.  \\

LOFAR, the Low Frequency Array designed and constructed by ASTRON, has facilities in several countries, that are owned by various parties (each with their own funding sources), and that are collectively operated by the International LOFAR Telescope (ILT) foundation under a joint scientific policy. The Open University is incorporated by Royal Charter (RC 000391), an exempt charity in England \& Wales and a charity registered in Scotland (SC 038302). The Open University is authorized and regulated by the Financial Conduct Authority.\\

Herschel is an ESA space observatory with science instruments provided by European-led Principal Investigator consortia and with important participation from NASA. Herschel/SPIRE has been developed by a consortium of institutes led by Cardiff University (UK) and including Univ. Lethbridge (Canada); NAOC (China); CEA, LAM (France); IFSI, Univ. Padua (Italy); IAC (Spain); Stockholm Observatory (Sweden); Imperial College London, RAL, UCL-MSSL, UKATC, Univ. Sussex (UK); and Caltech, JPL, NHSC, Univ. Colorado (USA). This development has been supported by national funding agencies: CSA (Canada); NAOC (China); CEA, CNES, CNRS (France); ASI (Italy); MCINN (Spain); SNSB (Sweden); STFC, UKSA (UK); and NASA (USA). HIPE is a joint development (are joint developments) by the Herschel Science Ground Segment Consortium, consisting of ESA, the NASA Herschel Science Center, and the HIFI, PACS and SPIRE consortia. 
\\
This research has made use of data from HerMES project (http://hermes.sussex.ac.uk/). HerMES is a Herschel Key Programme utilising Guaranteed Time from the SPIRE instrument team, ESAC scientists and a mission scientist. The HerMES data was accessed through the Herschel Database in Marseille (HeDaM - http://hedam.lam.fr) operated by CeSAM and hosted by the Laboratoire d'Astrophysique de Marseille.
\\
The National Radio Astronomy Observatory is a facility of the National Science Foundation operated under cooperative agreement by Associated Universities, Inc. CIRADA is funded by a grant from the Canada Foundation for Innovation 2017 Innovation Fund (Project 35999), as well as by the Provinces of Ontario, British Columbia, Alberta, Manitoba and Quebec.
\\
This publication makes use of data products from the Wide-field Infrared Survey Explorer, which is a joint project of the University of California, Los Angeles, and the Jet Propulsion Laboratory/California Institute of Technology, funded by the National Aeronautics and Space Administration.

\end{acknowledgements}

\begin{appendix} \section{Median-stacked infrared and radio maps of  LOFAR radio-detected and LOFAR radio-undetected quasars}

\begin{figure*}[tp]
\centering{}\includegraphics[scale=0.33]{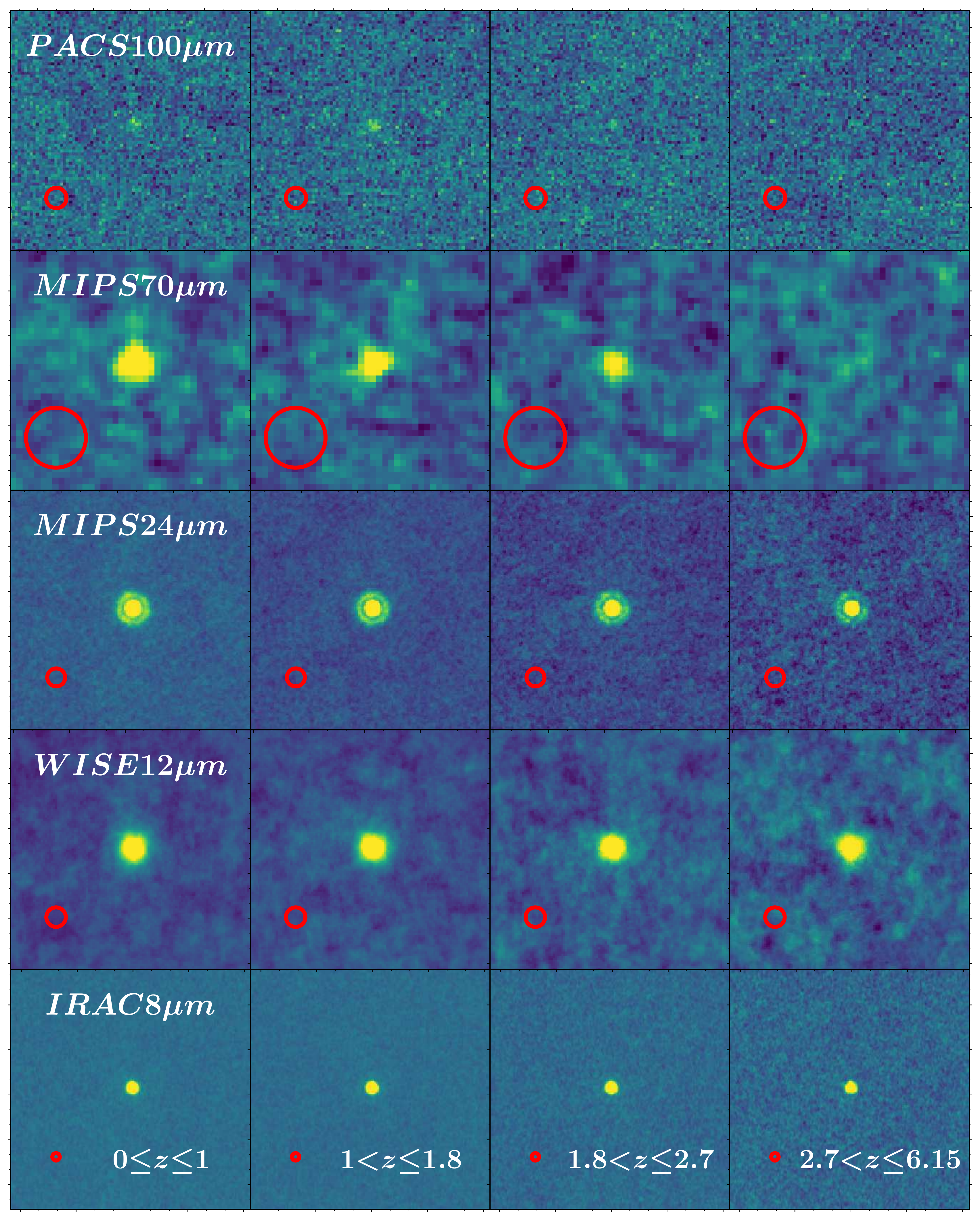}
\centering\caption{\label{fig:infrared_postages_und_1} Central area of the median stacked
infrared maps (IRAC, WISE, MIPS, and PACS) of LOFAR radio-undetected
quasars. The number of quasars stacked in each is indicated in Table
\ref{fig:noise_median_qsos}. The size of the maps is $80^{\prime\prime}\times80$.
The corresponding synthesized beam size is indicated in the left corner.
The color scale ranges from $-3\sigma_{L}$ to $5\sigma_{L}$ , where
$\sigma_{L}$ is the local rms noise.}
\end{figure*}
 
\begin{figure*}[tp]
\centering{}\includegraphics[bb=0bp 0bp 981bp 981bp,clip,scale=0.33]{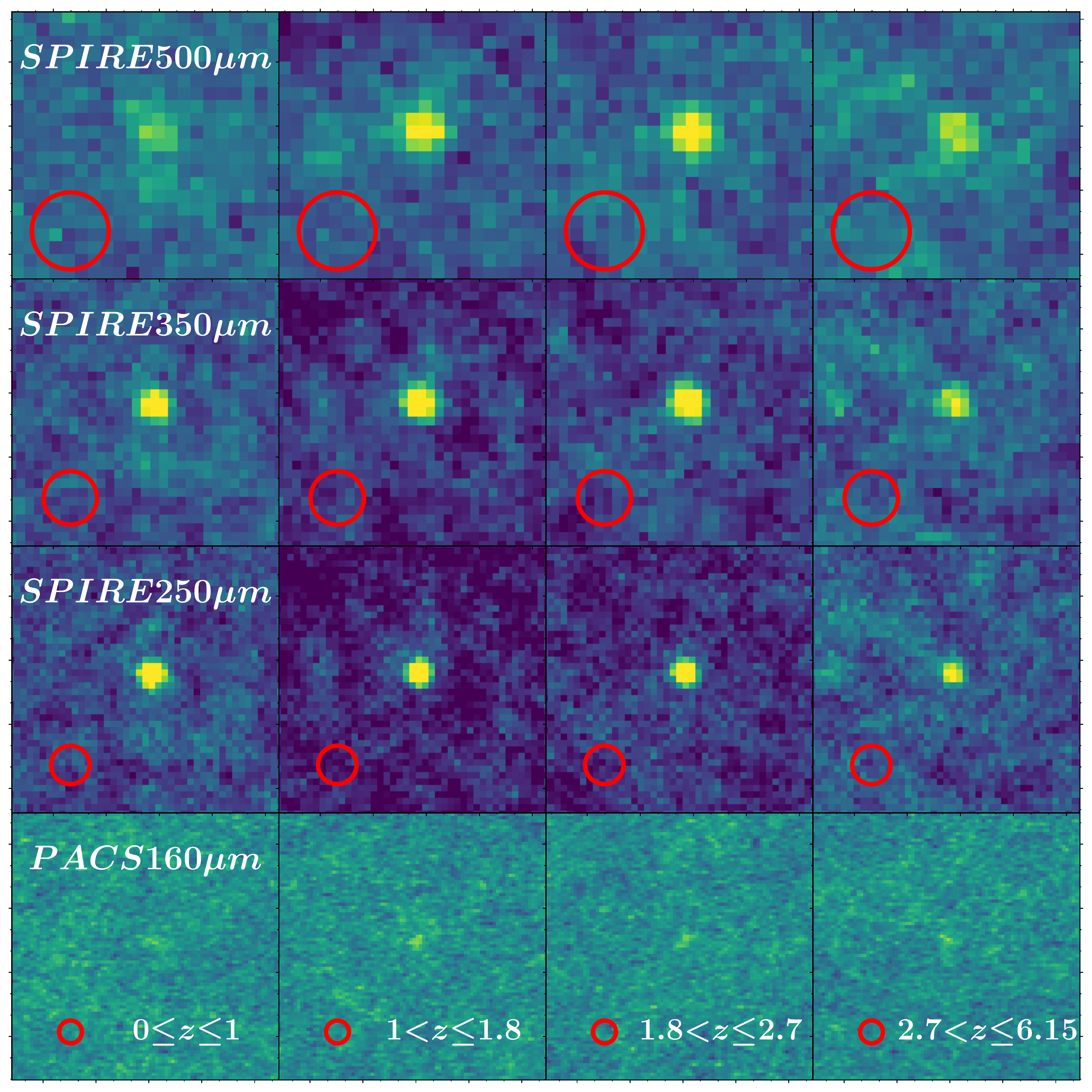}\centering\caption{\label{fig:infrared_postages_und_2} Central area of the median stacked
infrared maps (PACS, and SPIRE) of LOFAR radio-undetected quasars.
The number of quasars stacked in each is indicated in Table \ref{fig:noise_median_qsos}.
The size of the maps is $125^{\prime\prime}\times125^{\prime\prime}$.
The corresponding synthesized beam size is indicated in the left corner.
The color scale varies from $-3\sigma_{L}$ to $5\sigma_{L}$ , where
$\sigma_{L}$ is the local rms noise.}
\end{figure*}
 
\begin{figure*}[tp]
\centering{}\includegraphics[clip,scale=0.33]{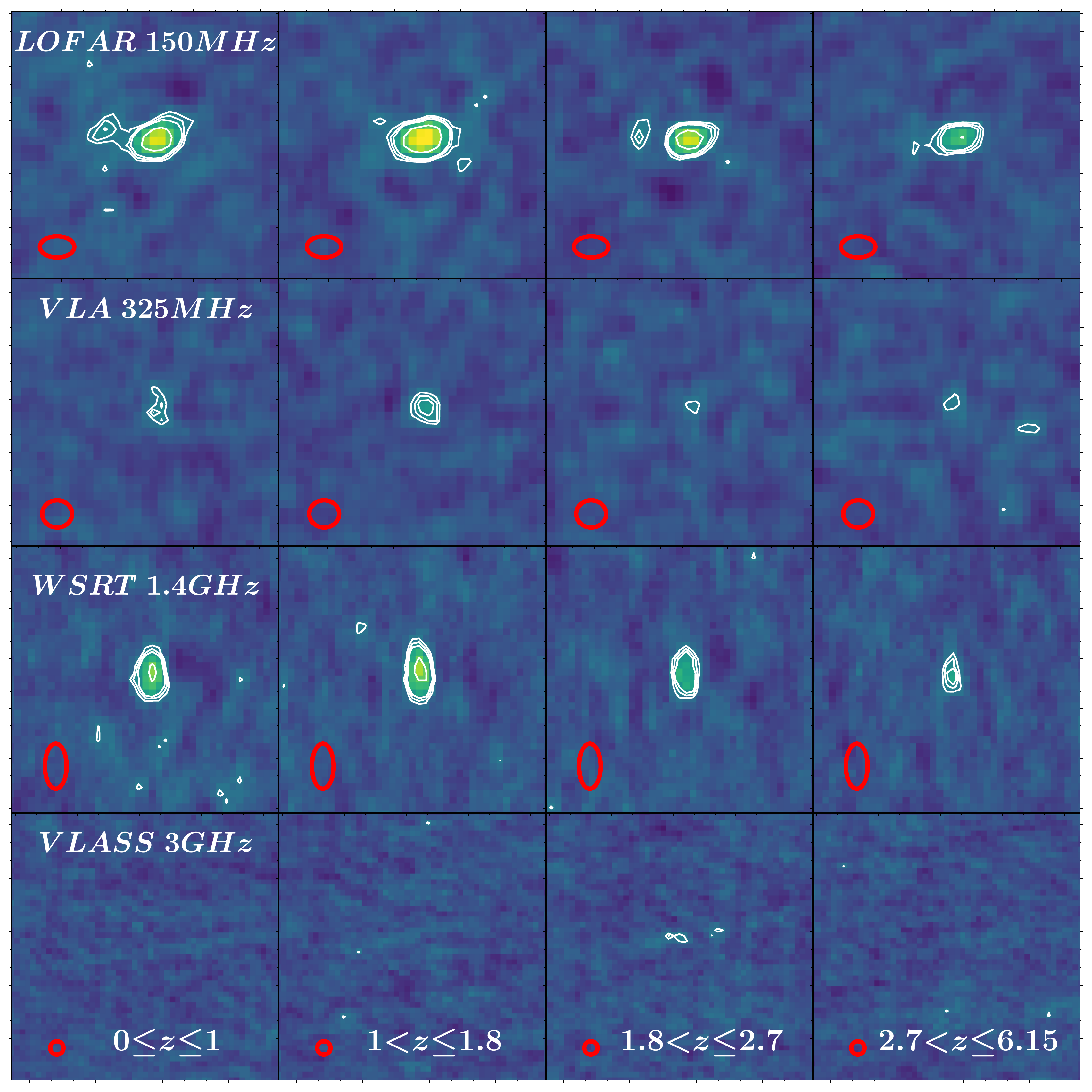}\centering\caption{\label{fig:radio_postages_und} Central area of the median stacked
radio maps of LOFAR radio-undetected quasars. The number of quasars
stacked in each is indicated in Table \ref{fig:noise_median_qsos}.
The size of the maps is $25^{\prime\prime}\times25^{\prime\prime}$,
except the WSRT map which has a size of $80^{\prime\prime}\times80^{\prime\prime}$.
The corresponding synthesized beam size is indicated in the left corner.
The color scale varies from $-5\sigma_{L}$ to $15\sigma_{L}$ , where
$\sigma_{L}$ is the local rms noise. The contours are drawn at $[3,4,5,10]\times\sigma_{L}$
times the local rms noise level.}
\end{figure*}
 
\begin{figure*}[tp]
\centering{}\includegraphics[clip,scale=0.33]{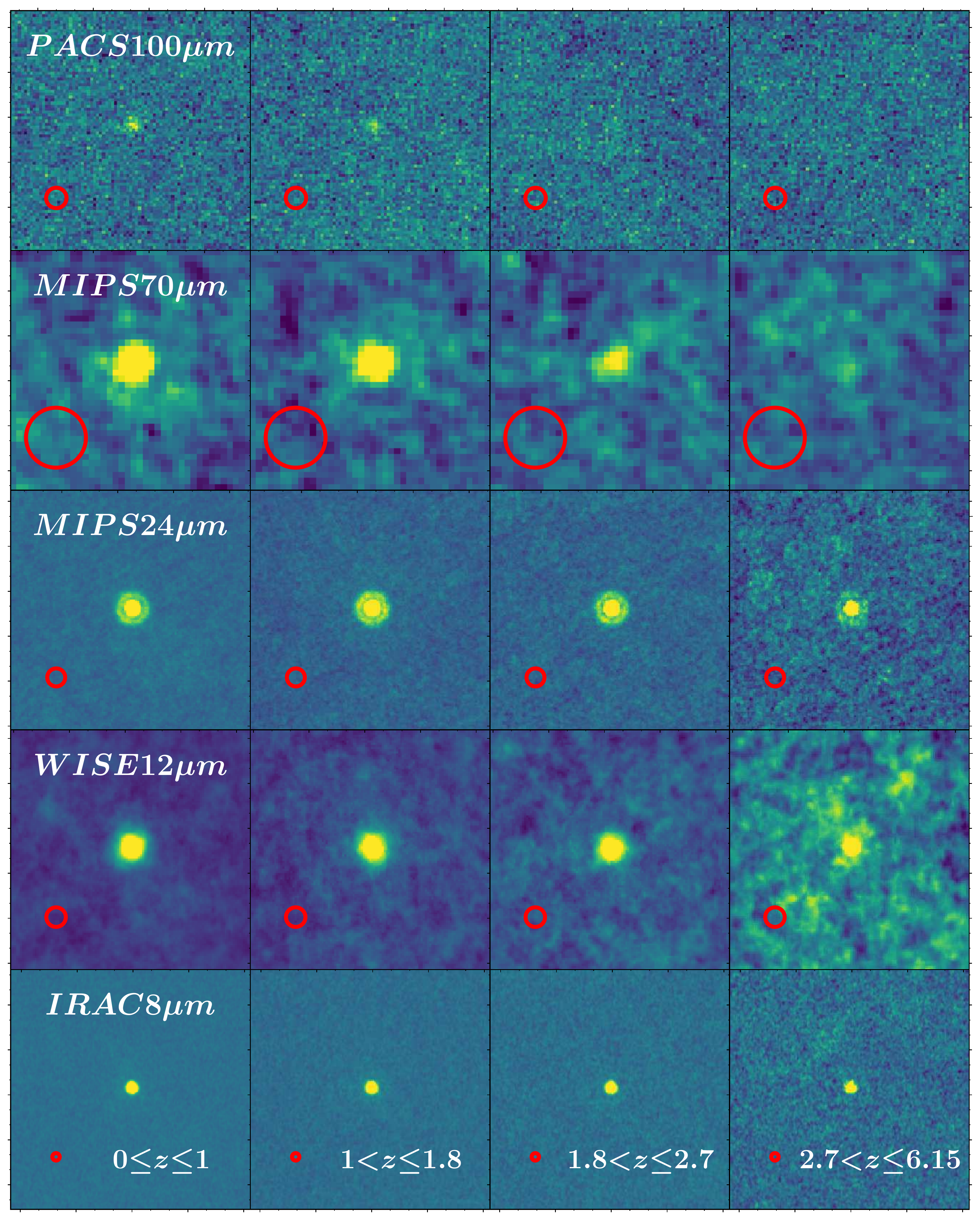}\centering\caption{\label{fig:infrared_postages_dect_1} Central area of the median stacked
infrared maps (IRAC, WISE, MIPS, and PACS) of LOFAR radio-detected
quasars. The number of quasars stacked in each is indicated in Table
\ref{fig:noise_median_qsos}. The size of the maps is $80^{\prime\prime}\times80$.
The corresponding synthesized beam size is indicated in the left corner.
The color scale ranges from $-3\sigma_{L}$ to $5\sigma_{L}$ , where
$\sigma_{L}$ is the local rms noise.}
\end{figure*}
 
\begin{figure*}[tp]
\centering{}\includegraphics[bb=0bp 0bp 981bp 981bp,clip,scale=0.33]{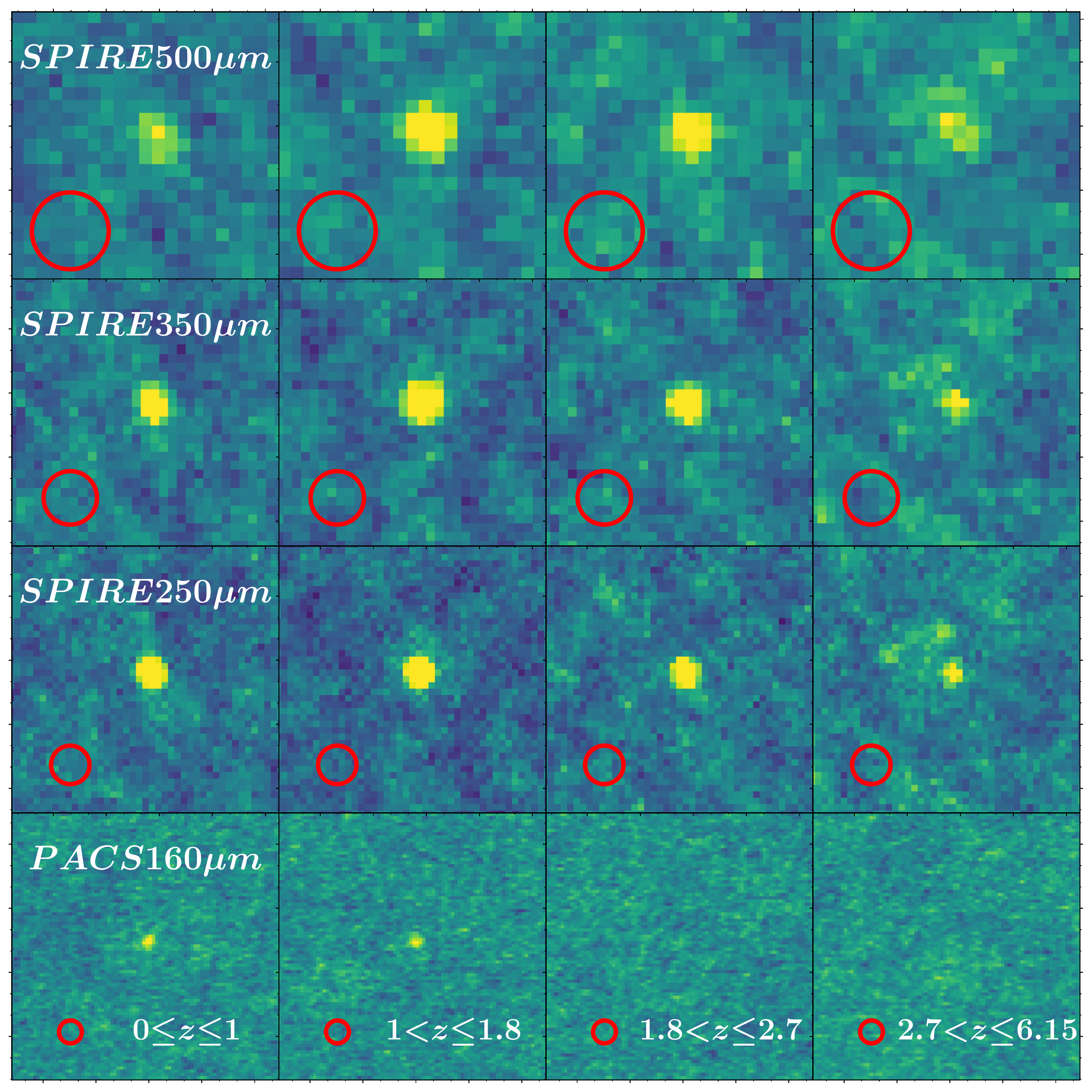}\centering\caption{\label{fig:infrared_postages_dect} Central area of the median stacked
infrared maps (PACS, SPIRE) of LOFAR radio-detected quasars. The number
of quasars stacked in each is indicated in Table \ref{fig:noise_median_qsos}..
The size of the maps is $125^{\prime\prime}\times125^{\prime\prime}$.
The corresponding synthesized beam size is indicated in the left corner.
The color scale ranges from $-3\sigma_{L}$ to $5\sigma_{L}$ , where
$\sigma_{L}$ is the local rms noise.}
\end{figure*}
 
\begin{figure*}[tp]
\centering{}\includegraphics[clip,scale=0.33]{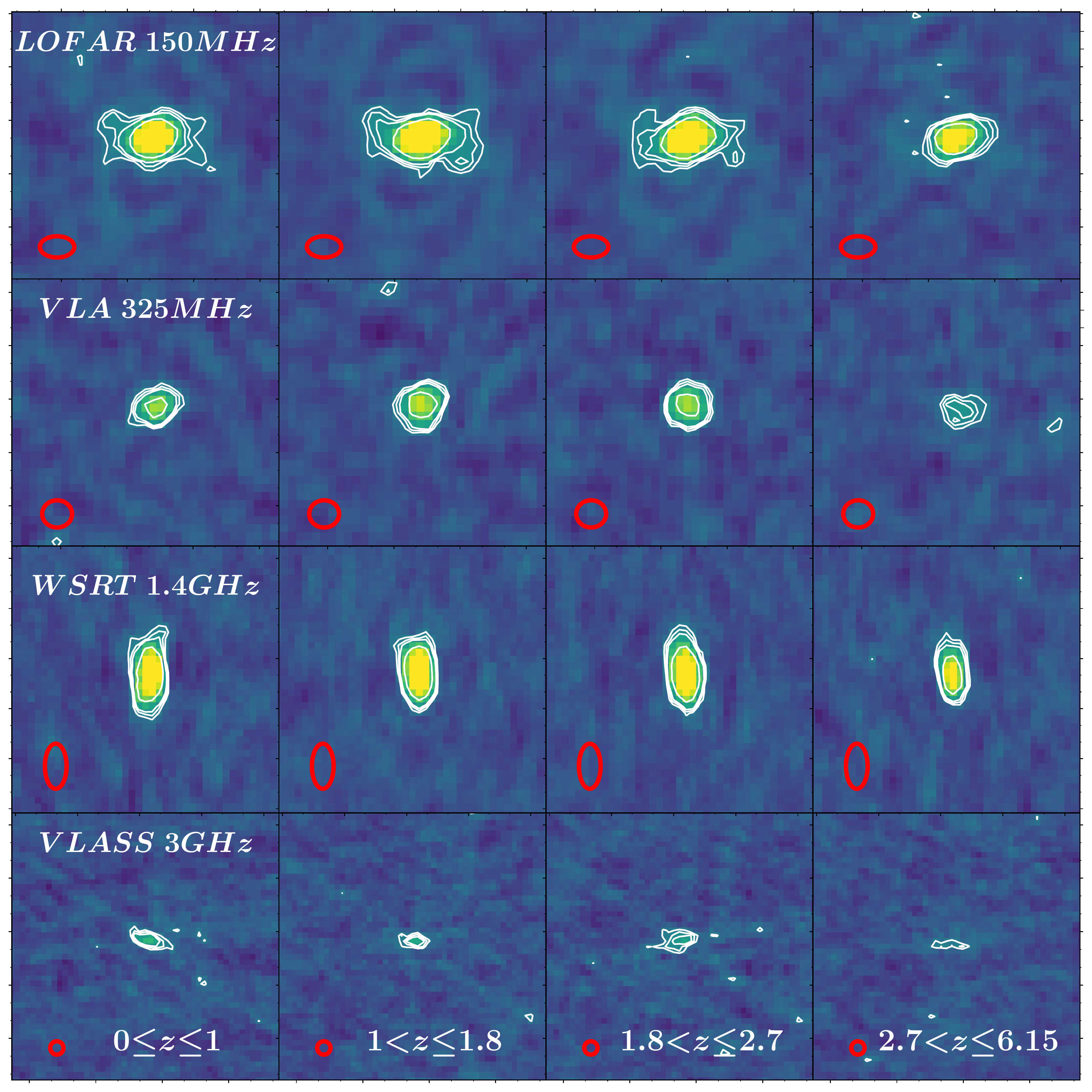}\centering\caption{\label{fig:radio_postages_dect}Central area of the median stacked
radio maps of LOFAR radio-detected quasars. The number of quasars
stacked in each is indicated in Table \ref{fig:noise_median_qsos}.
The size of the maps is $25^{\prime\prime}\times25^{\prime\prime}$,
except the WSRT map which has a size of $80^{\prime\prime}\times80^{\prime\prime}$.
The corresponding synthesized beam size is indicated in the left corner.
The color scale ranges from $-5\sigma_{L}$ to $15\sigma_{L}$ , where
$\sigma_{L}$ is the local rms noise.}
\end{figure*}

\section{Infrared photometry of median LOFAR radio-detected and LOFAR radio-undetected quasars stacked according to redshift}
 \label{sec:appendix_A}

\noindent 
\begin{sidewaystable*}[tp]
\begin{centering}
\begin{tabular}{cccccccccc}
\hline 
$z_{bin}$  & ${8{\mu}\textrm{m}}$  & ${12{\mu}\textrm{m}}$  & ${24{\mu}\textrm{m}}$  & ${70{\mu}\textrm{m}}$  & ${100{\mu}\textrm{m}}$  & ${160{\mu}\textrm{m}}$  & ${250{\mu}\textrm{m}}$  & ${350{\mu}\textrm{m}}$  & ${500{\mu}\textrm{m}}$ \tabularnewline
\hline 
RDQs & {[}mJy{]} & {[}mJy{]} & {[}mJy{]} & {[}mJy{]} & {[}mJy{]} & {[}mJy{]} & {[}mJy{]} & {[}mJy{]} & {[}mJy{]}\tabularnewline
\hline 
{\small{}$0.0\leq z\leq1.0$} & 0.61$\pm$0.27  & 0.87$\pm$0.03  & 1.80$\pm$1.1  & 8.00$\pm$1.0  & 9.43$\pm$1.4  & 22.52$\pm$5.4  & 20.67$\pm$1.5  & 13.65$\pm$2.0  & 7.20$\pm$1.2 \tabularnewline
{\small{}$1.0<z\leq1.8$} & 0.32$\pm$0.08  & 0.40$\pm$0.03  & 0.91$\pm$0.3  & 5.44$\pm$0.7  & 6.15$\pm$1.2  & 16.47$\pm$1.7  & 19.93$\pm$2.3  & 16.44$\pm$2.4  & 10.91$\pm$0.5 \tabularnewline
{\small{}$1.8<z\leq2.7$} & 0.26$\pm$0.08  & 0.36$\pm$0.03  & 0.92$\pm$0.2  & $<$5.07  & $<$7.60  & 4.23$\pm$3.6  & 18.97$\pm$2.3  & 19.04$\pm$1.0  & 11.79$\pm$3.9 \tabularnewline
{\small{}$2.7<z\leq6.15$} & 0.13$\pm$0.05  & 0.20$\pm$0.13  & 0.53$\pm$0.2  & $<$3.67  & $<$5.65  & 3.14$\pm$4.4  & 15.92$\pm$11.6  & 9.20$\pm$3.6  & 10.36$\pm$1.5 \tabularnewline
\hline 
RUQs &  &  &  &  &  &  &  &  & \tabularnewline
\hline 
{\small{}$0.0\leq z\leq1.0$} & 0.21$\pm$0.04  & 0.30$\pm$0.02  & 0.60$\pm$0.1  & 2.87$\pm$0.4  & 3.09$\pm$0.7  & 4.98$\pm$1.6  & 8.34$\pm$0.6  & 6.28$\pm$1.2  & 2.36$\pm$0.9 \tabularnewline
{\small{}$1.0<z\leq1.8$} & 0.12$\pm$0.01  & 0.16$\pm$0.01  & 0.34$\pm$0.1  & 1.21$\pm$0.2  & 2.20$\pm$0.3  & 3.63$\pm$1.0  & 5.16$\pm$0.4  & 4.69$\pm$0.5  & 2.83$\pm$0.7 \tabularnewline
{\small{}$1.8<z\leq2.7$} & 0.09$\pm$0.01  & 0.14$\pm$0.01  & 0.27$\pm$0.1  & 0.97$\pm$0.2  & 0.97$\pm$0.5  & 3.10$\pm$1.2  & 5.51$\pm$0.5  & 5.65$\pm$0.6  & 4.11$\pm$0.5 \tabularnewline
{\small{}$2.7<z\leq6.15$} & 0.09$\pm$0.01  & 0.16$\pm$0.01  & 0.27$\pm$0.1  & 0.55$\pm$0.3  & $<$2.11  & 7.04$\pm$1.5  & 5.66$\pm$1.0  & 5.46$\pm$1.2  & 4.65$\pm$0.8 \tabularnewline
\hline 
\end{tabular}
\par\end{centering}
\centering{}\centering\caption{Infrared fluxes of the median LOFAR radio-detected quasars (RDQs)
and LOFAR radio-undetected quasars (RUQs) stacked according to redshift
between 8$\mu\textrm{m}$ and 500$\mu\textrm{m}$. Values without
error bars are upper limits estimated at the center of the median
stacked images. \label{fig:infrared_fluxes_median_qsos} }
\end{sidewaystable*}

\end{appendix}

\end{document}